\documentclass[twocolumn,amsmath,amssymb,10pt,a4paper]{revtex4}
\pdfoutput=1

\usepackage{hyperref}

\hypersetup{
	colorlinks=true,
	linktoc=all,
	linkcolor=violet,
	urlcolor=blue
}

\usepackage{geometry}
\usepackage{amsmath, amssymb}
\usepackage{graphicx}
\usepackage{bbold}
\usepackage{algorithm}
\usepackage{algpseudocode}
\usepackage{amsthm}
\usepackage{mathtools}
 
\theoremstyle{definition}
 
\theoremstyle{remark}

\newcommand{\be}{\begin{equation}}
\newcommand{\ee}{\end{equation}}
\newcommand{\bd}{\begin{displaymath}}
\newcommand{\ed}{\end{displaymath}}
\newcommand{\BE}{\begin{eqnarray}}
\newcommand{\EE}{\end{eqnarray}}

\newcommand{\bra}{\left\langle}
\newcommand{\ket}{\right\rangle}

\newcommand{\id}{{\rm 1\!\!I}}

\newcommand{\bu}{\ensuremath{\mathbf{u}}}
\newcommand{\bv}{\ensuremath{\mathbf{v}}}

\newcommand{\bx}{\ensuremath{\mathbf{x}}}

\newcommand{\bp}{\ensuremath{\mathbf{p}}}

\newcommand{\tr}{\mathrm{Tr}}

\graphicspath{{../../figs/}}

\usepackage{epigraph,varwidth}
\usepackage{etoolbox}
\usepackage{lipsum}

\usepackage{tcolorbox}
\newtcolorbox{mybox}{colback=green!5!white,colframe=green!75!black}

\begin{document}

\title{Modeling observers as physical systems representing the world from within: Quantum theory as a physical and self-referential theory of inference}

\author{John Realpe-G\'omez$^1$}\email{john.realpe@gmail.com}
\affiliation{Theoretical Physics Group, School of Physics and Astronomy, The University of Manchester\footnote{A substantial part of this work was developed while being Research Associate at The University of Manchester.}, Manchester M13 9PL, United Kingdom}
\affiliation{Instituto de Matem\'aticas Aplicadas, Universidad de Cartagena, Bol\'ivar 130001, Colombia}

\date{\today}

\begin{abstract}
In 1929 Szilard pointed out that the physics of the observer may play a role in the analysis of experiments. The same year, Bohr pointed out that complementarity appears to arise naturally in psychology where both the objects of perception and the perceiving subject belong to `our mental content'. Here we argue that the formalism of quantum theory can be derived from two related intuitive principles:  (i) inference is a classical physical process performed by classical physical systems, observers, which are part of the experimental setup---this implies non-commutativity and imaginary-time quantum mechanics; (ii) experiments must be described from a first-person perspective---this leads to self-reference, complementarity, and a quantum dynamics that is the iterative construction of the  observer's subjective state. This approach suggests a natural explanation for the origin of Planck's constant as due to the physical interactions supporting the observer's information processing, and sheds new light on some conceptual issues associated to the foundations of quantum theory. It also suggests that fundamental equations in physics are typically of second order, instead of the more parsimonious first-order equations, due to the physical nature of the observer. It furthermore suggests some experimental conjectures: (i) the quantum of action could be understood as the result of the additional energy required to transition from unconscious to conscious perception---this is consistent with available experimental data; (ii) humans can observe a single photon of visible light---this is related to (i) and is consistent with existing psychophysics experiments; (iii) the neural correlates of the self are composed of two complementary sub-processes that essentially model each other, much like the DNA molecule is composed of two strands that essentially produce a copy of each other---this may help explain why the brain is divided into hemispheres and suggests self-aware systems should have a similar architecture. Moreover, by explicitly and consistently incorporating us observers and our everyday first-person perspective into the foundations of physics, this approach may help bridge the gap between science and human experience. We discuss the potential implications of these ideas for the modern research program on consciousness championed by Nobel laureate Francis Crick and the emerging field of contemplative science. As side results: (i) we show that message-passing algorithms and stochastic processes can be written in a quantum-like manner---this may suggest novel ways to simulate quantum systems with message-passing algorithms or to naturally implement these powerful distributed algorithms on quantum computers; (ii) we provide evidence that non-stoquasticity, a quantum computational resource, in some cases may be related to non-equilibrium phenomena---this suggests that some of the potential advantage of quantum computers associated to non-stoquasticity may be related to the type of computational advantages recently observed in non-equilibrium Monte Carlo methods where detailed balance is broken; (iii) we provide a different Hamiltonian function for a quantum particle in a classical electromagnetic field---this may suggest a probabilistic interpretation of electromagnetic phenomena. 
\end{abstract}

\maketitle

\begin{epigraphs}
\qitem{\it ``Describe the real factual situation.''}{\textup{Albert Einstein}}
\end{epigraphs}

\tableofcontents

\section{Introduction}\label{s:intro}
Perhaps some of the most difficult transitions in the evolution of our understanding of the universe have been those that removed our special status in some way---like the resistance against the concept that our planet is not the center of the universe, attributed to Copernicus, or against the concept that we are not as different as we thought from other animals, attributed to Darwin. Yet history has taught us again and again that once we surrender and accept the new status a previously hidden simplicity suddenly emerges. 

In part because our subjective biases are often misleading, we have usually made an effort to keep the subjective, ourselves, out of our picture of the universe in search of an objective reality. Even studies of the human brain have mostly focused on a third-person perspective (see Fig.~\ref{f:perspectives}a), i.e. scientist usually study others' brains, not their own. This has granted us the special status of being able to understand the world as if we were not part of it, independently of our everyday human experience. However, at the same time that we gained the special status of doing science without the scientist, we also created a deep tension between science and human experience (see below).

This work is a kind invitation to reconsider the resistance that mainstream physics has understandably developped against the role that human experience might play on the foundations of science. Such type of invitation is not new, of course. Indeed, a similar invitation made twenty five years ago by Varela, Thompson, and Rosch~\cite{varela2017embodied} have proved very fruitful for cognitive science. Let Varela, Thompson, and Rosch clearly express the tension between science and human experience mentioned above (see Fig.~\ref{f:hierarchy}):

\epigraph{\em ``In our present world science is so dominant that we give it the authority to explain even when it denies what is most immediate and direct---our everyday, immediate experience. Thus most people would hold as a fundamental truth the scientific account of matter/space as collections of atomic particles, while treating what is given in their immediate experience, with all of its richness, as less profound and ture. Yet when we relax into the immediate bodily well-being of a sunny day or of the bodily tension of anxiously running to catch a bus, such accounts of space/matter fade into the background as abstract and secondary. [...] 

``To deny the truth of our own experience in the scientific study of ourselves is not only unsatisfactory; it is to render the scientific study of ourselves without a subject matter. But to suppose that science cannot contribute to an understanding of our experience may be to abandon, within the modern context, the task of self-understanding. Experience and scientific understanding are like two legs without which we cannot walk.''}{F. Varela, E. Thompson, E. Rosch, Ref.~\cite{varela2017embodied} (pag. 12-13)}

Such type of invitation is not new in physics either: the role of the observer in physics, for instance, has been explored at least since Maxwell by many scientist working in subjects such us the physics of information and the foundations of quantum theory (see Sec.~\ref{s:overview}). There have also been several discussion about the relationship that some peculiar aspects of quantum theory might have with the admittedly fuzzy concept of `consciousness' (see Sec.~\ref{s:overview}). Such explorations, however, have not yet become mainstream nor, in our opinion, gone far enough. We hope to make a case here for why we consider the time is ripe and the stakes are high to bring this debate to the forefront.

For more than a century, much has been debated about what is the actual content of quantum theory. Although substantial progress has been done (see e.g. Refs.~\cite{d2017quantum, Rovelli-1996, hardy2013reconstructing, leifer2013towards, brukner2014quantum, coecke2012picturing, fuchs2013quantum,goyal2010origin,mermin1998quantum} and Sec.~\ref{s:overview}), no general consensus has been reached~\cite{jennings2016no,Zeilinger-Nature2005}. This elusive character of quantum theory contrasts with its outstanding success. Here we argue that the resistance we have developed against human experience as a key aspect for the scientific understanding of nature has prevented us from better grasping the essential message of quantum theory~\cite{mermin2014physics,fuchs2013quantum}. Indeed, there has usually been an understandable skepticism of any suggestion that observers or consciousness might play a special role in quantum theory. However, we are witnessing today a radical shift in our understanding and control of aspects that we previously thought were intrinsically human, perhaps even unreachable to the powerful methods of science (see Appendix~\ref{s:we}). 

It is already common to read in the news that artificial intelligence has outperformed humans in yet another task we had deemed intractable before~\cite{mnih2015human, lecun2015deep}. Brain research scientists have now managed to read and control thoughts, sensations and other aspects of human experience, making the idea of living in a virtual world, as depicted in the movie {\em The Matrix}, apparently just a matter of technological maturity~\cite{losey2016navigating,salazar2017correcting,cohen2014controlling,cohen2012fmri} (see Fig.~\ref{f:subjective}). Recent theoretical and experimental developments, as well as a new respect for the subjective (see Fig.~\ref{f:subjective} and Appendix~\ref{s:easy}), have brought the fuzzy concept of consciousness into the lab and allowed scientists to start cracking some aspects of it in ways that were unthinkable before~\cite{koch2016neural,dehaene2014consciousness}. Today it is not strange to find collaborations between world-class research institutions and monks of different spiritual traditions. Such collaborations have led, for instance, to find evidence that some practices previously labelled `spiritual', such as mindfulness meditation, can radically transform our brain and significantly improve the quality of our lives~\cite{tang2015neuroscience,hanson2009buddha}---these studies are mostly concerned with the so-called neural correlates of consciousness~\cite{crick1990towards,koch2016neural,dehaene2014consciousness}, they are studies from a third-person perspective (see Fig.~\ref{f:perspectives}a). 

There have also been important developments on the understanding of our subjective experience, our first-person perspective (see Fig.~\ref{f:perspectives}b and Appendix~\ref{s:hard}). An interesting experiment in this regard is the so-called rubber-hand illusion~\cite{botvinick1998rubber,ehrsson2004s} which shows that we can experience a fake hand, disconnected from us, as if it were part of our own body. This simple experiment, which can be carried out at home, requires that we focus our attention on a rubber hand while our real hand is concealed. Both the artificial hand and the invisible real hand are stroked repeatedly and synchronously with a probe. About one or two minutes later the experience that the rubber hand is our own emerges. We keep feeling strokes which are given only to the rubber hand as if they were actually given to our real hand. Furthermore, we feel as if there were a connection between our shoulder and the artificial hand. Other experiments have extended the illusion to the full body~\cite{lenggenhager2007video,ehrsson2007experimental,blanke2009full}.

Metzinger~\cite{metzinger2004being,metzinger2009ego}  argues that these experiments are consistent with the idea that our experience of reality is actually a mental simulation of the world taking place in our brains and that our phenomenal self, i.e. what we call `I', is a representational structure in our brains, a self-model (see chapter 9 of Ref.~\cite{edelman2008computing} for a review; for a short introduction to the most central ideas see Metzinger's talk {\em `The transparent avatar in your brain'} at TEDxBarcelona). To avoid the infinite regress of trying to represent a system that represents a system that represents a system, and so on {\em ad infinitum}, the model of the world, which includes the self-model, is taken to be the ultimate reality. Metzinger refers to this feature as `transparency' (see also Refs.~\cite{metzinger2004being,edelman2008computing}):

\epigraph{\em ``Transparency simply means that we are unaware of the medium
through which information reaches us. We do not see the window but
only the bird flying by. We do not see neurons firing away in our brain
but only what they represent for us. A conscious world-model active in
the brain is transparent if the brain has no chance of discovering that it
is a model---we look right through it, directly onto the world, as it were.
The central claim of [...] the self-model
theory of subjectivity [...] is that the conscious experience of being a self
emerges because a large part of the [phenomenological self-model] in your brain is transparent.''}{T. Metzinger, Ref.~\cite{metzinger2009ego} (page 7)}

Metzinger also argues that the self-model implemented in our brain gives rise to the first-person perspective (see Fig.~\ref{f:perspectives}b; see also Refs.~\cite{metzinger2004being,edelman2008computing}):

\epigraph{\em ``By placing the self-model within the world-model, a center is created. That center
is what we experience as ourselves [...] It is the origin of what
philosophers often call the first-person perspective. We are not in direct
contact with outside reality or with ourselves, but we do have an inner
perspective. We can use the word `I.'''}{T. Metzinger, Ref.~\cite{metzinger2009ego} (page 7)}

These scientific advances are often implicitly grounded on the scientific worldview prevalent today, i.e. on the idea that there is an objective mechanical world and that we have the special status of understanding such a world as if we were an abstract entity independent of it (see Fig.~\ref{f:paradigms}).  Today it is almost taken for granted that physics, and in particular quantum physics, provides the objective laws that lie at the very foundation of the skyscraper of science. The remaining scientific disciplines therefore emerge from it (see Fig.~\ref{f:hierarchy}). 

For instance, chemistry is often considered as an application of physics describing the effective laws that emerge at the molecular scale. In turn, biology is often considered as an application of chemistry describing the effective laws that emerge at the cellular scale. And so on. At the end of such a hierarchy, according to the mainstream paradigm, we find human experience as an illusion generated by the incessant activity of billions of neurons distributed throughout our brain and body. This worldview is nicely summarized in Crick's `Astonishing Hypothesis':

\epigraph{\em ``The Astonishing Hypothesis is that “You,” your joys and your sorrows, your memories and your ambitions, your sense of personal identity and free will, are in fact no more than the behavior
of a vast assembly of nerve cells and their associated molecules.''}{F. Crick, Ref.~\cite{crick1995astonishing} (page 3)}

Yet, similar in spirit to the so-called `science of science'~\cite{fortunato2018science}, which uses the tools of science to study the mechanisms underlying the doing of science, we can ask what these recent advances on the understanding of our human nature have to say about the scientists doing the science. If we take the view that the doing of science relies in part on the physical processes running on our brains, a natural question arises (cf. ~\cite{varela2017embodied}, page 10): shouldn't our scientific description of the universe be influenced by the structure of our own cognitive system? We are convinced that in the current state of affairs there is an opportunity to more rigorously investigate the role that concepts that have been largely considered taboos in physics to date might play on the foundations of science.  

In 1929 Szilard~\cite{szilard1929entropieverminderung} already pointed out that the physics of the observer may play a role in the analysis of experiments. The same year, Bohr~\cite{bohr1929quantum} pointed out that complementarity appears to arise naturally in psychology where both the objects of perception and the perceiving subject belong to `our mental content'. About a year ago~\cite{realpe2017quantum} we argued that quantum theory could be understood from two related principles. While in the mainstream scientific paradigm we expect the observer to induce decoherence and so destroy any potential quantum phenomena, here we work in the reverse paradigm, where the world is thought of as fundamentally classical and quantum phenomena arises as a consequence of the physicality of the observer, considered as another classical system (see Fig.~\ref{f:paradigms}). Here we provide a more detailed, hopefully clearer exposition of these ideas, as well as a more thorough discussion of their potential implications (see Ref.~\cite{realpe2018cognitive} for a more compact and formal discussion). In particular, we discuss why we consider these ideas hold the potential to bring physics and human experience closer together. 

Quantum dynamics can be described by the von Neumann equation~\cite{schwabl2005quantum}
\begin{equation}\label{e:vN}
i\hbar\frac{\partial\rho}{\partial t} = [H,\rho],
\end{equation}
where $\rho$, $H$, $\hbar$, and $i$ are the density matrix, Hamiltonian operator, Planck constant, and imaginary unit, respectively; furthermore $[H,\rho] = H\rho - \rho H$. Additionally, the diagonal elements of $\rho$ encode the probability of observing the corresponding outcomes in an experiment. More generally, if the Hermitian operator $O$ represents the physical observable of interest, its expected value $\bra O\ket$, when the system is in state $\rho$, is given by the Born rule
\begin{equation}\label{e:Born}
\bra O\ket = \tr\left[\rho O\right].
\end{equation}
Key questions to understand quantum theory are: Why is $\rho$ a matrix? Why is $\rho$ complex? Why does $\rho$ satisfies Eq.~\eqref{e:vN}? Why expected values are given by Eq.~\eqref{e:Born}?

Let us now introduce the two principles put forward in Ref.~\cite{realpe2017quantum}, and discuss more precisely what we actually mean:
\begin{description}
\item{\bf Principle I:} Inference is a classical physical process performed by classical physical systems, observers, which are part of the experimental setup.
\item{\bf Principle II:} Experiments must be described from a first-person perspective.
\end{description}
First of all, by  `physical' here we only refer to the textbook notion that there are certain events that can be described by certain mathematical variables. We do not attempt to make any claims beyond this strictly operational notion. Indeed, we will argue elsewhere, where we will compare our approach to the more common information-theoretical approach to quantum foundations, that we could also use the label `information' instead of the label `physical'. What really matters for our approach is that we treat nature as a whole in a consistent manner, i.e. either {\em everything is information} or {\em everything is physical}. The keyword in these expressions, that we shall argue elsewhere are equivalent, is neither the label `information' nor the label `physical', but rather the term `everything', which implies universality and self-reference.

The term `observer' here stands for a physical system, e.g. a robot, that can carry out experiments in a lab. While everything discussed in this manuscript can be considered as referring only to artificial observers, i.e. robots, we will often refer to human observers too. Although using terms like `humans', `we', `ourselves', etc, instead of terms like `robots' throughout our analysis might give the impression to some that we are doing philosophy rather than physics, we emphasize that both artificial and human observers are considered here exclusively as physical systems and nothing more---studying humans as physical systems is routinely done in neuroscience, for instance. There are two main reasons why we insist in referring to humans in our analysis. On the one hand, we think that the best place to find intuition about the first-person perspective (see Fig.~\ref{f:perspectives}) mentioned in {\bf Principle II} is our own subjective experience. On the other hand, we consider that the main implications of our work are related to us. 

{\bf Principle I} and {\bf Principle II} can be considered as two more assumptions added to our current physical description of the world. This manuscript can be read in its entirety as an analysis of the implications of such `additional' assumptions. However, we would like to argue that these two principles are better thought of as two assumptions {\it less}. 

Indeed, overwhelming experimental evidence suggests that any observation requires an underlying physical process. For instance, the electromagnetic radiation reflected from this page interact with the electric charges in our eyes and launch a highly complex physical process in our brains that essentially constitute the neural correlates of our experience of reading these words (see Figs.~\ref{f:perspectives}a, \ref{f:subjective} and \ref{f:third}). Nevertheless, our physical description of experiments has largely neglected the physical processes related to the observer. Even in experiments that explicitly deal with the phyics of the observer, such as those related to Maxwell's demon, the physics of the scientists performing the experiment is neglected (see e.g. Refs.~\cite{toyabe2010experimental,radin2012consciousness}). There are good reasons for this, of course. An accurate physical description of humans seems to be overwhelmingly complicated, and scientists have managed to do amazing progress anyways.

In this respect, {\bf Principle I} asks us to drop the assumption that we can neglect the physics of the scientists doing the experiments. For the purposes of this work, we can account for the observer as part of the experimental setup by only adding an effective interaction that essentially turns the linear chain of cause-effect relationships into a circle. From this perspective, the interactions associated to the observer could be considered colloquially as a missing link to quantum theory. 

The discussion above analyses the observer from a third-person perspective, i.e. from the perspective of an external observer that is {\it not} included in the analysis (see Figs.~\ref{f:perspectives} and \ref{f:first}).

However, overwhelming experimental evidence suggests that we can only do science from a subjective or first-person perspective. At the risk of stating the obvious we mention here a few examples. Indeed, to the best of our knowledge, Galileo, Newton, Einstein, Bohr, and all scientists we are aware of carried out their analysis and wrote their scientific reports from their own subjective perspective. When we read their works and try to reproduce their results, we do it from our own subjective perspective. Automated experiments carried out by robots (see e.g. Ref.~\cite{melnikov2018active}) can actually be considered as larger experiments where the robots are part of the experimental setup. Such larger experiments are carried out by scientists from their own subjective perspective. If scientists launch such automated experiments and never collect the results, whatever they claim that happened or did not happen would be just an assumption made from their own subjective perspective.  When scientists perform experiments where they study other humans observing a physical system~\cite{tinsley2016direct}, they do it from their own subjective perspective. Even the feeling experienced by some people~\cite{blanke2005out} of being out of their own bodies, which may appear as the phenomenon more consistent with the assumption that we can observe the world from the outside, is experienced from their own subjective perspective (see Fig.~\ref{f:subjective}).

As a by-product of the understandable and highly successful assumption of neglecting the observer, a further assumption has usually been made in physics: that we can somehow describe the world from an objective or third-person perspective, as if we were not part of it, even though every second of our lives, from birth to death, we can only experience it from a first-person perspective. In this respect, {\bf Principle II} asks us to drop such an assumption and be consistent with what we observe in our everyday lives until there is experimental evidence that suggests otherwise. From this perspective, the question would rather be how the perception of objectivity emerges out of the intersection of our subjectivities, i.e. out of the set of perceptions that are common to all of us. On this matter, all research done on the so-called quantum-to-classical transition may have much to say. 

In this sense we might consider that {\bf Principle I} and {\bf Principle II} are in line with Einstein's suggestion that we should describe the `real factual situation'~\cite{schilpp1949albert} (page 85; see also Ref.~\cite{Zeilinger-Nature2005}), a motto we would call {\em model what is}, i.e. what we actually experience, not what we assume it is (see Fig.~\ref{f:inter} and Appendix~\ref{s:we}). Interestingly, this motto is consistent with the perspective from some contemplative traditions which suggests that reality is like the blue sky, which is obscured by the clouds of the vast amount of conceptual constructs we have acquired during the course of our lives. From this perspective, such clouds or `conceptual baggage' makes difficult for us to see the `real factual situation'. In this view, a scientific theory should therefore be about how we do inferences and build abstract concepts out of our direct human experience that allow us to reach inter-subjective agreements with our peers about our shared human experience.

A related question we find of interest is why the mathematical structure of fundamental physics equations are typically second-order differential equations and not the most parsimonious first-order differential equations? We will argue that the observer has much to do with it.

\section{Overview and related work}\label{s:overview}
This work is organized as follows. In Secs.\ref{s:big}-\ref{s:MP_recasted} we set the framework and introduce the main conceptual tools. In Sec.~\ref{s:big} we present a general discussion of the ideas involved and why we consider they make sense; in particular, we thoroughly discuss how we interpret {\bf Principle I} and {\bf Principle II}---in Appendices~\ref{s:respected} and \ref{s:recursion} we summarize, respectively, some relevant scientific insights obtained via the modern approach to consciousness and the formal analysis of self-reference via the recursion theorem, for the reader who is not familiar with these. In particular, we emphasize that the main conceptual tool in the recursion theorem is a pair of complementary Turing machines that essentially print each other. In Sec.~\ref{s:QMrecasted} we rewrite Eq.~\eqref{e:vN} as a pair of complementary matrix equations, which are those that will be derived in Sec.~\ref{s:first} from {\bf Principle I} an {\bf Principle II}; furthermore, we provide a couple of examples to illustrate that the kernels involved hold potential to be interpreted in probabilistic terms, as we will discuss in Secs.~\ref{s:MP_recasted}, \ref{s:3rd}, and \ref{s:first}---Appendix~\ref{s:details} provides some relevant technical details. In Sec.~\ref{s:MP_recasted} we show how stochastic process can be re-casted in an Euclidean quantum-like manner; in particular, we show how message-passing algorithms can be interpreted as an instance of imaginary-time quantum mechanics. More precisely, if properly normalized, the so-called cavity messages can be considered as imaginary-time wave functions evolving forward and backward in time, and the corresponding belief propagation equations as an instance of imaginary-time Schr\"odinger equation and its adjoint. Yet, in this case the phase is just an optional artificial construct which can be taken equal to zero, as discussed in Appendix~\ref{s:Q-MP}; moreover, the initial and final conditions are completely specified by the interactions in the chain. We argue this is not true anymore once we have stochastic process on a cycle rather than a chain, where the na\"ive belief propagation algorithm is not exact anymore. 

In Sec.~\ref{s:3rd} we discuss the implementation of {\bf Principle I}; in particular, we show that considering the observer as a physical system turns the traditional chain of cause-effect relationships into a loop. Furthermore, we show that the class of stochastic processes on a cycle relevant for this work, derived via the so-called principle of maximum caliber, can be described via the imaginary-time version of the von Neuman equation. In Sec.~\ref{s:first} we show that shifting from the third-person perspective assumed in Sec.~\ref{s:3rd} to a first-person perspective leads to the pair of matrix equations derived in Sec.~\ref{s:QMrecasted}. So, the shift from the third- to the first-person perspective effectively implements a Wick rotation, turning the imaginary-time von Neuman equation into Eq.~\eqref{e:vN}. Although our discussion is based on transition kernels with non-negative entries, in Appendix~\ref{s:neg_prob} we discuss this is not a restriction in our approach. In Sec.~\ref{s:occam} we compare the mainstream paradigm to the reverse paradigm assumed here. Based on the results obtained before, we argue that Occam's razor favors the reverse paradigm over the mainstream paradigm. In Sec.~\ref{s:consciousness} we discuss some third-person perspective psychophysics experiments and argue that we can estimate Planck constant from them. Furthermore, comparing to results from the first-person perspective briefly described in Appendix~\ref{s:respected}, we suggest that the neural architecture of self-aware systems and the self should be composed of two complementary neural sub-systems that essentially model each other, similar to the double-stranded structure of the DNA molecule~\cite{watson1953molecular}; we conjecture this principle may underly the division of our brains into hemispheres, i.e. for the brain to be able to implement a self-model and refer to itself. Finally, in Sec.~\ref{s:discussion} we discuss some of the potential implications of this work.

The ideas presented here have been explored by many authors even before the inception of quantum theory. An exhaustive discussion is out of the capabilities of the author. We here mention some authors we are aware of. 

The idea that the observer can play a role in physics have been explored by Maxwell, Szilard, Landauer, among many others (see e.g. Ref.~\cite{leff1990maxwell} and references therein). The question on whether the observer and consciousness can play any role on quantum theory have been discussed since the discovery of the theory by Wigner~\cite{wigner1995remarks}, von Neumann~\cite{von1955mathematical}, Bohm~\cite{bohm1991changing}, Penrose~\cite{penrose1994shadows,penrose1999emperor}, Hameroff~\cite{hameroff2014consciousness} among many others. Explorations on the mechanics of the observer have been done by Bennett, Hoffman, and Prakkash~\cite{bennett2014observer}, as well as Fields~\cite{fields2018conscious, fields2012if,fields2016building} and Mueller~\cite{mueller2017could}; these authors also pointed out that modeling the observer leads to some quantum-like phenomena. The idea that quantum theory could be related to seeing the world from the inside has been explored by R\"ossler~\cite{rossler1998endophysics}. The idea that a combination of forward and backward stochastic processes could be described via quantum-like equations has been explored by McKeon and Ord~\cite{mckeon1992time}. The idea of deriving aspects of quantum theory via the principle of maximum caliber has been explored by Caticha~\cite{caticha2009entropic,caticha2011entropic}. The idea that non-equlibrium phenomea could play a role on the derivation of quantum theory has been explored by Gr\"ossing~\cite{grossing2010sub}. The idea that quantum theory can be derived from pairs of complementary variables has been explored by Goyal~\cite{goyal2008information}, Kunth, and Skilling~\cite{goyal2010origin}. Explorations on the potential relationships between quantum theory and self-reference has been done by Kauffman~\cite{kauffman2010reflexivity}, who has also explored the idea that self-reference may underly the fundamental equations of physics. More indirect explorations of this relationship through the G\"odel theorem and incompleteness have been done by Dalla Chiara~\cite{dalla1977logical}, Penrose~\cite{penrose1994shadows,penrose1999emperor} Brukner~\cite{brukner2009quantum}, Breuer~\cite{breuer1995impossibility}, Calude~\cite{calude2004algorithmic}. The idea that quantum theory and cognitive science may be related have been explored by Bohr~\cite{bohr1929quantum}, Aerts~\cite{aerts2009quantum}, Khrennikov~\cite{khrennikov2015quantum}, Bruza, Wang, and Bussemeyer~\cite{bruza2015quantum}. The idea that the self is composed of complementary systems has been explored by Maturana and Varela~\cite{maturana1991autopoiesis} through the concept of autopoiesis and more recently by Deacon~\cite{deacon2011incomplete}; Hofstadter~\cite{hofstadter1980godel,hofstadter2013strange} has also explored the relationship between the concept of self and the mathematical formalism of self-reference via the G\"odel theorem. The idea that taking into account the observer can help resolve some conceptual difficulties of quantum theory has been explord by Fuchs, Schack~\cite{fuchs2014introduction}, and Mermin~\cite{mermin2014physics}. The idea that the subject may play a fundamental role in our description of reality have been explored by many, among which we have Siddharta Gautama, best known a Buddha, about twenty six centuries ago, by Varela, Thompson, and Rosch~\cite{varela2017embodied} about twenty five years ago, by Fuchs and Schack~\cite{fuchs2013quantum} as well as Rovelli~\cite{rovelli2015relative} a few years ago, by Mueller~\cite{mueller2017could}, Brukner~\cite{brukner2018no}, and Chiribella~\cite{chiribella2018agents} a few months ago.

\section{The big picture}\label{s:big}
\subsection{Inference as a physical process}\label{s:inference_physical}
In this section we discuss how we interpret {\bf Principle I}. We essentially propose to upgrade the Maxwell demon, a physical system with memory that interacts with an experimental apparatus, with a {\it classical} computer, a physical realization of a Turing machine, that allows it to perform inferences about the environment and implement self-reference. However, our focus is not on the computations carried out on such computer, but on the minimal physical requirements necessary to implement them (see Figs. \ref{f:realisticobserver} and \ref{f:circular}, as well as item (vi) in Sec.~\ref{s:easy}; see also Fig.~\ref{f:third} and Appendix~\ref{s:we} for the extension of this discussion to human observers).

To gain some early intuition on the ideas discussed here, let us consider the physical requirements that allow a computer hardware to determine that two bits, $x$ and $y$, are equal, i.e. $x = y$ (see Fig.~\ref{f:infophysics}). First, there should be physical systems $\mathcal{S}_x$ and $\mathcal{S}_y$ representing bits $x$ and $y$, respectively. For instance, each physical system, say $\mathcal{S}_x$, could be a magnet whose north pole can point either upwards, representing $x=0$, or downwards, representing $x=1$. Second, the computer hardware needs to perform the comparison `$=$'. Such comparison requires a physical interaction, in hardware, between the two physical systems $\mathcal{S}_x$ and $\mathcal{S}_y$ representing the corresponding bits. For instance, we could perform this operation by implementing a pairwise interaction with energy $E(\mathcal{S}_x,\mathcal{S}_y) = -\mathcal{S}_x\mathcal {S}_y$. Such energy achieves its minimum value if and only if the two magnets point in the same direction, i.e. if $x=y$ (see Fig.~\ref{f:infophysics}). So, depending on the value of the energy the computer can determine whether the two bits are equal or not. 

More generally, the physical implementation of any non-trivial gate or function requires the physical interaction between the physical systems that represent the bits that participate of the computation. 

Similarly, how can a computer or a robot determine that there exists a correlation between the position of a switch and whether a lamp shines or not? (See Fig.~\ref{f:circular}) First, the computer hardware needs two physical systems representing the states of the switch, i.e. on or off, and the lamp, i.e. shining or not. Let the bit $x=1$ if the switch is on, and $x=0$ otherwise; analogously, let the bit $y=1$ if the lamp shines, and $y=0$ otherwise. Let us assume the robot has performed $n$ experiments with the switch and the lamp, obtaining a dataset of pairs $(x_i, y_i)$ with $i=1,\dotsc , n$. The robot can then compute, for instance, the Pearson correlation coefficient
\be\label{e:rxy}
r_{xy} = \frac{\sum_{i=1}^n x_i\, y_i - n\, \bar{x}\,\bar{y}}{(n-1)\sigma_x\sigma_y},
\ee
where $\bar{x} = \sum_{i=1}^n x_i / n$ is the sample mean and ${\sigma_x = \sqrt{\sum_{i=1}^n (x_i - \bar{x})^2 /(n-1)}}$
is the sample standard deviation, both corresponding to the state of the switch. The sample mean $\bar{y}$ and sample standard deviation $\sigma_y$ corresponding to the state of the lamp are defined in a similar way.

Again, the computation of $r_{xy}$ requires the interaction, whether direct or indirect, between the physical systems representing in hardware all variables involved. Such computation can be performed, for instance, by first saving all data in memory using physical systems $\mathcal{S}_{x_i}$ and $\mathcal{S}_{y_i}$, for $i=1,\dotsc , n$, and then carrying out the corresponding physical interactions; this is an instance of {\it offline} learning. Alternatively, the robot can receive each pair of data ($x_i,\, y_i)$ one by one and use it to update on the fly the estimation of $r_{xy}$ by performing the corresponding physical interactions; this is an instance of {\it online} learning.

Now, how can a robot determine that turning the switch on {\it causes} the lamp to shine? One way to do this is via interventions~\cite{winn2012causality,pearl2009causality,wood2015lesson}. For instance, the robot can force the switch to be on or off, which is denoted as $\textsc{do}(x=x^\ast)$ , where $x^\ast$ represents the action taken by the robot (see for instance example $2$ in Ref.~\cite{winn2012causality}). Then the robot can estimate the corresponding distribution $P(y|\textsc{do}(x=x^\ast))$ from experimental data. If the condition
\be\label{e:causal}
P(y=1|\textsc{do}(x=1))\neq P(y=1|\textsc{do}(x=0)),
\ee
is satisfied, for instance, the robot can infer that the position of the switch causes the lamp to shine. In this example we are assuming a simple scenario where there are no latent variables and interventions are possible; causal inference can be highly non-trivial in more general situations~\cite{winn2012causality,pearl2009causality,wood2015lesson}. The point we want to make here, however, is that any computation made above, including the comparison `$\neq$' in Eq.~\eqref{e:causal}, involves a physical interaction in the hardware implementation.  

In summary, there are two main physical requirements for a robot to determine the existence of causal (or acausal) influences between two physical systems $X$ and $Y$ (see Fig.~\ref{f:circular}). First, there must be internal physical systems $\mathcal{S}_X$ and $\mathcal{S}_Y$ in the hardware of the robot capable of representing the states of the external systems $X$ and $Y$, respectively. Second, the two internal systems $\mathcal{S}_X$ and $\mathcal{S}_Y$ must interact in one way or another for the robot to be able to detect any potential correlation between the external systems $X$ and $Y$. Now, notice that each internal system allows the robot to detect or `percieve' the corresponding external system that it represents. In this sense, roughly speaking, while the robot is using internal system $\mathcal{S}_X$ to represent or percieve external system $X$, the robot cannot detect or percieve the internal system $\mathcal{S}_X$ itself--- a simple analogy of this is the fact that an eye cannot see itself. This already suggests that there is a `resolution restriction' ~\cite{jennings2016no} that allows the robot to percieve only one out of two physical systems. Such resolution restriction alone already leads to several features of quantum theory~\cite{jennings2016no,spekkens2016quasi,bartlett2012reconstruction,spekkens2007evidence} (see Sec.~\ref{s:foundations}). 

Furthermore, the external systems $X$ and $Y$ can reffer to different times, like when turning the switch on at a given time causes the lamp to shine at a later time (see Fig.~\ref{f:circular}). In this case the corresponding internal systems $\mathcal{S}_X$ and $\mathcal{S}_Y$ refer to different times, say `past' (initial state) and `future' (final state). The interaction between these two internal physical systems, whose state can be described by variables which are hidden to the robot in the sense discussed above, is therefore an effective interaction between `past' and `future'. In this sense, the state of the internal systems are described by hidden variables which interact non-locally. This already suggests how the locality condition underlying Bell theorem can be broken (see Sec.~\ref{s:foundations}).

Finally, the physical interactions supporting the observer's information processing define an intrinsic energy scale which the external system being observed should provide. If the energy of the external system is smaller than the energy associated to the observer's internal physical processes necessary to generate a perception, the observer may not be able to perceive anything. This might explain the origin of the quantization of energy and suggests Planck constant might be measured from psychophysics experiments (see Sec.~\ref{s:consciousness} and Fig.~\ref{f:experiment}).

As early as 1929, Leo Szilard \cite{szilard1929entropieverminderung} had already considered observers as physical systems to argue that the Maxwell demon~\cite{leff1990maxwell} could not violate the second law of thermodynamics as Maxwell had suggested back in 1871~\cite{maxwell1871theory}. Landauer argued~\cite{landauer1961dissipation} a few decades later that the physical process of erasing information from the demon's memory could account for the non-decrease in entropy postulated by the second law of thermodynamics. A first experimental demonstration of a Maxwell demon was reported a few years ago in Ref.~\cite{toyabe2010experimental}.

\

\subsection{First-person perspective and self-reference}\label{s:big_1st_person}
The previous discussion summarizes the intuitive picture that can be derived from {\bf Principle I} (see Figs.~\ref{f:circular} and \ref{f:third}). While this is enough to derive the formalism of imaginary-time quantum theory, it is not enough to obtain the full formalism of quantum theory (see Secs.~\ref{s:MP_recasted} and \ref{s:3rd}). This is because our analysis was from the perspective of an external observer: we have been describing the robot from our own perspective (see Fig.~\ref{f:first}). However, according to {\bf Principle II} experiments must be described from the perspective of the robot itself as an internal observer. This is a more subtle problem that involves self-reference (see Figs.~\ref{f:first}, \ref{f:self-observer}, \ref{f:noneq_self-observer}; see also Appendix~\ref{s:recursion} and Figs.~\ref{f:self}, \ref{f:quine}, \ref{f:recursion}). 

To illustrate the kind of ideas involved in {\bf Principle II}, let us consider the example of a program that prints itself. A na\"ive attempt would be to print the program \texttt{print ``Hello world!'';} by simply doing \texttt{print `print ``Hello world!'';'}. But the latter does not coincide with the former; a new \texttt{print} operator has appeared. We can try to solve this problem by adding a new \texttt{print} operator, but in this way we actually end up in an infinite regress. An example of a program in Phyton that does print itself is~\cite{wiki:quine}

\

\hspace{1.3cm}\texttt{s = `s = \%r$\backslash$nprint(s \%\% s)'}

\hspace{1.3cm}\texttt{print(s \% s)}

\

This program is composed of two parts that, roughly speaking, print each other (see Figs. \ref{f:self}, \ref{f:quine}, \ref{f:recursion} and Sec.~\ref{s:recursion}). Indeed, the first line defines a string \texttt{s} that contains the  second line, while the second line prints the string \texttt{s} defined in the first one. This is a general feature of self-printing programs~\cite{sipser2006introduction} (see chapter 6), as we will discuss in more detail in Sec.~\ref{s:self}. In this respect, self-reference leads to complementary pairs. In Appendix \ref{s:recursion} we briefly review the recursion theorem of computer science that formalize the construction of self-referential programs.

Another example is the sentence~\cite{sipser2006introduction}:
\begin{widetext}
\begin{eqnarray}
&\textrm{\em Print two copies of the sentence below, the second copy in quotes}\\
&\textrm{\em ``Print two copies of the sentence below, the second copy in quotes''}
\end{eqnarray}
\end{widetext}

If we do what is asked in this sentence we end up printing the sentence itself. This again is composed of two `complementary' pairs: On the one hand, the first sentence plays an active role by instructing to print the second one. On the other hand, the second sentence plays a passive role by representing the first sentence in quotation marks; in a sense, the second sentence could be considered as information about the first. Both sentences, however, are made of the same `stuff': characters in an alphabet.

As we will discuss in Sec.~\ref{s:quantum}, something similar happens when a robot describes the world from within, including itself: there should exist a pair of complementary systems that, in a sense, mutually observe each other (see Figs.~\ref{f:first-third}, \ref{f:self-observer}, \ref{f:noneq_self-observer}). This does not imply that there is a misterious entity observing the robot, rather the corresponding architecture of the robot should support such a reflexive feature--- a simple analogy of this is the fact that although an eye cannot see itself if it is alone, it can do so with the aid of a complementary system, like a mirror, that implements the required reflexive feature. Figure~\ref{f:first-third} shows a simple experiment that we can do at home to get some intuition of these ideas. (See Appendix~\ref{s:we} for the extension of this discussion to human observers.)

\section{Quantum mechanics recasted}\label{s:QMrecasted}

\subsection{Von Neumann equation as a pair of real matrix equations}\label{s:vNreal}
Here we focus on finite-dimensional systems for simplicity; so we can represent the adjoint operation $\dagger$ by the combination of transpose $T$ and complex conjugate $\ast$ operations, i.e. if $\mu$ is a generic finite-dimensional matrix with complex entries, then $\mu^\dagger = (\mu^T)^\ast = (\mu^\ast)^T$. Notice that a generic Hermitian matrix can be written as $\mu = M_s + i M_a$, where $M_s = M_s^T$ is a real symmetric matrix and $M_a = -M_a^T$ is a real antisymmetric matrix; indeed ${\mu^\dagger = M_s^T - i M_a^T = M_s + i M_a = \mu}$. Furthermore, since any generic real matrix $M$ can be decomposed into symmetric and antisymmetric parts, i.e. $M = M_s + M_a$, then we can write $M_s = (M+M^T)/2$ and $M_a = (M-M^T)/2$. From this perspective, we can consider a generic Hermitian matrix $\mu$ as a convenient representation of a generic real matrix $M$ that allows us to keep explicit track of the symmetric and antisymmetric parts of the latter via the the real and imaginary parts of the former, respectively. 

So, we can write $\rho $ and $H$ in Eq.~\eqref{e:vN} as $\rho = P_s + i P_a$ and $H = \hbar J_s + i \hbar J_a$. Here the symmetric and antisymmetric matrices corresponding to $\rho$ and $H$ can be written in terms of real matrices $P$ and $J$, respectively, as done for the generic matrix $M$ above. Since $\tr \rho = 1$ and the diagonal elements of an antisymmetric matrix are zero, we have $\tr\rho=\tr P_s = \tr P$, so $\tr P = 1$. We have written $H$ in terms of a real matrix $\hbar J$ so we do not have to worry about $\hbar$ in the equations below. We will refer to $J$ as the {\em dynamical matrix}. 

In this way, Eq.~\eqref{e:vN} can be written as
\begin{equation}\label{e:PsPa}
i \frac{\partial}{\partial t}(P_s + i P_a) = [J_s + iJ_a , P_s + i P_a],
\end{equation}
where $\hbar$ has been absorved in $J=H/\hbar$. Equating the real and imaginary parts of Eq.~\eqref{e:PsPa} we get a pair of equations
\begin{eqnarray}
\frac{\partial P_s}{\partial t} &=& [J_a , P_s] + [J_s , P_a],\label{e:DtPs} \\
\frac{\partial P_a}{\partial t} &=& [J_a , P_a] - [J_s , P_s].\label{e:DtPa}
\end{eqnarray} 

By adding and substracting Eqs.~\eqref{e:DtPs} and \eqref{e:DtPa}, we obtain an equivalent pair of equations in terms of the real matrix $P$, i.e.
\BE
\frac{\partial P}{\partial t} &=& [J_a , P] - [J_s , P^T],\label{e:DtP} \\
\frac{\partial P^T}{\partial t} &=& [J_a , P^T] + [J_s , P].\label{e:DtPT}
\EE 
While Eq.~\eqref{e:DtPT} is the transpose of Eq.~\eqref{e:DtP}, we can also write these two equations as corresponding to two different observers $A$ and $B$ who describe the experiment with probability matrices $P_A$ and $P_B$, respectively; i.e. 
\BE
\frac{\partial P_A}{\partial t} &=& [J_a , P_A] - [J_s , P_B],\label{e:DtPA} \\
\frac{\partial P_B}{\partial t} &=& [J_a , P_B] + [J_s , P_A],\label{e:DtPB}
\EE 
under the condition that $P_A = P_B$ at time $t=0$, which guarantees that at all next time steps we have $P_A = P_B^T$ (see Eqs.~\eqref{e:DtP} and \eqref{e:DtPT}). This condition is satisfied for any experiment since any initial quantum state $\rho_0 = U\rho_{\rm diag}U^\dag$ can be prepared by applying a quantum operation $U$ to a diagonal state $\rho_{\rm diag}$, and such diagonal state would lead to diagonal probability matrices $P_A=P_B$ which are equal to each other. So, without loss of generality the initial state of a quantum experiment can always be considered to be a diagonal density matrix, as long as we include the quantum operation $U$ as part of the experiment; since the initial state is diagonal, the condition $P = P^T$ or $P_A = P_B$ is automatically satisfied at the beginning of the experiment.

As we will describe in Sec.~\ref{s:first}, $A$ and $B$ can indeed be considered as two complementary third-person sub-observers that essentially observe each other to mutually build a first-person observer, much as the two photographers in Fig.~\ref{f:first-third}b, or the self-printing programs described in Sec.~\ref{s:big_1st_person}, or the more general programs described by the recursion theorem (see Appendix~\ref{s:recursion}). In Sec.~\ref{s:examples} we recast some typical examples of quantum dynamics to show that these equations can be formulated in terms of real non-negative kernels that therefore can in principle be interpreted probabilistically (see also Secs.~\ref{s:MP_recasted}, \ref{s:3rd}, \ref{s:first}).

\

\noindent{\bf Remark 1:} Notice that since for small time steps $\epsilon$ we can write unitary evolution operators as ${U_\epsilon = I - i\epsilon H}/\hbar$. So, we can write commutators with $H$ in terms of commutators with $U_\epsilon$ since ${[U_\epsilon,\rho] = -i\epsilon [H,\rho]/\hbar}$; this is also true for other types of evolution kernels. We can then write the von Neumann equation, Eq.~\eqref{e:vN}, as 
\be\label{e:vN_Ueps}
\frac{\partial\rho}{\partial t} =\frac{1}{\epsilon}[U_\epsilon , \rho], 
\ee
where the limit $\epsilon\to 0$ is understood. 
 
This observation is useful when dealing with Gaussian kernels (see Secs.~\ref{s:path} and \ref{s:EM} as well as Appendices~\ref{s:App_path} and \ref{s:App_EM}), for instance, because a Gaussian kernel $\mathcal{K}(x,x^\prime)$ with vanishing variance $\sigma^2$ cannot be expanded in a Taylor series as $\mathcal{K}(x, x^\prime)\approx \delta(x- x^\prime) + {O}(\sigma^2)$, where $\delta(x)$ is the Dirac delta. However, we can straightforwardly write $\mathcal{K}(x, x^\prime) = \delta(x- x^\prime) - \mathcal{L}(x,x^\prime)$, where $\mathcal{L}(x,x^\prime) = \delta(x-x^\prime) - \mathcal{K}(x, x^\prime) $. Although $\mathcal{L}$ is not ${O}(\sigma^2)$, its convolution with a smooth function yields a term ${O}(\sigma^2)$. In Secs.~\ref{s:path} and \ref{s:EM} (see also Appendices~\ref{s:App_path} and \ref{s:App_EM}) we obtain von Neumann-like equations similar to Eq.~\eqref{e:vN_Ueps}, where commutators are directly written in terms of real Gaussian kernels $\mathcal{K}$ with variance $\sigma^2 = O(\epsilon)$.

\

\noindent{\bf Remark 2:} When the Hamiltonian is real, i.e. ${H = \hbar J_s = \hbar J}$ (since $J_a = 0$), Eqs.~\eqref{e:DtP} and \eqref{e:DtPT} become
\BE
\frac{\partial P}{\partial t} &=& - [J_s , P^T],\label{e:realDtP} \\
\frac{\partial P^T}{\partial t} &=& [J_s , P].\label{e:realDtPT}
\EE 
Furthermore, if the Hamiltonian is independent of time, we can take the partial time derivative of Eq.~\eqref{e:realDtP} and replace $\partial P^T/\partial t$ in its right hand side by the right hand side of Eq.~\eqref{e:realDtPT} to obtain
\be
\frac{\partial^2 P}{\partial t^2} = - [J_s , [J_s, P ]],\label{e:DDtP}
\ee
which is a real second order differential equation in the probability matrix $P$. This contrasts with the first-order equations typically obtained for the evolution of probabilities in Markov processes, e.g. the master equation, which is usually a reflection of the linearity of the Bayesian update~\cite{caticha2009entropic,caticha2011entropic}. 

Equation~\eqref{e:DDtP} is similar to the equation describing the second law of Newtonian mechanics. Insisting in the current paradigm (see Fig.~\ref{f:paradigms}), we could attempt to interpret this as a reflection of the physical nature of the probabilities associated to a physical observer embedded in the system under study. In other words, such physical probabilities  not only should represent the subjective beliefs the observer has about the physical system being studied, but they should also be objectively implemented in the observer's `hardware' (e.g. as a population of neurons). We will argue, however, that the structure of physical laws themselves could be considered a result of the self-referential problem of describing the world from a first-person perspective.

\

\subsection{Some examples of quantum dynamics in terms of non-negative real kernels}\label{s:examples}
Here we briefly discuss how some well-known examples of quantum systems can be described in terms of non-negative real kernels, including systems associated to complex non-stoquastic Hamiltonian operators. In Appendix~\ref{s:details} we provide the details of the derivations. In Appendix~\ref{s:neg_prob} we discuss why the non-negativity of the kernels is not a restriction in our approach. Therein we show how we can obtain effective kernels with negative entries, associated to real non-stoquastic Hamiltonian operators.
\subsubsection{Non-relativistic Schr\"odinger equation}\label{s:non-relativistic}
The Hamiltnoian of a single particle of mass $m$ in a one-dimensional non-relativistic potential $V(x)$ is given by $H=-({\hbar^2}/{2 m}){\partial^2}/{\partial x^2} + V(x)$ which, after a suitable space discretization of $x=\ell\delta$ with latice constant $\delta$ and $\ell = \dotsc,-2, -1,0,1,2,\dotsc$, can be represented by the matrix (see Appendix~\ref{s:App_Schrodinger} for all technical details)
\begin{widetext}
\be
H \equiv \hbar J = \begin{pmatrix} \ddots & \ddots & \ddots & \ddots & \ddots & & & &  \\ \cdots & 0 & -\frac{\hbar^2}{2 m\delta^2}  & {\frac{\hbar^2}{m\delta^2} + V_{-1}} & -\frac{\hbar^2}{2 m\delta^2}  & 0 & \cdots &  & \\ &\cdots & 0 & -\frac{\hbar^2}{2 m\delta^2} & {\frac{\hbar^2}{m\delta^2} + V_0} & -\frac{\hbar^2}{2 m\delta^2} & 0 & \cdots & \\ & &\cdots & 0 & -\frac{\hbar^2}{2 m\delta^2} & {\frac{\hbar^2}{m\delta^2} + V_1} & -\frac{\hbar^2}{2 m\delta^2} & 0 & \cdots  \\  & & & & \ddots & \ddots & \ddots & \ddots & \ddots\end{pmatrix},
\ee
\end{widetext}

\subsubsection{From non-relativistic path integrals to real convolutions}\label{s:path}

In terms of the short-time path integral representation~\cite{feynman1948space}, with time step $\epsilon\to 0$, the evolution equation of the example in Sec.~\ref{s:non-relativistic} can be written as (see Appendix~\ref{s:App_path} for all technical details)
\be\label{e:Main_GaussianKernel_vN}
\frac{\partial \rho}{\partial t} = \frac{i}{\epsilon}\left[(\mathcal{K}\ast\rho)-(\rho\ast\mathcal{K})\right],
\ee
where $\rho(x, x^\prime , t)$ is the density matrix and $\mathcal{K}(x,x^\prime)$ is a {\em real} kernel given by
\be\label{e:Main_K_T+V}
\mathcal{K}(x, x^{\prime}) = \frac{1}{|\mathcal{A}|}\exp{\left[-\frac{\mathcal{H}(x, x^{\prime})\epsilon}{\hbar} \right]},
\ee
with
\be\label{e:Main_classicalH}
\mathcal{H}(x, x^\prime) = \frac{m}{2}\frac{(x - x^\prime)^2}{\epsilon^2} + V(x),
\ee
the corresponding {\em Hamiltonian function} and $\mathcal{A} = \sqrt{i 2\pi\hbar\epsilon /m }$ (here we consider the Hamiltonian function as a function of position only).

Here we have introduced the convolutions 
\begin{widetext}
\BE
\left[\mathcal{K}\ast\rho\right] (x,x^\prime) &=& \frac{1}{|\mathcal{A}|}\int \exp{\left[-\frac{\mathcal{H}(x, x^{\prime\prime})\epsilon}{\hbar} \right]}\rho(x^{\prime\prime},x^\prime, t)\mathrm{d} x^{\prime\prime},\label{e:int_K*rho}\\
\left[\rho\ast\mathcal{K}\right] (x, x^\prime ) &=& \frac{1}{|\mathcal{A}|}\int \rho(x, x^{\prime\prime}, t)\exp{\left[-\frac{\mathcal{H}(x^{\prime\prime}, x^\prime )\epsilon}{\hbar} \right]}\mathrm{d} x^{\prime\prime}. \label{e:int_rho*K} 
\EE
\end{widetext}
Notice that the integration variables in $\mathcal{K}\ast\rho$ and $\rho\ast\mathcal{K}$ are, respectively, the first and second arguments of $\rho$, which yields the analogous of left and right matrix multiplication. 

Following the discussion in Sec.~\ref{s:vNreal}, Eq.~\eqref{e:Main_GaussianKernel_vN} can be written as a pair of real matrix equations (see more general example in Sec.~\ref{s:EM}). The point we want to make here is that the kernel appearing in such pair of equations can be real and non-negative, as we can see in Eq.~\eqref{e:Main_K_T+V}. In the next section we show this is also true in more general cases where the Hamiltonian is complex.

\

\noindent{\bf Remark:} While the kernel $\mathcal{K}$ defined in Eq.~\eqref{e:Main_K_T+V} is real and non-negative, it is {\em not} normalized. Indeed, we have (e.g. take $\psi(x^\prime, t) = 1$ in Eq.~\eqref{e:realGaussian})
\be\label{e:intK_T+V}
\int \mathcal{K}(x,x^\prime) \mathrm{d} x = 1-\epsilon V(x)/\hbar + O(\epsilon^2).
\ee
This fact has sometimes been used to argue against the viability of any probabilistic interpretation of the Euclidean, or imaginary-time, Schr\"odinger equation \cite{Zambrini-1987, zambrini1986stochastic,zambrini1986variational}. 

However, we will show in Sec.~\ref{s:MP_recasted} and Appendix~\ref{s:Q-MP} that the proper probabilistic analogous of $\mathcal{K}$ is not a transition probability but something closer to the squared root of the product of forward and backward transition probabilities (cf. Eqs.~\eqref{e:K_P+*P-} and \eqref{e:intKepsilon}). Furthermore, we will show there that such a type of kernel arises naturally in a less common, more symmetric representation of {\em standard} Markov process.


\subsubsection{Particle in an electromagnetic field via asymmetric real kernels}\label{s:EM}

The Schr\"odinger equation of a particle of charge $e$ interacting with an electromagnetic field can be written as 
\be\label{e:Main_SchrodingerEM}
\begin{split}
i\hbar\frac{\partial \psi(\bx,t)}{\partial t}=& -\frac{\hbar^2}{2m}\left(\nabla - i \frac{e}{\hbar c}\mathbf{A}\right)^2\psi(\bx,t)\\
 &+ e V(\bx,t)\psi(\bx,t),\\
\end{split}
\ee
%

\

\noindent where $\bx$ denotes the position vector in three dimensional space, while $V$ and $\mathbf{A}$ denote the scalar and vector fields respectively. Notice that the Hamiltonian associated to Eq.~\eqref{e:Main_SchrodingerEM} now contains an imaginary part given by the terms linear in $\mathbf{A}$ arising from the expansion of ${(\nabla - i e\mathbf{A} /\hbar c)^2\psi(\bx,t)}$.

As shown in full detail in Appendix~\ref{s:App_EM}, and following Sec.~\ref{s:vNreal}, the von Neumann equation corresponding to Eq.~\eqref{e:Main_SchrodingerEM} can be written as a pair of real matrix equations
\BE
\frac{\partial P_A}{\partial t} &=& -\frac{1}{\epsilon}[\mathcal{K}_a, P_A] + \frac{1}{\epsilon}[\mathcal{K}_s, P_B],\\
\frac{\partial P_B}{\partial t} &=& -\frac{1}{\epsilon}[\mathcal{K}_a, P_B] - \frac{1}{\epsilon}[\mathcal{K}_s, P_A],
\EE
where the two probability matrices satisfy $P_A = P$, $P_B = P^T$, and ${\rho = (P + P^T)/2 + i (P-P^T)/2}$. Here, $\mathcal{K}_s$ and $\mathcal{K}_a$, are the symmetric and anti-symmetric parts of a real kernel ${\mathcal{K}=\mathcal{K}_s+\mathcal{K}_a}$ given by 
\be\label{e:Main_KrealEM}
\mathcal{K}(\bx , \bx^\prime) = \frac{1}{|\mathcal{A}_{EM}|}\exp{\left[-\frac{\epsilon}{\hbar}\mathcal{H}_{\rm EM}(\bx , \bx^\prime) \right]},
\ee
where the real electromagnetic Hamiltonian function is given by (here we consider the Hamiltonian function as a function of position only)
\begin{widetext}
\be\label{e:Main_HrealEM}
\begin{split}
\mathcal{H}_{\rm EM}(\bx , \bx^\prime) &= \frac{m}{2}\left(\frac{\bx - \bx^\prime}{\epsilon}\right)^2 + V\left(\frac{\bx+\bx^\prime}{2}, t\right) - \frac{e}{c}\left(\frac{\bx-\bx^\prime}{\epsilon}\right)\cdot\mathbf{A}\left(\frac{\bx+\bx^\prime}{2}, t\right) +\frac{e^2}{m c^2} \left[\mathbf{A}\left(\frac{\bx+\bx^\prime}{2}, t\right)\right]^2\\
&= \frac{m}{2}\left[\frac{\bx - \bx^\prime}{\epsilon} - \frac{e}{m c}\mathbf{A}\left(\frac{\bx+\bx^\prime}{2}, t\right)\right]^2 + V\left(\frac{\bx+\bx^\prime}{2}, t\right)
+ \frac{e^2}{2 m c^2} \left[\mathbf{A}\left(\frac{\bx+\bx^\prime}{2}, t\right)\right]^2,
\end{split}
\ee
\end{widetext}

As we will argue in Sec.~\ref{s:first} these equations can be interpreted in probabilistic terms. So, even in the case of a charged particle in an electromagnetic field, whose Hamiltonian operator is complex (and so non-stoquastic), can be thought of as arising from a real non-negative kernel $\mathcal{K}$.  It is no clear at this point, though, how to interpret $\mathcal{H}_{\rm EM}$ defined in Eq.~\eqref{e:Main_HrealEM} nor the real kernel $\mathcal{K}$ defined in Eq.~\eqref{e:Main_KrealEM}. It seems to suggests a probabilistic interpretation of electromagnetic phenomena. We leave this for future work. 

\

\noindent{\bf Remark:} We can see that the antisymmetric part of the Hamiltonian funcion $\mathcal{H}_{\rm EM}$ defined in Eq.~\eqref{e:Main_HrealEM} comes from the term linear in $\mathbf{A}$, which changes sign when we transpose $\bx$ and $\bx^\prime$. Intuitively, this anti-symmetric term is related to non-equilibrium irreversible phenomena since $\mathbf{A}$ can be thought of as an effective interaction generated by the collective motion of charged particles, while $V$ can be generated by charged particles at rest. (We shall argue in Sec.~\ref{s:first} this interpretation is valid in general.) 

Indeed, from the classical electromagnetic Lagrangian 
\be
\mathcal{L} = \frac{m}{2}\dot{\bx}^2 - V + \frac{e}{c}\dot{\bx}\cdot\mathbf{A},
\ee
we get the expression for the canonical momentum
\be
\bp = m \dot{\bx} + \frac{e}{c}\mathbf{A},
\ee
which implies that the Hamiltonian can be written as
\be\label{e:Main_stdEM}
\begin{split}
\mathcal{H} &= \frac{1}{2m}\left(\bp - \frac{e}{c}\mathbf{A}\right)^2 + V\\
&= \frac{m}{2}{\dot{\bx}}^2 + V,
\end{split}
\ee
which is the sum of the standard kinetic energy and the static potential energy $V$. The term $\mathbf{A}$ appears only when we write the kinetic energy in terms of the canonical momentum.

\section{Markov processes recasted}\label{s:MP_recasted}
\subsection{Principle of maximum caliber and factor graphs}\label{s:MaxCal}
The principle of maximum Shannon entropy introduced by Jaynes~\cite{jaynes2003probability} to derive some common equilibrium probability distributions in statistical physics can be extended to the so-called principle of maximum caliber to deal with non-equilibrium distributions on trajectories~\cite{presse2013principles}. In particular Markov processes can be derived from the principle of maximum caliber (see e.g. Sec. IX B in Ref.~\cite{presse2013principles}). We introduce this principle here with an example relevant for our discussion. 

Consider the probability distributions $\mathcal{P}(x_1,\dotsc ,x_n)$ on (discretized) paths $(x_1,\dotsc , x_n)$, where $x_\ell$ refers to the position at time $t=\ell\epsilon$. Assume that we only have information about the average energy on the (discretized) paths given by
\be\label{e:Eav}
\mathcal{H}_{\rm av}[\mathcal{P}] = \int \mathcal{P}(x_1,\dotsc ,x_n)\left[\frac{1}{T}\sum_{\ell=1}^{n-1}\mathcal{H}(x_\ell,x_{\ell+1})\epsilon\right]\prod_{\ell = 1}^n\mathrm{d}x_\ell  ,
\ee
where $T=n\epsilon$ is the total time duration of the path, and $\mathcal{H}$ is the Hamiltonian function which, without loss of generality, we will assume is given by Eq.~\eqref{e:Main_classicalH}; all results in this section are valid for general Hamiltonian functions like, for instance, the one given by Eq.~\eqref{e:Main_HrealEM}.

The principle of maximum caliber tells us that among all possible probability distributions we should choose the one that both maximizes the entropy
\be\label{e:entropy}
\mathcal{S}[\mathcal{P}] = -\int \mathcal{P}(x_1,\dotsc ,x_n)\ln \mathcal{P}(x_1,\dotsc ,x_n)\mathrm{d} \prod_{\ell = 1}^n\mathrm{d}x_\ell,
\ee
and is conistent with the information we have, i.e. $\mathcal{H}_{\rm av}[\mathcal{P}] = E_{\rm av}$, where $E_{\rm av}$ is the fixed value of the average energy. Introducing a Lagrange multiplier $\lambda$ to enforce the constraint on the average energy, the constrained maximization of $\mathcal{S}[\mathcal{P}]$ becomes equivalent to the maximization of the Lagrangian $\mathcal{S} - \lambda \mathcal{H}_{\rm av}[\mathcal{P}]$. The solution to this problem is the distribution
\be\label{e:MaxCal}
\mathcal{P}(x_1,\dotsc , x_n) = \frac{1}{\mathcal{Z}}\exp\left[-\frac{\lambda}{T}\sum_{\ell=1}^{n-1}\mathcal{H}(x_\ell , x_{\ell+1})\epsilon\right],
\ee
where $\mathcal{Z}$ is the normalization factor. 

Notice that $\mathcal{P}$ in Eq.~\eqref{e:MaxCal} can be written as a product of factors
\be\label{e:P_prodF}
\mathcal{P}(x_1,\dotsc , x_n) = \frac{1}{Z}\prod_{\ell=1}^{n-1} F_\ell(x_\ell , x_{\ell+1}).
\ee
To keep the analogy with Eq.~\eqref{e:Main_K_T+V} as close as possible, we can choose the factors as
\be\label{e:F}
F_\ell(x_\ell , x_{\ell+1}) = \frac{1}{|\mathcal{A}|}\exp\left[-\frac{\lambda}{T} \mathcal{H}(x_\ell , x_{\ell+1})\epsilon\right],
\ee
with $|\mathcal{A}|= \sqrt{2\pi T\epsilon / m\lambda}$, so $Z = \mathcal{Z}/|\mathcal{A}|^{n-1}$ in Eq.~\eqref{e:P_prodF}.  Indeed, by writting $\hbar = T/\lambda$ Eq.~\eqref{e:F} is the exact analogous of Eq.~\eqref{e:Main_K_T+V}. Since $F_\ell$ is {\em not} a probability distribution, it is not normalized in general either.

The probability distribution in Eq.~\eqref{e:MaxCal}, or Eq.~\eqref{e:P_prodF}, can also be interpreted as the Boltzmann distribution of a system of particles interacting on a chain. Now, any probability distribution of particles interacting on a chain can be parametrized in terms of the pairwise marginals $\mathcal{P}_{\ell}(x_\ell, x_{\ell+1})$ of neighboring variables $x_\ell$ and $x_{\ell+1}$ on the chain and the single marginals $p_\ell(x_\ell)$ as (see e.g Eq.~(14.22) in Ref.~\cite{Mezard-book-2009})
\be\label{e:P()}
\mathcal{P}(x_1,\dotsc ,x_n) = \left.\prod_{\ell=1}^{n-1}\mathcal{P}_\ell(x_\ell, x_{\ell+1})\middle/\prod_{\ell=2}^{n-1}{p_\ell(x_\ell)}\right. ;
\ee
notice the single marginals of the first and last node in the chain are excluded because they have only one neighbor. 

Now, using the product rule of probability theory we can write the pairwise marginals in the three different ways
\be\label{e:MarkovBayes}
\begin{split}
\mathcal{P}_\ell(x_\ell, x_{\ell+1}) &=  \mathcal{P}^+_\ell(x_{\ell+1}| x_\ell) p_\ell(x_\ell)\\ 
&= \mathcal{P}^-_{\ell}(x_\ell|x_{\ell+1})p_{\ell+1}(x_{\ell+1})\\
&= \theta_\ell(x_\ell) K_\ell(x_\ell, x_{\ell+1})\theta_{\ell+1}(x_{\ell+1})
\end{split}
\ee
where $\mathcal{P}^+_\ell$ and $\mathcal{P}^-_\ell$ denote the forward and backward transition probabilities, respectively, and
\BE
\theta_\ell(x_\ell) &=& \sqrt{p_\ell(x_\ell)},\label{e:theta}\\
K_\ell(x_\ell, x_{\ell+1}) &=& \sqrt{\mathcal{P}^+_{\ell}(x_{\ell+1}|x_\ell)\mathcal{P}^-_{\ell}(x_{\ell}|x_{\ell+1})}.\label{e:K_P+*P-}
\EE

The less common, more symmetric alternative in the third line of Eq.~\eqref{e:MarkovBayes} is obtained by multiplaying the first two lines in Eq.~\eqref{e:MarkovBayes} and taking the square root. While this is just a more symmetric description of a Markov process, we show in Appendix~\ref{s:Q-MP} that $\theta_\ell$ and $K_\ell$, respectively, are {\em similar} to real wave functions and to the real transition kernels appearing in the modified formulation of quantum theory introduced in Sec.~\ref{s:QMrecasted} (see e.g. Eqs.~\eqref{e:K_T+V} and~\eqref{e:KrealEM}). Indeed, the third line in Eq.~\eqref{e:MarkovBayes} could be considered as a slightly more general presentation of the type of stochastic processes originally studied by Schr\"odinger \cite{schrodinger1931umkehrung,schrodinger1932theorie}, known sometimes as Schr\"odinger bridges. In such Schr\"odinger bridges we not only know the initial probability distribution $p_1$, but also the final one $p_n$. We show in Appendix~\ref{s:Q-MP} that there are indeed some analogies with the so called Euclidean (or imaginary-time) quantum mechanics and Bernstein processes~\cite{Zambrini-1987,zambrini1986stochastic, zambrini1986variational} built on Schr\"odinger's work. Although $\theta_\ell$ does not contain any `phase' information, we will see in Sec.~\ref{s:QBP} that the analogous of a phase can arise by using a formulation based on the cavity method. 

\subsection{Quantum-like formulation of stochastic processes via the cavity method}\label{s:QBP}
\subsubsection{Cavity messages as imaginary-time wave functions}
Here we show how the belief propagation algorithm obtained via the cavity method~\cite{Mezard-book-2009} (see chapter 14) can be naturally written in terms of the imaginary-time Schr\"odinger equation and its conjugate. 

First, notice that by marginalizing the probability distribution defined in Eq.~\eqref{e:P_prodF} over all variables except $x_\ell$ and $x_{\ell+1}$ we obtain
\begin{widetext}
\BE
\mathcal{P}_\ell(x_\ell , x_{\ell +1}) &=& \frac{1}{Z} F_\ell(x_\ell, x_{\ell +1}) Z_{\to\ell}(x_\ell)Z_{\ell+1\leftarrow}(x_{\ell +1}),\label{e:P_BP}\\
p_\ell(x_\ell ) &=& \frac{1}{Z} Z_{\to\ell}(x_\ell)Z_{\ell\leftarrow}(x_{\ell }),\label{e:p_BP}
\EE
\end{widetext}
where the partial parition functions $Z_{\to\ell}(x_\ell)$ and $Z_{\ell\leftarrow}(x_{\ell })$ of the original factor graph are given by the partition functions of the modified factor graphs that contain all factors $F_{\ell^\prime}$ to the left (i.e. $\ell^\prime < \ell$) and to the right (i.e. $\ell^\prime \geq \ell$) of variable $x_\ell$, respectively; i.e. (see Fig.~\ref{f:cavity}a,b; cf. Eq. (14.2) in Ref.~\cite{Mezard-book-2009}).
\BE
Z_{\to\ell}(x_\ell) &=& \int \prod_{\ell^\prime = 1}^{\ell-1}F_{\ell^\prime}(x_{\ell^\prime}, x_{\ell^\prime +1})\mathrm{d}x_{\ell^\prime},\label{e:Z->}\\
Z_{\ell\leftarrow}(x_{\ell }) &=& \int \prod_{\ell^\prime = \ell}^{n-1}F_{\ell^\prime}(x_{\ell^\prime}, x_{\ell^\prime +1})\mathrm{d}x_{\ell^\prime +1}.\label{e:Z<-}
\EE
$Z_{\to\ell}(x_\ell)$ and $Z_{\ell\leftarrow}(x_{\ell })$ can be interpreted as information that arrives to variable $\ell$ from the left and from the right side of the graph, respectively. 

By separating factor $F_{\ell -1}$ and $F_\ell$ in Eqs.~\eqref{e:Z->} and \eqref{e:Z<-}, respectively, we can write these equations in a recursive way as (see Fig.~\ref{f:cavity}c; cf. Eq. (14.5) in Ref.~\cite{Mezard-book-2009})
\BE
Z_{\to\ell}(x_\ell) &=& \int F_{\ell -1}(x_{\ell -1}, x_{\ell})Z_{\to\ell-1}(x_{\ell-1})\mathrm{d}x_{\ell-1},\label{e:BP->}\\
Z_{\ell\leftarrow}(x_{\ell }) &=& \int F_{\ell}(x_{\ell}, x_{\ell +1})Z_{\ell+1\leftarrow}(x_{\ell+1 })\mathrm{d}x_{\ell+1}.\label{e:BP<-}
\EE
These recursive equations are usually referred to as the {\em belief propagation algorithm}. Since the partial partition functions are typically exponentially large, Eqs.~\eqref{e:BP->} and \eqref{e:BP<-} are commonly written in terms of normalized {\em cavity messages} ${\nu_{\to\ell}(x) = Z_{\to\ell}(x)/Z_{\to\ell}}$ and ${\nu_{\ell\leftarrow}(x)=Z_{\ell\leftarrow}(x)/Z_{\ell\leftarrow}}$, where $Z_{\to\ell}$ and $Z_{\ell\leftarrow}$ are the corresponding normalization constants. This choice of normalization has at least two advantages: (i) it allows us to interpret the messages as probability distributions and (ii) it keeps the information traveling from left to right separated from the information traveling from right to left. 

We will now show that a different choice of normalization, i.e. 
\be\label{e:mu_def}
\mu_{\to\ell}(x) = \frac{Z_{\to\ell}(x)}{\sqrt{Z}},\hspace{0.5cm} \mu_{\ell\leftarrow}(x) = \frac{Z_{\ell\leftarrow}(x)}{\sqrt{Z}},
\ee
which violates the features (i) and (ii) mentioned above, allows us to connect the belief propagations equations, i.e. Eqs.~\eqref{e:BP->} and \eqref{e:BP<-}, with those of Euclidean quantum mechanics. Indeed, let us write
\BE
\mu_{\to\ell}(x) \mu_{\ell\leftarrow}(x) &=& p_\ell(x),\label{e:mu->*<-mu} \\
\frac{\mu_{\to\ell}(x)}{\mu_{\ell\leftarrow}(x)} &=& e^{2\phi_\ell(x)}, \label{e:mu->/<-mu}
\EE
where Eq.~\eqref{e:mu->*<-mu} comes from Eq.~\eqref{e:p_BP} and Eq.~\eqref{e:mu->/<-mu} is a definition of the `effective field' or `phase' $\phi_\ell$. Equations~\eqref{e:mu->*<-mu} and \eqref{e:mu->/<-mu} imply that we can parametrize the cavity messages in terms of $p_\ell$ and $\phi_\ell$ as
\BE
\mu_{\to\ell}(x)  &=& \sqrt{p_\ell(x)}e^{\phi_\ell(x)},\label{e:mu->} \\
\mu_{\ell\leftarrow}(x)  &=& \sqrt{p_\ell(x)}e^{-\phi_\ell(x)},\label{e:<-mu}
\EE
which are the exact analog of a `wave function' in imaginary time.

\

\noindent{\bf Remark:} When $x\in\{-1,+1\}$ is a binary variable, the cavity messages are usually parametrized in terms of the the ratio  ${\mu_{\to\ell}(+1)}/{\mu_{\to\ell}(-1)} = e^{2 u_{\to\ell}}$, where $u_{\to\ell}$ is considered as an effective `cavity field' (cf. Eq.~(14.6) in Ref.~\cite{Mezard-book-2009}). This choice follows the custom of not mixing information flowing in opposite directions. Instead, the phase $\phi_\ell$ in Eq.~\eqref{e:mu->/<-mu} {\em does} mix information flowing in opposite directions. 

\

\subsubsection{Belief propagation as imaginary-time quantum dynamics}
In terms of the quantum-like cavity messages $\mu_{\to\ell}$ and $\mu_{\ell\leftarrow}$, the belief propagation equations \eqref{e:BP->} and \eqref{e:BP<-} become
\BE
\mu_{\to\ell}(x) &=& \int F_{\ell -1}(x^\prime, x)\mu_{\to\ell-1}(x^\prime)\mathrm{d}x^\prime,\label{e:muBP->}\\
\mu_{\ell\leftarrow}(x) &=& \int F_{\ell}(x, x^\prime)\mu_{\ell+1\leftarrow}(x^\prime)\mathrm{d}x^\prime,\label{e:muBP<-}
\EE
where we have done $x_\ell = x$, $x_{\ell - 1} = x^\prime$ in Eq.~\eqref{e:muBP->}, and $x_{\ell + 1} = x^\prime$ in Eq.~\eqref{e:muBP<-}. This contrasts with the standard formulation in terms of the $\nu$-messages described after Eq.~\eqref{e:BP<-}, where the messages must be renormalized at each iteration of the belief propagation equations (cf. Eq. (14.2) in Ref.~\cite{Mezard-book-2009}). Such iterative renormalization is avoided here because the normalization constant $\sqrt{Z}$ is the same for {\em all} quantum-like cavity messages. Equations~\eqref{e:muBP->} and \eqref{e:muBP<-} are the analogous of Eqs.~\eqref{e:quantum-like} and \eqref{e:quantum-like*} in Appendix~\ref{s:Q-MP}, and the exact equivalent of Eq. (2.16) in Ref.~\cite{zambrini1986stochastic} and its adjoint, respectively. (Indeed, the integrals in the right hand side of Eqs.~\eqref{e:muBP->} and \eqref{e:muBP<-} are the imaginary-time analogous to the integral in Eq.~\eqref{e:realGaussian}, where the cavity messages play the role of wave functions and the kernel in Eq.~\eqref{e:F} correspond to the kernel in Eq.~\eqref{e:K_T+V}; alternatively, we can also use the kernel in Eq.~\eqref{e:Main_KrealEM} or any generic kernel.) 

Due to the Gaussian term in the factors $F$ (see Eqs.~\eqref{e:F}, \eqref{e:Main_K_T+V} and \eqref{e:Main_classicalH}), the integrals in Eqs.~\eqref{e:muBP->} and \eqref{e:muBP<-} can be approximated to first order in $\epsilon$ (in a way similar to that of the integral in Eq.~\eqref{e:path_step1}). Indeed, since $\epsilon\to 0$, the real Gaussian factor associated to the kinetic term in Eq.~\eqref{e:Main_classicalH} is exponentially small except in the region where ${x - x^\prime = O(\sqrt{\hbar\epsilon/m})} $. This allow us to estimate the integral to first order in $\epsilon$ by expanding the $\mu$ terms in Eqs.~\eqref{e:muBP->} and \eqref{e:muBP<-} around $x$ up to second order in ${x - x^\prime}$. Consistent with this approximation to first order in $\epsilon$, we can also do $\exp{[- V(x)\epsilon / \hbar]} = 1 - V(x)\epsilon / \hbar + O(\epsilon^2) $ in factors $F$ (see Eqs.~\eqref{e:F}, \eqref{e:Main_K_T+V} and \eqref{e:Main_classicalH}). In this way we get the equations (cf. Eqs.~\eqref{e:realGaussian} and~\eqref{e:C*psi})
\begin{widetext}
\BE
\mu_{\to\ell}(x) &=& \mu_{\ell-1}(x) -\frac{\lambda\epsilon}{T} V_\ell(x)\mu_{\to\ell}(x) + \frac{T\epsilon}{2m\lambda}\frac{\partial^2\mu_{\to\ell}(x)}{\partial x^2} + O(\epsilon^2),\\
\mu_{\ell\leftarrow}(x) &=& \mu_{\ell+1\leftarrow}(x) -\frac{\lambda\epsilon}{T} V_\ell(x)\mu_{\ell\leftarrow}(x) + \frac{T\epsilon}{2m\lambda}\frac{\partial^2\mu_{\ell\leftarrow}(x)}{\partial x^2} + O(\epsilon^2). 
\EE
\end{widetext}
Let ${\mu_\to (x, \ell\epsilon) = \mu_{\to\ell}(x)}$ and ${\mu_\leftarrow (x, \ell\epsilon) = \mu_{\ell\leftarrow}(x)}$, and expand ${\mu_{\to}(x,t-\epsilon) = \mu_\to(x,t) - \epsilon\dot{\mu}_{\to}}(x,t)$ as well as ${\mu_{\leftarrow}(x,t+\epsilon) = \mu_{\leftarrow} + \epsilon\dot{\mu}_{\leftarrow}(x,t)}$, where $t = \ell\epsilon$ and the dot operator stands for time derivative. So, taking $\epsilon\to 0$ yield
\begin{widetext}
\BE
-\frac{T}{\lambda} \frac{\partial\mu_{\to}(x, t)}{\partial t} &=& -\frac{T^2}{2m\lambda^2}\frac{\partial^2\mu_{\to}(x, t)}{\partial x^2} +V(x, t)\mu_{\to}(x, t)  ,\label{e:EQBP->} \\
\frac{T}{\lambda}\frac{\partial\mu_{\leftarrow}(x, t)}{\partial t} &=& -\frac{T^2}{2m\lambda^2}\frac{\partial^2\mu_{\leftarrow}(x, t)}{\partial x^2} +V(x,t)\mu_{\leftarrow}(x, t)  ,\label{e:EQBP<-}
\EE
\end{widetext}
which yields precisely the imaginary-time Schr\"odinger equation and its adjoint, with $T/\lambda$ playing the role of Planck constant $\hbar$. Indeed, Eqs.~\eqref{e:EQBP->} and \eqref{e:EQBP<-} are equivalent to Eqs. (2.1) and (2.17) in Ref.~\cite{Zambrini-1987}; the analogous of $\theta$ and $\theta^\ast$ therein are here $\mu_{\leftarrow}$ and $\mu_{\to}$, respectively. 

We can also use the kernel in Eq.~\eqref{e:Main_KrealEM} and obtain the imaginary-time Schr\"odinger equation for a particle in an electromagnetic field, or use any generic kernel and obtain the corresponding Schr\"odinger equation.

Using Eq.~\eqref{e:mu_def} we can write Eqs.~\eqref{e:P_BP} and \eqref{e:p_BP} as
\BE
\mathcal{P}_\ell(x_\ell , x_{\ell +1}) &=& F_\ell(x_\ell, x_{\ell +1}) \mu_{\to\ell}(x_\ell)\mu_{\ell+1\leftarrow}(x_{\ell +1}),\label{e:P_muBP}\\
p_\ell(x_\ell ) &=& \mu_{\to\ell}(x_\ell)\mu_{\ell\leftarrow}(x_{\ell }).\label{e:p_muBP}
\EE
So, the forward and backward transition probabilities are given by
\begin{widetext}
\BE
\mathcal{P}^+_\ell(x_{\ell +1}| x_\ell ) &=& \frac{\mathcal{P}_\ell(x_\ell , x_{\ell +1})}{p_\ell(x_\ell )} = F_\ell(x_\ell, x_{\ell +1}) \frac{\mu_{\ell+1\leftarrow}(x_{\ell +1})}{\mu_{\ell\leftarrow}(x_\ell)},\label{e:P+_muBP}\\
\mathcal{P}^-_\ell(x_\ell | x_{\ell +1}) &=& \frac{\mathcal{P}_\ell(x_\ell , x_{\ell +1})}{p_{\ell+1}(x_{\ell+1} )} = F_\ell(x_\ell, x_{\ell +1}) \frac{\mu_{\to\ell}(x_\ell)}{\mu_{\to\ell+1}(x_{\ell +1})},\label{e:P-_muBP}\\
\EE
\end{widetext}
which are analogous to Eqs.~\eqref{e:KEQM+} and~\eqref{e:KEQM-} in Appendix~\ref{s:Q-MP}, and the exact equivalent of Eqs.~(2.12) and (2.11) in Ref.~\cite{Zambrini-1987}, respectively. 

\subsection{Euclidean quantum mechanics: From linear chains to cycles}\label{s:EQM}

As indicated by the third line of Eq.~\eqref{e:MarkovBayes}, and discussed in more detail in Appendix~\ref{s:Q-MP}, it is always possible to choose the factors as $F_\ell = K_\ell$ and obtain an equivalent description where the imaginary-time wave functions $\theta_\ell$ have no phase. Moreover, in the case of a chain the initial $\mu_{\to 1}$ and final messages $\mu_{n\leftarrow}$ are completely determined by the factors $F_1$ and $F_{n-1}$ through Eqs.~\eqref{e:muBP->} and \eqref{e:muBP<-}. This is not true, however, when the stochastic process takes place on a cycle, i.e. (see Fig.~\ref{f:cavity_cycle}; cf. Eq.~\eqref{e:P_prodF})
\begin{equation}\label{e:ring_EQM}
\mathcal{P}_{\rm cycle}(x_1,\dotsc x_n) = \prod_{\ell=1}^{n} F_\ell(x_{\ell},x_{\ell+1}),
\end{equation}
where $x_{n+1}=x_1$ and the factor $F_n$ closes the chain. In contrast to what happen on a chain, the na\"ive belief propagation equations on a cycle, Eqs.~\eqref{e:muBP->} and \eqref{e:muBP<-}, are not exact anymore~\cite{Weiss-2000}. Indeed, while in a chain there are two nodes that can be clearly identified as initial and final points, in a cycle all nodes are topologically equivalent, i.e. there is no intrinsic distinction between first and last because the whole process is cyclical. 

So, it is not possible in general to decompose the joint distribution $\mathcal{P}_{\rm cycle}(x_1,\dotsc x_n)$ as a Markov chain. The best we can do in general is
\be\label{e:Bernstein} 
\mathcal{P}_{\rm cycle}(x_1,\dotsc ,x_n) = \mathcal{P}_1(x_1, x_n) \prod_{\ell = 1}^{n-2} \mathcal{P}^+_{\ell}(x_{\ell + 1}|x_\ell , x_n),
\ee
which has the structure of a Berstein process (see e.g. the integrand in Eq.~(2.7) in Ref.~\cite{Zambrini-1987}, where $\mathcal{P}(x_1 , x_n) \to m(x, y)$ therein and $\mathcal{P}^+_{\ell}\to h$ therein). Equation~\eqref{e:Bernstein} is related to the fact that we can turn a cycle into a chain by removing a factor, say $F_n(x_n , x_1)$ (see Fig.~\ref{f:cavity_cycle}), which requires to condition on the two arguments of the factor, i.e. $x_1$ and $x_n$. 

Furthermore, running the belief propagation dynamics described by Eqs.~\eqref{e:EQBP->} and \eqref{e:EQBP<-} on a cycle does not lead to exact results anymore~\cite{Weiss-2000}. However, since removing a factor $F_n(x_n,x_1)$ (see Fig.~\ref{f:cavity_cycle}) turns the cycle into a chain (with variables $x_1$ and $x_n$ clamped), Eqs.~\eqref{e:EQBP->} and \eqref{e:EQBP<-} become exact again for fixed values of $x_1$ and $x_n$. We can implement the clamping of variables $x_1$ and $x_n$ by keeping one of the messages associated to each of the two variables equal to a Dirac delta peaked at the value $x_1^\ast$ and $x_n^\ast$ to which the corresponding variables are clamped, i.e. by keeping the messages always equal to 
\BE
\mu_{\to 1}(x_1) &=& \mu^{x_1^\ast}_{\to 1}(x_1) \equiv \delta(x_1 - x_n^\ast),\\
\mu_{n\leftarrow }(x_n) &=& \mu^{x_n^\ast}_{n\leftarrow }(x_n) \equiv \delta(x_n - x_n^\ast).
\EE 
Now, messages $\mu_{\to 1}^{x_1^\ast}$ and $\mu_{n\leftarrow}^{x_n^\ast}$ are the imaginary-time analogous of the quantum position eigenstates $\left|x_1^\ast\ket$ and $\bra x_n^\ast\right|$, respectively, where ${\bra x\right|\left. x^\prime\ket = \delta(x-x^\prime)}$. Furthermore, any quantum pure state $\left| \psi\ket = U \left|x^\ast\ket$ can be written as the product of an eigenstate $\left|x^\ast\ket$ and a unitary operator $U = \left|\psi\ket\bra x^\ast\right| + \sum_{x\neq x^\ast}\left|\phi_x\ket\bra x\right|$, where  $\left|\psi\ket\bra \psi\right| + \sum_{x\neq x^\ast}\left|\phi_x\ket\bra \phi_x \right| = I$, which can be implemented via a suitable Hamiltonian. Similarly, any imaginary-time pure state can be written as the product of an eigenstate, like $\mu_{\to 1}^{x_1^\ast}$ or $\mu^{x_n^\ast}_{n\leftarrow}$ above, and a kernel given by some suitable factors $F$. 
 
To summarize, for graphical models with circular topology Eqs.~\eqref{e:EQBP->} and \eqref{e:EQBP<-} are exact as long as the initial and final states are the eigenstates $\mu_{\to 1}^{x_1^\ast}$ and $\mu_{n\leftarrow}^{x_n^\ast}$. Furthermore, since it is not generally possible to factorize $\mathcal{P}_{\rm cycle}(x_1, \dotsc ,x_n)$ as a Markov chain, the phase is non-trivial. So, this is somehow similar to the imaginary-time version of the two-state vector formalism of quantum mechanics~\cite{reznik1995time, aharonov2010time}. In Sec.~\ref{s:3rd} we will use a different approach to show that a graphical model with circular topology also leads to the imaginary-time version of von-Neumann equation, which only requires an initial condition.

\section{Third-person perspective}\label{s:3rd}
Here we will more thoroughly discuss how we can interpret observations as internal representations (see Fig.~\ref{f:realisticobserver}) and how taking into account the observer leads to an interpretation of experiments as circular interactions. If you dear reader already agree with this view, you could skip straight to Sec.~\ref{s:imaginary_vN}, where we more thoroughly discuss the arguments in Ref.~\cite{realpe2017quantum} that show that circularity entails non-commutativity and the imaginary-time von Neumann equation (cf. Sec.~\ref{s:EQM}). Although our discussion here is restricted to kernels, or factors, with non-negative entries, in Appendix~\ref{s:neg_prob} we discuss how this discussion can be extended to kernels with negative entries, associated to real non-stoquastic Hamiltoinan operators.
\subsection{Observations as internal representations}

To fix ideas, consider an artificial observer, Alice, whose `brain' is a computer, i.e. a physical realization of a Turing machine. A possible architecture of such an artificial observer based on current machine learning technology is discussed in detail in Fig.~\ref{f:realisticobserver}. Generally speaking, such an artificial observer has two major components: (i) a feature-extraction algorithm which allows the observer to trim the raw data provided by the external world to extract relevant patterns from it, reducing its dimensionality; (ii) a Turin machine that operates on the relevant features extracted by the component described in (i), allowing the artificial observer to detect potential relationships between features, manipulate such features, generate actions based on those features to control the external world, and implement self-reference (see Appendix~\ref{s:recursion}).

According to recent research~\cite{dehaene2014consciousness} we expect components (i) and (ii) to be associated with unconscious and conscious information processing, respectively, in the case of humans (see Appendix~\ref{s:easy} for a summary of some relevant insights, particularly items (i) and (vi)). More specifically, component (i) quickly pre-process the external information to extract the relevant features that have the potential to become conscious percepts for the observer (see item (i) in Appendix~\ref{s:easy}); component (ii) provides a slower but more powerful way to process the relevant information extracted by component (i) (see item (vi) in Appendix~\ref{s:easy}). 

Suppose now that Alice receives, through her vision input channel, raw data generated by both the light scattered from a switch and the light radiated by a lamp (see Fig.~\ref{f:circular}a). Suppose also that Alice has access to some algorithm that allows her to extract the feature that both objects have two relevant states, which can be labeled {\sc On} and {\sc Off} for the switch, {\sc Light} and {\sc Dark} for the lamp---for a concrete example of a feature-extraction algorithm based on deep learning see Fig.~\ref{f:realisticobserver}. We say that Alice has observed the external system when she has acquired an internal representation of it, denoted in Fig.~\ref{f:circular}a by enclosing a replica of the system within (green) quotation marks. 

Such internal representation requires a physical implementation in Alice's hardware. Furthermore, assume that Alice can have a causal model of the external world, i.e. whether turning the switch {\sc On} causes the lamp to radiate {\sc Light}. This information is represented in Fig.~\ref{f:circular}a by the green line joining Alice's internal representations of the switch and the lamp. Such a line actually stands for an arrow whose direction depends on whether we interpret Alice's internal model as a simulation of the external system (see Fig.~\ref{f:circular}b) or in the so-called {\it ideomotor view}~\cite{pezzulo2006actions}, where the cause-effect relationship is, in a sense, reversed: Alice's representation of the intended effect (e.g. lamp in state {\sc Light}) of her action (e.g. turn switch {\sc On}) is the cause of the action. In other words, it is not the action that produces the effect, but rather the internal representation of the effect that produces the action~\cite{pezzulo2006actions}. 

To be more precise, let $v\in \mathcal{V}$ denote the state of the external system under investigation and $u\in \mathcal{U}$ denote Alice's internal representation of it (see Fig.~\ref{f:circular}d). Here $\mathcal{U}$ and $\mathcal{V}$ denote the corresponding sets of states for the systems that are internal and external to Alice, respetively. For simplicity, we will assume that $\mathcal{U}=\mathcal{V}$, i.e. we will describe only the features of the external system considered relevant by Alice, rather than the whole raw data (see Fig.~\ref{f:realisticobserver}). Indeed, both the external world and Alice's internal representation of it are here described from the perspective of an external observer that can also extract the same features from both Alice's internal and external world (see Fig.~\ref{f:first}). 

\subsection{Experiments as circular interactions}\label{s:exps_circular}
\subsubsection{Simulation interpretation}\label{s:simulation}
In this section we formalize {\bf Principle I} and show that treating the observer as a physical system which is part of the experimental setup implies that experiments can be represented by graphical models with circular topology (see Figs.~\ref{f:third}, \ref{f:cavity_cycle}, and \ref{f:circular}; cf. Refs.~\cite{hoffman2014objects,fields2018conscious}). We will also show that such a circular topology naturally leads to a non-commutative probability theory. Here, we are only interested in the formal probabilistic structure of the theory, not in the specific representation of each probability distribution in terms of physical quantities, such as mass, charge, etc. Finally, we will also show that when the interactions associated to the observer are neglected, we recover the standard representation of physical systems by linear graphical models, i.e. chains. 

Assume that external system is in an initial state $v_\textrm{i}$ and that Alice's corresponding internal representation is $u_\textrm{i}$ (see Fig.~\ref{f:circular}d). Similarly, assume the final state of the external system is $v_\textrm{f}$ and that Alice's corresponding internal representation is $u_\textrm{f}$. Let $\mathcal{P}_\textrm{ext}(v_\textrm{f}|v_\textrm{i})$ be the probability for the external system to evolve from the initial state $v_\textrm{i}$  to the final state $v_\textrm{f}$ (pink arrow in Fig.~\ref{f:circular}d). Let $\mathcal{P}_\textrm{prep}(v_\textrm{i}|u_\textrm{i})$ be the probability that the system is in state $v_\textrm{i}$ when Alice's internal representation is $u_\textrm{i}$ (red arrow in Fig.~\ref{f:circular}d); this can be interpreted as Alice's (possibly noisy) preparation of the initial state. Let $\mathcal{P}_\textrm{meas}(u_\textrm{f}|v_\textrm{f})$ be the probability that, when the final state of the external system is $v_\textrm{f}$, Alice's corresponding internal representation is $u_\textrm{f}$ (blue arrow in Fig.~\ref{f:circular}d); this can be interpreted as a (possibly noisy) measurement of the final state. 

Now, in what we here refer to as the `simulation interpretation' (see Fig.~\ref{f:circular}b) we can define $\mathcal{P}_\textrm{sim}(u_\textrm{f}|u_\textrm{i})$ as the probability that Alice's internal representation $u_\textrm{i}$ of the initial state evolves towards the representation $u_\textrm{f}$ of the final state (green arrow). This dynamics is expected to be a faithful simulation of the dynamical evolution of the external system. The joint probability of all variables can therefore be written as
\begin{widetext}
\begin{equation}\label{e:PSI}
\begin{split}
\mathcal{P}_\textrm{SI}(u_\textrm{i},v_\textrm{i},u_\textrm{f},v_\textrm{f}) &= \mathcal{P}_\textrm{prep}(v_\textrm{i}|u_\textrm{i}) p_\textrm{prep}(u_\textrm{i}) \mathcal{P}_\textrm{ext}(v_\textrm{f}|v_\textrm{i}) \mathcal{P}_\textrm{meas}(u_\textrm{f}|v_\textrm{f}) \mathcal{P}_\textrm{sim}(u_\textrm{f}|u_\textrm{i})\\
&= F_{\rm prep}(u_{\rm i},v_i) F_{\rm ext}(v_{\rm i},v_{\rm f}) F_{\rm meas}(v_{\rm f}, u_{\rm f}) F_{\rm sim}(u_{\rm f}, u_{\rm i})
\end{split}
\end{equation}
\end{widetext}
where $p_\textrm{sim}(u_\textrm{i})$ stands for the probability that Alice's representation of the initial state is $u_\textrm{i}$. 

The second line of Eq.~\eqref{e:PSI} emphasizes the factor graph representation, where the factors correspond to the probabilities with the same subscript. The most relevant feature of Eq.~\eqref{e:PSI} is that it represent a graphical model on a {\em circular} topology (see Figs.~\ref{f:third}, \ref{f:cavity_cycle}, and \ref{f:circular}). As  first discussed in Ref.~\cite{realpe2017quantum}, and reviewed in Sec.~\ref{s:circulatiry_noncommutativity} below, circularity entails non-commutativity.

\subsubsection{Ideomotor interpretation}\label{s:ideomotor}

The simulation interpretation discussed in the previous section assumes the observer passively models the external world; she does not have the opportunity to interact with it, to control it. Here we describe a more active interpretation of the observer, where she can intervene the system (cf. Refs.~\cite{hoffman2014objects,fields2018conscious}). We expect this to be a more faithful representation of what actually happens in an experiment and, as we shall see in Sec.~\ref{s:first} it is the one that is consistent with quantum dynamics.

So, suppose that Alice performs an experiment to {\it learn} whether the state of the switch (i.e. {\sc On} or {\sc Off}) has a causal influence on the state of the lamp (i.e. {\sc Light} or {\sc Dark}). Again, we say Alice has observed the external system when she has acquired an internal representation of it, denoted by enclosing a replica of the system within quotation marks (see Fig.~\ref{f:circular}a). And such an internal representation requires a physical implementation in Alice's hardware, as we have already mentioned. 

To perform the experiment, Alice first `decides' which intervention to do, i.e. where to position the switch, and then `acts' by moving the switch accordingly; such action requires a physical interaction represented by a red arrow in Fig.~\ref{f:circular}a (cf. Fig.~\ref{f:third}). After preparing the system via her interventions, Alice leaves the system evolve (see pink arrow in Fig.~\ref{f:circular}a) and measure the state of the lamp. Such measurement also requires a physical interaction represented by the blue arrow in Fig.~\ref{f:circular}a (cf. Fig.~\ref{f:third}). 

Now, to test a probabilitic theory we need to repeat an experiment a number of times large enough to have statistically significant results; we need to do so even if the theory is deterministic because otherwise how can we be sure it is indeed deterministic? By running the experiment $m$ times Alice can obtain a dataset $\mathcal{D} = \{(u_\textrm{i}^{(1)},u_\textrm{f}^{(1)}),\dotsc ,u_\textrm{i}^{(m)},u_\textrm{f}^{(m)})\}$, where $u_\textrm{i}^{(d)}$ and $u_\textrm{f}^{(d)}$ respectively stand for Alice's internal representations of the state of the switch and the lamp at the $d$-th run of the experiment. Alice can use $\mathcal{D}$ to build a causal model represented by the green line in Fig.~\ref{f:circular}a joining the two representations; this line actually stands for an arrow whose direction depends on whether we assume the simulation interpretation discussed in the previous section or the ideomotor interpretation discussed in this section. If we interpret Alice's causal model as a simulation of the external system (see Fig.~\ref{f:circular}b), then the arrow corresponding to the green line should point in the same direction of the external (pink) arrow.

Alternatively, as we already mentioned above, in the so-called {\it ideomotor view}~\cite{pezzulo2006actions}, Alice's causal model is reversed (see Fig.~\ref{f:circular}c): Alice's representation of the intended effect (e.g. lamp in state {\sc Light}) of her action (e.g. turn switch {\sc On}) is the cause of the action. In other words, it is not the action that produces the effect, but rather the internal representation of the effect that produces the action~\cite{pezzulo2006actions}. We expect this to be a more faithful representation of the situation in an experiment. However, this leads to a graphical model that is a directed loop representing reciprocal causation, a subject that to the best of our knowledge is not as developed as the most standard models of causality based on directed acyclical graphs, i.e. with no loops (see e.g. Ref.~\cite{spirtes1993causation}, chapter 12.1). Nonetheless, circular causality is a common theme in cognitive science~\cite{varela2017embodied}.

However, we can use the principle of maximum caliber introduced in Sec.~\ref{s:MaxCal} to extend the derivation of Markov chains~\cite{presse2013principles} to cycles (see Sec.~\ref{s:MaxCal}).  This yields a factor graph like the one in Fig.~\ref{f:cavity_cycle}a. Alternatively, if $\mathcal{P}_{\rm dec}(u_{\rm i}|u_{\rm f})$ is the probability for the agent to decide to prepare $u_{\rm i}$ if she wants to observe $u_{\rm f}$ we could write (cf. Ref.~\cite{fields2018conscious})
\begin{widetext}
\begin{equation}\label{e:IMV1}
\begin{split}
\mathcal{P}_\textrm{IM}(u_\textrm{i},v_\textrm{i},u_\textrm{f},v_\textrm{f}, u_\textrm{i}^\prime) &= \mathcal{P}_\textrm{prep}(v_\textrm{i}|u_\textrm{i})  \mathcal{P}_\textrm{ext}(v_\textrm{f}|v_\textrm{i}) \mathcal{P}_\textrm{meas}(u_\textrm{f}|v_\textrm{f}) \mathcal{P}_\textrm{dec}(u_\textrm{i}^\prime|u_\textrm{f})\\
&= \tilde{F}_{\rm prep}(u_\textrm{i},v_\textrm{i}) F_{\rm ext}(v_\textrm{i},v_\textrm{f}) F_{\rm meas}(v_\textrm{f}, u_\textrm{f}) F_{\rm dec}(u_\textrm{f}, u_\textrm{i}^\prime),
\end{split}
\end{equation}
\end{widetext}
for the probability to observe a path ${u_\textrm{i} \to v_\textrm{i}\to v_\textrm{f} \to u_\textrm{f}\to u_\textrm{i}^\prime}$ in one cycle. However, if we assume that for the observer to prepare the external system in state $v_\textrm{i}$ she has to always have the same internal representation $u_\textrm{i}$, e.g. $\mathcal{P}_{\rm prep}(v_\textrm{i}|u_\textrm{i}) = \delta (v_\textrm{i}-u_\textrm{i})$, then $u_\textrm{i} = u_\textrm{i}^\prime$ once the experiment has stabilized; this is consistent with von Foerster's view~\cite{foerster1981observing} that the objects we percieve can be considered as tokens for the behavior of the organism that apparently creates stable forms~\cite{kaufman2016cybernetics}. Let us considere a thought experiment to better illustrate this point. 

\

\noindent{\bf Thought experiment:} Imagine, for instance, recording all the research process that led to the detection of the statistical regularity, or pattern, that we call `Higgs boson' in the massive dataset generated at the LHC~\cite{aad2012observation,chatrchyan2012observation}. Imagine that we now play such recording at a much faster speed to compress years of work into a few minutes of video. To avoid any potential prejudice to interfere, imagine that all researchers appearing in the video are instead robots (see e.g. Ref.~\cite{melnikov2018active}). We could say that what we observe in the video is itself a physical process where some physical systems, to some of which we perhaps attribute some notion of `agency', interact with others. We could also say that one of the outcomes of such myriad interactions is that the robots `learn' about the statistical regularities in the systems they investigate. Finally, we could also say that, in general, the robots would have to learn even what to call a `system', which experimental devices to built and how to build them, from their repeated interactions with the other `systems', including the interactions associated to the message passing or `discussions' among themselves. We are not concerned here with the whole learning process, but only with the last stage when the robots have already identified some regularities, or stabilities, in their interaction with the experimental devices.

An instance of such regularities or stabilities could be that the agent is involved in a circular path ${u_\textrm{i} \to v_\textrm{i}\to v_\textrm{f} \to u_\textrm{f}\to u_\textrm{i}}$ which always returns to the same initial state $u_{\rm i}$. The corresponding probability in this case would be given by
\begin{widetext}
\begin{equation}\label{e:IMV2}
\begin{split}
\mathcal{P}_\textrm{IM}(u_\textrm{i},v_\textrm{i},u_\textrm{f},v_\textrm{f}) &= \mathcal{P}_\textrm{prep}(v_\textrm{i}|u_\textrm{i})  \mathcal{P}_\textrm{ext}(v_\textrm{f}|v_\textrm{i}) \mathcal{P}_\textrm{meas}(u_\textrm{f}|v_\textrm{f}) \mathcal{P}_\textrm{dec}(u_\textrm{i}|u_\textrm{f})\\
&= \widetilde{F}_{\rm prep}(u_\textrm{i},v_\textrm{i}) F_{\rm ext}(v_\textrm{i},v_\textrm{f}) F_{\rm meas}(v_\textrm{f}, u_\textrm{f}) F_{\rm dec}(u_\textrm{f}, u_\textrm{i}).
\end{split}
\end{equation}
\end{widetext}
We can see that either the simulation interpretation or the ideomotor view produce the same factor graph topology (see Fig.~\ref{f:circular}d). From now on we will assume we can find such a cyclic Markov process via the maximum caliber principle (see Sec.~\ref{s:MaxCal}).

In summary, the most relevant feature from this analysis is that, once we take into account the observer as part of the experimental set-up, the topology of the interactions taking place in an experiment is circular (see Figs.~\ref{f:third}, \ref{f:cavity_cycle}, and \ref{f:circular}; cf. Refs.~\cite{hoffman2014objects,fields2018conscious}). In contrast, when the interactions associated to the observer are neglected, the aparent topology of interactions taking place in an experiment is that of a chain (see Fig.~\ref{f:circular}e). Notice that the initial and final nodes of a chain (e.g. $v_1$ and $v_3$ in Fig.~\ref{f:circular}e) interact with only one single node (e.g. $v_2$ in Fig.~\ref{f:circular}e), while the rest of the nodes (e.g. $v_2$ in Fig.~\ref{f:circular}e) interact with two nodes instead (e.g. $v_1$ and $v_3$ in Fig.~\ref{f:circular}e). This allows us to specify well-defined initial and final states and propagate them forward and backward in time, respectively, through the chain via the transition probabilities. This contrasts with the case of a circular topology (see Fig.~\ref{f:circular}d) where no single node is special in this sense: there is neither beginning nor end on a circle (see Sec.~\ref{s:foundations}). 

We will show in the following that this point of view is indeed useful and allows us to derive the formalism of quantum theory. Although these ideas could in principle be formulated using the language of active or reinforcement learning, we will stick here to the more traditional and more general language of probabilistic graphical models~\cite{Mezard-book-2009,Koller-2009,Kschischang-2001,Weiss-2000}.

\subsection{Imaginary-time von Neumann equation}\label{s:imaginary_vN}
\subsubsection{Circularity entails non-commutativity}\label{s:circulatiry_noncommutativity}

We already discussed in Sec.~\ref{s:EQM} that circularity leads to something that looks somehow similar to the imaginary-time version of the two-state vector formalism of quantum mechanics~\cite{reznik1995time, aharonov2010time}. Here we will review the derivation of the imaginary time von Neumann equation provided in Ref.~\cite{realpe2017quantum} for factor graphs with circular topologies (see Fig.~\ref{f:cavity_cycle}). In this case, the probability of observing a path of external visible variables $(v_1,\dotsc , v_n)$ is given by Eq.~\eqref{e:ring_EQM}, i.e. (see Fig.~\ref{f:cavity_cycle}a; cf. Eq.~\eqref{e:P_prodF}):
\begin{equation}\label{e:ring}
\mathcal{P}_{\rm cycle}(v_1,\dots ,v_n) = \prod_{\ell=1}^{n} F_\ell(v_{\ell},v_{\ell+1}),
\end{equation}
where $v_{n+1}=v_1$ and the factor $F_n(v_n,v_1)$ summarizes the chain of interactions usually ignored between initial and final states through the observer represented by the green chain in Figs.~\ref{f:cavity_cycle} and \ref{f:circular}---here we are assuming the normalization constant has been absorbed into the factors $F_\ell$ to simplify the notation.  For instance, in the case illustrated in Fig.~\ref{f:circular}d we can write $F_n(v_n , v_1)$ as the product of all pairwise factors involving at least one of the variables $a_1, u_\textrm{i},u_1,u_\textrm{f}, o_1$, marginalized over all those variables. This leaves only the dependence on the variables $v_\textrm{i}$ (through the factor involving variables $v_\textrm{i}$ and $a_1$) and $v_\textrm{f}$ (through the factor involving variables $v_\textrm{f}$ and $o_1$), summarized by the factor $F_n(v_n , v_1)$.

With some abuse of notation we will denote by $F_\ell$ the matrix with elements $F_\ell(v_\ell,v_{\ell+1})$; so, the marginal probability of a variable, say $v_1$, 
\be
p_1(v_1) = \sum_{v_2,\dotsc , v_n}\mathcal{P}_{\rm cycle}(v_1,\dots ,v_n),
\ee
according to Eq.~\eqref{e:ring} is given by the diagonal elements of the matrix product
\begin{equation}\label{e:P_1}
P_1 = F_1\cdots F_n,
\end{equation}
which defines the probability matrix $P_1$; i.e. $p_1(v_1) = P_1(v_1,v_1)$, much like the Born rule of quantum mechanics. Notice that $P_1$, being a matrix product, also contains off-diagonal terms $P_1(v_1, v_1^\prime)$. Similarly, the marginal probability $p_2(v_2)$ that the state of the external system is $v_2$ at the next time step is given by the diagonal elements of the probability matrix 
\be\label{e:P_2}
P_2=F_2\cdots F_n F_1,
\ee
and so on. In other words, the dynamical evolution of the probability matrices $P_\ell$ is given by the cyclical permutation of the factor matrices $F_\ell$ (cf. Ref.~\cite{svozil2018evolution}).

Now, if $F_1$ is invertible we can multiply Eq.~\eqref{e:P_1} by $F_1^{-1}$ and $F_1$ from the left and from the right, respectively (see Sec.~\ref{s:pure_causality} for an alternative approach that also works when some factors $F_\ell$ are non-invertible). In this way we can see that $P_2 = F_1^{-1} P_1 F_1$ and, extending the argument for any $\ell$, we can see that it is valid in general, i.e.
\begin{equation}\label{e:EQM}
P_{\ell+1} = F_\ell^{-1} P_\ell F_\ell.
\end{equation}
Equation~\eqref{e:EQM} is analogous to a Markovian update rule in the case of an open chain, in the sense that it allows us to compute the probability matrix $P_{\ell+1}$ at time step $\ell +1$ from the probability matrix $P_\ell$ at the previous time step $\ell$. So, it is mathematically convenient to interpret the state of the external system at time step $\ell$ by the matrix $P_\ell$. The off-diagonal elements of this probability matrix encode information necessary to reconstruct all states step by step. In this sense, the state contains kinematic and dynamical information as suggested by Spekkens~\cite{spekkens2015paradigm}. 

If we assume that changes in the system during time steps of size $\Delta t>0$ are small, so we can write ${F_\ell= I - J_\ell \Delta t+ O(\Delta t ^2)}$, where $I$ is the identity matrix and $J_\ell$ is a matrix with non-positive off-diagonal elements; without loss of generality, we will assume $J_\ell = J$ for all $\ell$ of interest, and will refer to $J$ as the {\it dynamical matrix}. Similarly, ${F_\ell^{-1} = I + J\Delta t + O(\Delta t ^2)}$. In this way, Eq.~\eqref{e:EQM} becomes {${P_{\ell+1} = P_\ell + [J\Delta t , P_\ell]}$} or, in the continuous limit
\be\label{e:EvN}
\frac{\partial P}{\partial t} = [J, P].
\ee
So, circularity entails non-commutativity.

\subsubsection{Pure states and causality}\label{s:pure_causality}
We will now present in more detail the argument outlined in Appendix A of Ref.~\cite{realpe2017quantum} related to the conditions enforced by the requirement that the probability matrix is (the imaginary-time version of) a pure state. Recall that any quantum pure state $\rho_\psi = \left|\psi\ket\bra\psi\right| = U \rho_v U^\dag $ can be written as the evolution of an eigenstate $\rho_v = \left| v\ket\bra v\right|$, with entries ${\bra v^\prime\right|\rho_v \left| v^{\prime\prime}\ket = \delta_{v v^\prime}\delta_{v v^{\prime\prime}}}$, via a unitary operator $U=\left| \psi\ket\bra v\right| + \sum_{v^\prime\neq v} \left| \chi_{v^\prime}\ket\bra v^\prime\right|$, where $\bra v\right|\left. v^\prime\ket = \delta_{v v^\prime}$ and $\left| \psi\ket\bra \psi\right| + \sum_{v^\prime\neq v} \left| \chi_{v^\prime}\ket\bra \chi_{v^\prime}\right| = I$. In the same way, any imaginary-time pure state $P_{\rm pure} = F^{-1} P^v F$ (see Eq.~\eqref{e:EQM}) can be written as the evolution of a pure state $P^v$ with only one single entry different from zero, i.e. an `eigenstate', via the imaginary-time version of $U$, which is $F$. Since the probability matrix $P^v$ has only one entry different from zero, at least one of the factors that defines it (see Eqs.~\eqref{e:P_1} and \eqref{e:P_2}) should be non-invertible, however, the approach that follows do not require invertibility of all factors $F_\ell$, only of the factor $F$ defining $P_{\rm pure}$ above.

So, without loss of generality, we will study the evolution from an initial probability matrix $P_1^v$ with entries 
\be\label{e:P=GFn}
P_1^v(v^\prime, v^{\prime\prime}) = \sum_{v_n} F_{\rm ext}(v^\prime, v_n) F_n({v_n, v^{\prime\prime}}) =  \delta_{v v^\prime}\delta_{v v^{\prime\prime}},
\ee
where the matrix $F_{\rm ext} = F_1\cdots F_{n-1}$ summarizes the product of the first $n-1$ factors in the left hand side of Eq.~\eqref{e:P_1}, which characterize the physical process taking place on the external system. So, the sum in Eq.~\eqref{e:P=GFn} is equivalent to the whole product in the left hand side of Eq.~\eqref{e:P_1}. 

Since all factors $F_\ell$ are non-negative so is $F_{\rm ext}$ and the sum in Eq.~\eqref{e:P=GFn} can be zero only if all its terms are zero. So, if $v^\prime\neq v$ then $F_{\rm ext}(v^\prime , v_n) = 0$ for all $v_n$, i.e. the corresponding matrix $F_{\rm ext}$ has all entries equal to zero except for those in row $v$. Similarly, if $v^{\prime\prime}\neq v$ then $F_n( v_n,v^{\prime\prime} ) = 0$ for all $v_n$, i.e. the corresponding matrix $F_n$ has all entries equal to zero except for those in column $v$. In other words, since the matrix $P_1^v = F_{\rm ext} F_n$ is a product of two matrices with non-negative entries and $P_1^v$ has only one entry different from zero, matrices $F_{\rm ext}$ and $F_n$ should only have one row and one column with non-zero entries, respectively. In other words, assuming the $v$ variables are discretized in steps of size $\xi$, we can write Eq.~\eqref{e:P=GFn} as
\begin{widetext}
\be\label{e:pure_matrices}
\begin{pmatrix} \ddots & \ddots  & \ddots & & & &  \\ \cdots & 0 & 0 & 0 & \cdots & & \\ & \cdots & 0 & 1 & 0 & \cdots & \\ & & \cdots & 0 & 0 & 0 & \cdots \\ & & & & \ddots & \ddots  & \ddots    \end{pmatrix} = \begin{pmatrix} & \vdots & \vdots & \vdots &  \\ \cdots & 0 & 0 & 0 & \cdots \\ \cdots & F_{\rm ext}(v, -\xi)  & F_{\rm ext}(v, 0) & F_{\rm ext}(v, \xi) & \cdots  \\ \cdots & 0 & 0 & 0 & \cdots \\ & \vdots & \vdots & \vdots &  \end{pmatrix} \begin{pmatrix} & & \vdots & \vdots & \vdots &  \\ & \cdots & 0 & F_n(-\xi, v) & 0 & \cdots  \\ & \cdots & 0 & F_n(0, v) & 0 & \cdots \\ & \cdots & 0 & F_n(\xi, v) & 0 & \cdots \\& & \vdots & \vdots & \vdots &   \end{pmatrix}
\ee
\end{widetext}
Now, since the dynamical evolution of $P_1^v$ is given by the permutation of factors $F_\ell$ (see Eqs.~\eqref{e:P_1} and \eqref{e:P_2}), the final probability matrix $P_n$ satisfies
\begin{widetext}
\be
P_n(v^\prime, v^{\prime\prime}) = \sum_{v_1}  F_n({v^\prime, v_1) F_{\rm ext}(v_1,  v^{\prime\prime}}) = F_n(v^\prime, v) F_{\rm ext}(v,  v^{\prime\prime});
\ee
\end{widetext}
the last equality follows because $F_{\rm ext}$ and $F_n$ are equivalent to a row and a column vector, respectively, and therefore only terms with $v_1 = v$ are different from zero (see Eq.~\eqref{e:pure_matrices}). $P_n$ is obtained by permuting the two matrices in the right hand side of Eq.~\eqref{e:pure_matrices}.

As the diagonal elements yield the corresponding probabilities, i.e. $P_n(v^\prime, v^\prime) = p_n(v^\prime)$ we can parametrize the factors via (cf. Eqs.~\eqref{e:mu->*<-mu} and \eqref{e:mu->/<-mu})
\BE
F_n(v^\prime, v) F_{\rm ext}(v,  v^{\prime}) &=& p_n(v^\prime),\\
\frac{F_n(v^\prime, v) }{F_{\rm ext}(v,  v^{\prime})} &=& e^{2\phi_n(v^\prime)}.
\EE
Hence the factors can be written as 
\BE
F_n(v^\prime, v) &=& \sqrt{p_n(v^\prime)} e^{\phi_n(v^\prime)},\\
F_{\rm ext}(v, v^\prime) &=& \sqrt{p_n(v^\prime)} e^{-\phi_n(v^\prime)},
\EE
which look as imaginary-time wave functions (cf. Eqs.~\eqref{e:mu->} and \eqref{e:<-mu}). So, the probability matrix can be written accordingly as 
\be\label{e:P_n_pure}
P_n(v^\prime, v^{\prime\prime}) = \sqrt{p_n(v^\prime)p_n(v^{\prime\prime})}, e^{\phi_n(v^\prime)-\phi_n(v^{\prime\prime})}. 
\ee
This analysis is valid for all initial eigenstates $P^v_1$ corresponding to all possible values of $v$. We can therefore write Eq.~\eqref{e:P_n_pure} in terms of the imaginary-time version of a uitary operator that includes the corresponding transformation for all possible values of $v$, in the same way it is usually done in quantum mechanics. 

It might be tempting to use the additional degree of freedom $\phi_n$ and apply the Madelung transformation~\cite{madelung1927quantentheorie} to get a perfect analogy with quantum mechanics (see e.g.~\cite{caticha2009entropic,caticha2011entropic}). However, this will add further mathematical structure associated with the periodicity of the wave function~\cite{Wallstrom-1994}, which we will argue in Sec.~\ref{s:first} arises from the implementation of the first-person perpective.

Now, following the notation introduced in Sec.~\ref{s:exps_circular} we can write $F_{\rm ext}(v_1, v_n) = \mathcal{P}_{\rm ext}(v_n|v_1)$. So, that $F_{\rm ext}(v_1, v_n) = 0$ for all $v_1\neq v$ could be interpreted as an intervention done {\em on} the external system enforcing that $v_1=v$. Using Pearl's {\sc do} calculus of causal inference, we can represent this intervention as $\textsc{do}(v_1=v)$; thus we can write $F_{\rm ext}(v_1, v_n) = {\mathcal{P}_{\rm ext}(v_n| \textsc{do} (v_1 = v))}$, using the convention that the right hand side, considered as a function of $v_1$, is zero for all $v_1\neq v$.

Similarly, since factor $F_n$ characterizes the physical processes related to the observer (see Figs.~\ref{f:cavity_cycle} and \ref{f:circular}), following the notation introduced in Sec~\ref{s:exps_circular} we can write $F_n(v_n , v_1) = \sum_{u_1}\mathcal{P}_{\rm prep}(v_1 | u_1) {G}(u_1, v_n)$, where $u_1$ is the observer's representation of $v_1$, and $G$ summarizes the remaining interactions regarding the observer, i.e. measurement, and decision or simulation. So, that factor $F_n(v_n , v_1) = 0$ for all $v_1\neq v$ could be interpreted as an intervention done {\em by} the agent which, extending Pearl's {\sc do} calculus notation, we here write as $\mathcal{P}_\textrm{prep}(\textsc{do}(v_1= v)| u_1)$. We can therefore write $F_n(v_n , v_1) = \sum_{u_1}\mathcal{P}_{\rm prep}(\textsc{do}(v_1 = v) | u_1) {G}(u_1, v_n)$, using the convention that the right hand side, considered as a function of $v_1$, is zero for all $v_1\neq v$. This could be interpreted as modeling the agent's intention of doing an intervention on the system. This also suggest how causality may be enforced by the need to break the intrinsic circularity of experiments (see Sec.~\ref{s:EQM} and Fig.~\ref{f:cavity_cycle}).

\section{First-person perspective}\label{s:first}
Here we discuss how turning from the third- to the first-person perspective leads to genuine, real-time quantum dynamics, effectively implementing a Wick rotation. We first discuss how the self-referential problem of describing the world from within, including ourselves, requires a complementary architecture of the first-person observer (Sec.~\ref{s:self}). Afterwards, we discuss how implementing such a complementary architecture leads to the von Neumann equation in real-time. Although here we restric our attention to transition kernels, or factors, with non-negative entries, in Appendix~\ref{s:neg_prob} we discuss how effective kernels with negative entries can also be included in this approach. 

\subsection{Self-reference and complementarity}\label{s:self}
Up to now our analysis has been based on Fig.~\ref{f:circular}, which is from the perspective of an external third-person observer not included in the figure: in this case you dear reader that has been looking at it from your own private first-person perspective. To include a model of you in the figure we would need a fourth observer and so on (see Fig.~\ref{f:first}); we get an infinite regress. A similar phenomenon happen when a system, such as the DNA molecule~\cite{watson1953molecular}, has to reproduce itself. A na\"ive approach may suggest that such a system must contain a copy of itself, and for such a copy to be able to reproduce too, the copy must also contain another copy; continuing with this argument we may soon conclude that a self-reproducing system must contain  infinite copies of itself. We know, however, that the architecture of the DNA molecule composed of two complementary strands, each generating a copy of the ohter strand, avoids such an infinite regress. Or like the painter, Alice, that wants to make a painting of the whole Universe, but cannot paint herself. A second painter, Bob, can paint the Universe, including Alice but not himself, and so on. Nevertheless, Alice and Bob could play complementary roles and mutually paint each other along with the rest of the Universe (see Fig.~\ref{f:first-third}). As we discussed in Sec.~\ref{s:big_1st_person} (see also Appendix~\ref{s:recursion}), using complementary pairs is a standard technique to deal with self-reference~\cite{sipser2006introduction, deacon2011incomplete, hofstadter2013strange}. 

Consider the example of self-printing programs, or quines, discussed in Sec.~\ref{s:big_1st_person} (see also Appendix~\ref{s:recursion} for a detailed description based on chapter 6 of Ref.~\cite{sipser2006introduction}): A Turing machine $\textsc{Self}=\textsc{Alice}\circ\textsc{Bob}$ that ignores its input and prints out a copy of its own description is composed of two parts, {\sc Alice} and {\sc Bob}. These two parts are complementary in the sense that the task of {\sc Alice} is to print out a description of {\sc Bob}, and vice versa. While {\sc Alice} directly prints out a description of {\sc Bob}, the latter has to compute or infer who {\sc Alice} is from its output to avoid a circular definition. In this sense, {\sc Alice} and {\sc Bob} run in opposite directions. We will see in Sec.~\ref{s:quantum} that something similar occurs in quantum theory. 

Furthermore, as we have already mentioned in Sec.~\ref{s:big_1st_person} (see also Appendix \ref{s:recursion}), an important aspect that allows a system to refer to itself is a `duality' between the active and passive roles that it can play. Consider, for instance, the active role played by a Turing machine {\sc TM} (the abstract version of an executable file) and the passive role played by a description of it {\sc ``TM''} (the abstract version of the code underlying the executable file). A universal Turing machine can be defined as $\textsc{UTM}(\textsc{``TM''}, w)= \textsc{TM}(w)$, where the first argument, {\sc ``TM''},  is interpreted as the description of a Turing machine {\sc TM} and the second argument, $w$, as an input to the Turing machine {\sc TM} (see Ref.~\cite{moore2011nature}, chapter 7). Such a universal Turing machine {\sc UTM} runs the Turing machine {\sc TM} described by its first argument on the input $w$ on its second argument; this is represented here as {\sc TM}$(w)$. If we chose $w = \textsc{``TM''}$ we obtain $\textsc{UTM}(\textsc{``TM''}, \textsc{``TM''})=\textsc{TM}(\textsc{``TM''})$, i.e. a program running on itself.

Something similar happens in logic where every logical formula can be indexed by a number, a `G\"odel number'~\cite{godel1931formal}. In this way, we can create a logical formula $\mathcal{F}(\bar n)$ that has a free variable $\bar n$, an integer, which can be interpreted by the mathematician as the G\"odel number of another logical formula. It is possible to set $\bar n = \bar n^\ast$ to be equal to the G\"odel number $\bar n^\ast$ of the formula $\mathcal{F}(\bar n)$ itself, and in this way $\mathcal{F}(\bar n^\ast)$ can be a self-referential logical statement. In quantum theory a physical observer can also play dual roles: active when observing another system, and passive when being observed.

Another example of self-reference is the so-called autocell~\cite{deacon2006reciprocal} or autogen~\cite{deacon2011incomplete} (chapter 10), which consists of two molecular processes: autocatalysis and self-assembly. These two processes are complementary in the following sense: the autocatalysis produces molecules that tend to spontaneously self-assemble, while such self-assembling molecules can form closed structures that isolate the catalizers from the environment and prevent them from getting exhausted. If such container breaks, the catalyzers can produce further self-assembling molecules and re-generate or replicate themselves, if they are in a suitable environment . This is a biological analogous of von Neumann self-replicating machines~\cite{neumann1966theory}. It is curious that von Neumann was aware of the infinite regress that plagued both the process of self-replication~\cite{neumann1966theory} and the process of measurement in quantum mechanics~\cite{von1955mathematical}. While he tamed self-reference in the former he did not tackle the latter, as far as we know.

\subsection{First-person observers and quantum dynamics}\label{s:quantum}
\subsubsection{General considerations}
Following the insights of self-printing programs, formalized by the recursion theorem (see Appendix~\ref{s:recursion}), we here explore in more detail the assumption that an observer capable to have a representation of itself should be composed of two parts, {\sc Alice} and {\sc Bob}, which play dual roles: active when observing and passive when being observed (see Figs.~\ref{f:self-observer} and \ref{f:noneq_self-observer}; cf. Fig.~\ref{f:first-third})---we are using here the ideomotor interpretation described in Sec.~\ref{s:ideomotor} which we will argue in Sec.~\ref{s:eq} is the one consistent with quantum theory, while the simulation interpretation described in Sec.~\ref{s:simulation} is not.  We will use the name \textsc{Self} to refer to a generic instance of such self-referential observers; using a notation similar to that in Appendix~\ref{s:recursion}, we could write $\textsc{Self}=\textsc{Alice}\circ\textsc{Bob}$. The intuition is that {\sc Bob} can help {\sc Alice} infer her subjective state by acting as a kind of mirror. In the case of humans this mirroring mechanism might be implemented by the so-colled mirror neuron system~\cite{hamzei2016dual,ramachandran2007neurology}. As suggested in Ref.~\cite{ramachandran2007neurology}, mirror neurons are usually activated when we temporarily `adopt' a third-person perspective of ourselves, and self-awareness may be understood as the use of these neurons for `looking at myself as if someone else is looking at me'. As discussed further in Sec.~\ref{s:architecture}, the insights from the recursion theorem suggest that the neural architecture of the self is expected to be composed of two complementary systems, which might explain why our brain is divided into hemispheres (see Fig.~\ref{f:brain_self}).

Figure~\ref{f:self-observer} describes the architecture of such a self-referential observer {\sc Self} (cf. Fig.~\ref{f:first-third}). As we said, {\sc Self} consists of two sub-observers, {\sc Alice} and {\sc Bob}, that mutually observe each other, playing complementary roles as either observing subjects or observed objects. Figures~\ref{f:self-observer}a, b show this mutual observation form the perspective of an external, or third-person observer, {\sc Chris}, while Fig.~\ref{f:self-observer}c shows the first-person observer \textsc{Self}  (cf. Fig.~\ref{f:first-third}b). Notice that the physical process is one and the same for {\sc Chris}, except for the interpretation {\sc Chris} does of the roles played by {\sc Alice} and {\sc Bob}. 

Here it is important to distinguish between the first- and third-person descriptions of the changes in an observer's state. For instance, {\sc Chris} in Fig.~\ref{f:self-observer} is an observer external to the system composed of {\sc Alice} and {\sc Bob}. So, {\sc Chris} can apply the results discussed in Sec.~\ref{s:3rd}; in particular, {\sc Chris} can describe the changes in the states of {\sc Alice} and {\sc Bob} using Eq.~\eqref{e:EvN} for each. To distinguish the third- and first-person perspectives, we here denote the changes described by Eq.~\eqref{e:EvN} with a superscript `3rd'. So, we will denote the changes in the state of observer $O \in\{ A, B\}$ as $\Delta^{3\rm rd} P_O$ which, according to Eq.~\eqref{e:EvN}, for small time-intervals $\Delta t$ are given by
\be\label{e:Delta3rdPO}
\Delta^{3\rm rd} P_O = \Delta t [J, P_O].
\ee

Consider now Fig.~\ref{f:first-third} again. When {\sc Alice} takes a picture of {\sc Bob}, which is a physical system, such a picture has to be physically represented in {\sc Alice}'s camera's hardware, i.e. both the system being observed (i.e. {\sc Bob}) and the corresponding representation in the observer's `brain' (i.e. the camera's hardware) are made of the same physical stuff. Similarly in Fig.~\ref{f:self-observer}, when {\sc Alice} observes the changes in {\sc Bob}'s state---which should coincide with those observed by {\sc Chris}', i.e. $\Delta^{3\rm rd} P_B = \Delta t [J, P_B]$, because they are both external to {\sc Bob}---such changes should be represented in {\sc Alice}'s hardware with a change in her own state. Such internal changes in {\sc Alice}'s state, which represent for her the external changes she observes in {\sc Bob}'s state, are {\sc Alice}'s first-person changes $\Delta^{1\rm st} P_A$, denoted here with a superscript `1st' . So, while {\sc Alice} cannot directly observe the changes in her own state, she can still infer them from the changes she observes in {\sc Bob}'s state. So, the changes in state associated to observer $O\in\{A, B\}$, thought of as a component of the first-person observer {\sc Self}, will be denoted by $\Delta^{1\rm st} P_O$. In the next two sections we will show how this strategy allows {\sc Alice} and {\sc Bob} to iteratively construct their subjective state. To do so, it will be useful to remember a couple of things we have already discussed. 

First, as we mentioned in Sec.~\ref{s:pure_causality} (see paragraph before Eq.~\eqref{e:P=GFn}), without loss of generality we can assume that the initial states $P_{A, 0}$ and $P_{B, 0}$ for both {\sc Alice} and {\sc Bob}, respectively, are diagonal. Furthermore, such diagonal states are the same for both {\sc Alice} and {\sc Bob} as they represent common-knowledge `classical' information, i.e. information accessible to all experimenters. In summary 
\be
{P_{A, 0} = P_{B, 0} = \mathrm{diag}( \dotsc , p_{-\xi}, p_{0}, p_{\xi}, \dotsc )},\label{e:diag}
\ee
where $\dotsc , p_{-\xi}, p_{0}, p_{\xi}, \dotsc$ are the probabilities of the variables $v$, which for illustrative purposes we assume are discretized in steps of size $\xi$ (cf. Sec.~\ref{s:pure_causality}).

Second, as illustrated in Fig.~\ref{f:realisticobserver} (see also Appendix~\ref{s:easy}) an observer has two components: (i) a feature-extraction algorithm that allows the observer to extract high-level features from the raw data provided by the external environment; (ii) a sort of Turing machine, or recursive neural network (see Sec.~\ref{s:architecture}), that can operate over those high-level features to create, for instance, a dynamical model of the external world, here characterized by the dynamical matrix $J$. According to recent research we expect that, in the case of humans, components (i) and (ii) are associated with unconscious and conscious information processing respectively (see items (i) and (vi) in Appendix~\ref{s:easy}). 

In particular, we assume here that observers can only perceive the high-level features extracted by component (i), as well as any relevant operations performed with component (ii). This is consistent with the fact that we do not perceive the raw data provided by the bombardment of electromagnetic radiation reflected from an image, but rather the high-level features that characterize the image. Figures~\ref{f:self-observer} and \ref{f:noneq_self-observer} only take into account component (ii), referring to the raw data along with all information processing carried out with component (i) as the environment. Now, the effective dynamical model that can be created by component (ii) can be reversible or irreversible, i.e. the corresponding dynamical matrix is symmetric $J=J^T$ or asymmetric $J\neq J^T$ respectively. Much as quasi-static thermodynamic processes, i.e. those that are near equilibrium, are reversible, we expect that an effective reversible dynamical model is generated by an environment at equilibrium. Similarly, much as nonequilirbium thermodynamic processes produce entropy and are therefore irreversible, we expect that an effective irreversible dynamical model is generated by an environment out of equilibrium. Although we will assume this is the case here (see Fig.~\ref{f:noneq_self-observer}), our derivations only depend on the distinction between reversible and irreversible effective dynamical models, and not on whether the environment is in or out of equilibrium.

In the next two sections we will describe in more detail the architecture of such self-referential observers (see Figs.~\ref{f:self-observer} and \ref{f:noneq_self-observer}) and argue they lead to the von Neumann equation of quantum mechanics. To build intuition we first discuss the special case of symmetric kernels, such as the kernel introduced in Eq.~\eqref{e:Main_K_T+V}, which we expect to be associated to equilibrium environments. Afterwards we discuss the general case where kernels can be asymmetric, such as the kernel introduced in Eq.~\eqref{e:Main_KrealEM}, which we expect to be associated to environments that can be out of equilibrium.

\

\noindent{\bf Remark:} Everything we discuss here is considered physical, and so is the state $P$ we associate to a physical system. Such state must be represented in the `hardware' of the physical observer. Although the state is subjective for Alice in that it is information she possesses, it is objective for Bob as it is physically represented in Alice's brain, e.g. as a population of neurons. In this sense, including the observer in the description of the experiment merges the Bayesian~\cite{fuchs2013quantum} and frequentist interpretation of probability theory~\cite{kass2011statistical}. 


\subsubsection{Equilibrium environment and symmetric kernels}\label{s:eq}

In this section we assume the environment the self-referential observer $\textsc{Self}=\textsc{Alice}\circ\textsc{Bob}$ is embedded in (see Fig.~\ref{f:noneq_self-observer}; cf Fig.~\ref{f:realisticobserver}) is in equilibrium. So, there are no irreversible contributions to the circular process of mutual observation between {\sc Alice} and {\sc Bob}, i.e. the dynamical matrix ${J = J^T = J_s}$ is symmetric. (see Ref.~\cite{realpe2018cognitive} for a more compact and formal discusion)



In this case the first-person changes in {\sc Alice}'s state, i.e. the changes in her state when she is the observing subject and {\sc Bob} is the observed object, are equal in magnitude but opposite in sign to the third-person changes observed by {\sc Chris} in {\sc Bob}'s state, i.e. $\Delta^{1\rm st} P_A = -\Delta^{3\rm rd} P_B$ (see Fig.~\ref{f:self-observer}a). This kind of action-reaction effect is similar to the energy exchanges that take place in an isolated system composed of two subsystems: any change in energy in one of the subsystems is equal and opposite to the energy change in the other subsystem. We can see this more clearly by noticing in Fig.~\ref{f:self-observer}a that the (green) vertical arrow associated to $\Delta^{1\rm st} P_A$ goes upwards, while the (pink) vertical arrow associated to $\Delta^{3\rm rd} P_B$ goes downwards. 

Something similar could be said for {\sc Bob}'s first-person changes when the roles are reversed (see Fig.~\ref{f:self-observer}b). However, while from {\sc Chris}' perspective {\sc Alice} sees the variable $x$ evolving forward from time step $\ell$ to time step $\ell + 1$, {\sc Bob} sees the variable $x$ evolving backwards from time step $\ell+ 1$ to time step $\ell$ (see Fig.~\ref{f:self-observer}c). Since the physical process $x_\ell\to x_{\ell + 1}$ is the same either way, for the first-person observer {\sc Self} to consistently describe the transition $x_\ell\to x_{\ell +1}$
we must have an additional change in sign between {\sc Alice}'s and {\sc Bob}'s descriptions, i.e. $\Delta^{1\rm st} P_B = \Delta^{3\rm rd} P_A$ (see Fig.~\ref{f:self-observer}c). Notice that this additional minus sign would not arise in the simulation interpretation described in Sec.~\ref{s:simulation} becuase the two arrows in Fig.~\ref{f:self-observer}c will point in the same direction.

In summary, the first-person changes in the state of {\sc Alice} and {\sc Bob}, i.e. when they are thought of as components of the first-person observer {\sc Self}, are given by:
\BE
\Delta^{1\rm st} P_A &=& -\Delta^{3\rm rd} P_B = -\Delta t [J_s, P_B],\label{e:Delta3rdPA}\\
\Delta^{1\rm st} P_B &=& \Delta^{3\rm rd} P_A = \Delta t [J_s, P_A].\label{e:Delta3rdPB}
\EE
Here we have used the notation $J_s$ instead of $J$ to emphasize that these equations are valid only when there are no irreversible contributions, i.e. when $J$ is symmetric. 

Equations~\eqref{e:Delta3rdPA} and \eqref{e:Delta3rdPB} coincide with Eqs.~\eqref{e:DtPA} and \eqref{e:DtPB} when $J_a = 0$ as we are assuming here.  Since the initial states of {\sc Alice} and {\sc Bob} are equal (see Eq.~\eqref{e:diag}), Eqs.~\eqref{e:Delta3rdPA} and \eqref{e:Delta3rdPB} enforce that $P_B = P_A^T$ at all times (see Eqs.~\eqref{e:DtP} and \eqref{e:DtPT}). So, by reversing the arguments in Sec.~\ref{s:vNreal} we can see that Eqs.~\eqref{e:Delta3rdPA} and \eqref{e:Delta3rdPB}, if written in terms of ${\rho = P_s + i P_a}$, with ${P_s = (P_A + P_B)/2}$ and ${P_a = (P_A - P_B)/2}$ the symmetric and anti-symmetric components of $P_A$, are equivalent to the von Neumann equation, Eq.~\eqref{e:vN}, restricted to real Hamiltonians because $J$ is symmetric. In the next section we discuss this derivation in more detail in the context of the general case when the environment can be out of equilibrium, so there can also be irreversible contributions captured by $J_a$.

\subsubsection{Nonequilibrium environment and asymmetric kernels}

If the environment is out of equilibrium, so $J\neq J^T$, there are irreversible contributions which are characterized by the antisymmetric part ${J_a = (J-J^T)/2}$ of the dynamical matrix $J$. In this case it is not true anymore that changes in {\sc Bob}'s state are equivalent to changes in {\sc Alice}'s state because we now have to take into account also the asymmetric contributions external to the self-referential observer {\sc Self}. Since $J_a$ characterizes influences that are external to {\sc Self}, the first-person change in the state of sub-observer $O\in\{A,B\}$ due to $J_a$ coincides with the third-person change in the state of sub-observer $O$ due to $J_a$, i.e. as described by external observer {\sc Chris} (see Figs.~\ref{f:self-observer} and \ref{f:noneq_self-observer}). (see Ref.~\cite{realpe2018cognitive} for a more compact and formal discussion)

Let us make explicit in the first- and third-person changes of the state of sub-observer $O$, 
\BE
\Delta^{1\rm st} P_O &=& \Delta_s^{1\rm st} P_O + \Delta^{1\rm st}_a P_O,\label{e:1st_comp}\\
\Delta^{3\rm rd} P_O &=& \Delta_s^{3\rm rd} P_O + \Delta^{3\rm rd}_a P_O,\label{e:3rd_comp}
\EE
the reversible and irreversible contributions due to $J_s$ and $J_a$, respectively; these are denoted in Eqs.~\eqref{e:1st_comp} and \eqref{e:3rd_comp}, respectively, by subscripts `$s$' and `$a$'. According to Eq.~\eqref{e:Delta3rdPO} the reversible and irreversible contributions to the third-person changes are given respectively by
\BE
\Delta_s^{3\rm rd} P_O &=& \Delta t[J_s , P_O],\label{e:3rdOs}\\
\Delta_a^{3\rm rd} P_O &=& \Delta t [J_a , P_O].\label{e:3rdOa}
\EE

Furthermore, as we said above, the first- and third-person changes associated to the irreversible contributions due to $J_a$ coincide because these are external to the first-person observer {\sc Self} (see description of Eqs.~\eqref{e:Delta1stPA_rel} and \eqref{e:Delta1stPB_rel} below for another way to look at this; see also Ref.~\cite{realpe2018cognitive}). So, 
\be
{\Delta_a^{1\rm st} P_O = \Delta_a^{3\rm rd} P_O = \Delta t [J_a , P_O]}.\label{e:1stOa}
\ee
On the other hand, the changes associated to the reversible contributions due to $J_s$ are internal to the first-person observer {\sc Self}, so we can apply the analysis of Sec.~\ref{s:eq} to these. In other words, according to the analysis of Sec.~\ref{s:eq} we have (cf. Eqs.~\eqref{e:Delta3rdPA} and \eqref{e:Delta3rdPB})
\BE
\Delta^{1\rm st}_s P_A &=& -\Delta^{3\rm rd}_s P_B = -\Delta t [J_s, P_B],\label{e:1stAs}\\ 
\Delta^{1\rm st}_s P_B &=& \Delta^{3\rm rd}_s P_A = \Delta t [J_s, P_A],\label{e:1stBs}
\EE
where the subscript `$s$' indicates that the analysis applies only to the changes due to the reversible contributions (see Eqs.~\eqref{e:3rdOs} and \eqref{e:3rdOa}). Using Eq.~\eqref{e:1st_comp} we can write $\Delta^{1\rm st}_s P_O = \Delta^{1\rm st} P_O - \Delta^{1\rm st}_a P_O$ for sub-observer $O\in\{A, B\}$, which according to Eq.~\eqref{e:1stOa} is equivalent to 
\be
\Delta^{1\rm st}_s P_O = \Delta^{1\rm st} P_O - \Delta t [J_a , P_O].\label{e:1stOfinal}
\ee
Replacing the left hand sides of Eqs.~\eqref{e:1stAs} and \eqref{e:1stBs} by the right hand side of Eq.~\eqref{e:1stOfinal} with $O=A$ and $O=B$, respectively, we obtain
%
\BE
\Delta^{1\rm st} P_A   -  \Delta t [J_a, P_A] &=&  -\Delta t [J_s, P_B],\label{e:Aalmost}\\
\Delta^{1\rm st} P_B   -  \Delta t [J_a, P_B] &=&  \Delta t [J_s, P_A],\label{e:Balmost}
\EE
%
where, again, the minus sign in the second equation takes into account that if {\sc Alice} observes the forward process, {\sc Bob} observes the backward (see Fig.~\ref{f:self-observer}). These equations allows us to iteratively construct the subjective states of {\sc Alice} and {\sc Bob}, and therefore of {\sc Self}. 

Reorganizing Eqs.~\eqref{e:Aalmost} and \eqref{e:Balmost} we finally obtain
%
\BE
\Delta^{1\rm st} P_A   =  \Delta t [J_a, P_A] - \Delta t [J_s, P_B],\label{e:Delta1stPA}\\
\Delta^{1\rm st} P_B   =  \Delta t [J_a, P_B] + \Delta t [J_s, P_A],\label{e:Delta1stPB}
\EE
%
which are equivalent to Eqs.~\eqref{e:DtPA} and \eqref{e:DtPB} in Sec.~\ref{s:vNreal}, after taking the continuous time limit. Again, the initial states  $P_{A, 0}$ and  $P_{B,0}$ of {\sc Alice} and {\sc Bob}, respectively, are diagonal and equal (see Eq.~\eqref{e:diag}) since the observed initial state in an experiment is common-knowledge `classical' information to all observers. So, as in Sec.~\ref{s:eq}, Eqs.~\eqref{e:Delta1stPA} and \eqref{e:Delta1stPB} enforce the condition $P_A = P_B^T$, for all times (see Eqs.~\eqref{e:DtP} and \eqref{e:DtPT}). 

Another way to look at this is as follows. Imagine that Chris in Fig.~\ref{f:first-third} observes Alice and Bob moving with velocity $v_{A|C}$ and $v_{B|C}$. However, Alice and Bob see each other moving with relative velocities ${v_{A|B} = v_{A|C} - v_{C}}$ and ${v_{B|A} = v_{B|C} - v_{C}}$, where ${v_C = (v_{A|C} + v_{B|C})/2}$ is the extrinsic velocity that Chris observes for the composed system of Alice and Bob. In short, Alice and Bob can only see their relative motion (see Ref.~\cite{realpe2018cognitive} for a more formal discusion).  Similarly, we can write Eqs.~\eqref{e:Delta1stPA} and \eqref{e:Delta1stPB} as 
\BE
\Delta^{1\rm st} P_A   -  \Delta t [J_a, P_A] =- \Delta t [J_s, P_B],\label{e:Delta1stPA_rel}\\
\Delta^{1\rm st} P_B   -  \Delta t [J_a, P_B] = \Delta t [J_s, P_A],\label{e:Delta1stPB_rel}
\EE
where we can interpret that we are removing the extrinsic motion of the system because Alice and Bob can only observe their relative motions. 

Now, by adding and substracting Eqs.~\eqref{e:Delta1stPA} and \eqref{e:Delta1stPB}, and dividing the so-obtained equations by half, we obtain the equivalent equations (see Eqs.~\eqref{e:DtPs} and \eqref{e:DtPa})
\BE
\Delta^{1\rm st} P_s   =  \Delta t [J_a, P_s] + \Delta t [J_s, P_a],\label{e:Delta1stPsa}\\
\Delta^{1\rm st} P_a   =  \Delta t [J_a, P_a] - \Delta t [J_s, P_s],\label{e:Delta1stPas}
\EE
in terms of the matrices $P_s = (P_A+P_B)/2$ and $P_a=(P_A-P_B)/2$, which are the symmetric and antisymmetric parts of $P_A$, since $P_B = P_A^T$ as we argued above. Furthermore, by multiplying Eq.~\eqref{e:Delta1stPas} by the imaginary unit $i$, adding the resulting equation to Eq.~\eqref{e:Delta1stPsa}, and writing the so-obtained complex equation in terms of $\rho = P_s + i P_a$ and $H = \hbar J_s + i \hbar J_a$, we finally get Eq.~\eqref{e:vN} after taking the continuous limit.\qedsymbol

\section{Occam's razor favors the reverse paradigm}\label{s:occam}
Apparently the analysis we have done here only changes the interpretation of quantum theory and by extension that of experiments too, as we will discuss in Sec.~\ref{s:consciousness}. However, we will briefly argue here that Occam's razor suggest we should favor the reverse paradigm over the mainstream one (see Fig.~\ref{f:paradigms}). To do so we explicitly state below some relevant assumptions made in the mainstream paradigm and contrast them with the corresponding interpretation in the reverse paradigm presupposed in this work.

\subsection{Some relevant assumptions in the mainstream paradigm}\label{s:mainstream}
Let us now state some of the relevant assumptions implicitly made in the mainstream paradigm (see Fig.~\ref{f:paradigms}):

\begin{enumerate}
\item The world is fundamentally quantum for some reason we do not yet understand. 

\item The associated discreteness in the quantities measured is characterized by a constant $\hbar$ whose origin we do not know yet. 

\item Although in this paradigm we admit that our body, brain, nervous system, etc, are made of the same stuff the rest of nature is made of, e.g. atoms, we implicitly assume that in an experiment we can always neglect the physical interactions associated to the experimenter.

\item Although every second of our lives we experience the world from a first-person perspective, our only perspective as far as we know, we assume that we can somehow look at the world from a supposedly objective third-person perspective, as if we were not part of it; or as if we could somehow exit the universe and look at it from the outside, as if we were abstract entities devoid of matter.
\end{enumerate}

Assumptions (i) and (ii) refer to aspects of quantum theory that the mainstream paradigm still does not have an answer to. Assumptions (iii) and (iv) are inconsistent with our everyday experience; moreover, assumption (iii) is inconsistent with the universality expected from physics as a general theory of nature that should therefore apply to everything, including observers. In the next section we show how the reverse paradigm could help resolve all of these issues.
\subsection{Corresponding interpretation in the reverse paradigm}\label{s:reverse}
Let us now see the potential explanations the reverse paradigm provides for the four assumptions implicit in the mainstream paradigm stated above.

\begin{enumerate}
\item The world can be modelled with classical probability; however, it appears to be quantum due to the interactions associated to the observer's information processing. 

\item Planck constant $\hbar$ characterizes the relevant interactions taking place in the experimenter performing the experiment, as we will discuss in more detail in Sec.~\ref{s:message}.

\item {\bf Principle I} asks us to treat consistently both experiment and experimenter as physical systems and let the analysis alone tells us whether the interactions associated to the experimenter can indeed be neglected. 

\item {\bf Principle II} asks us to describe the world from the only perspective we have, the first-person perspective. As a side result, {\bf Principle II} has the potential to explain the symplectic structure in the fundamental equations of physics, as we will discuss in Sec.~\ref{s:message}.
\end{enumerate}

Item (i) is consistent with our everyday experience of the world as a classical system, and provides a potential explanation as to why the world appears to be quantum. Item (ii) provides a potential explanation for the origin of Planck constant $\hbar$ that suggests it can be determined from psychophysics experiments, as we will argue in Sec.~\ref{s:consciousness}. These kind of experiments are qualitatively different to the kind of experiments physicist typically use to estimate $\hbar$. Items (iii) and (iv) are consistent with our everyday experience; moreover, item (iii) is consistent with the universality expected from a general theory of natural phenomena, which should also apply to observers. So, the reverse paradigm has both a higher explanatory power and a higher consistency both with our human experience and with itself, at least regarding the assumptions here analyzed. We discuss in Sec.~\ref{s:foundations} how the reverse paradigm holds the potential to explain some characteristic quantum phenomena.

\subsection{Quantum foundations in reverse mode}\label{s:foundations}
We now sumarize how we implemented the {\bf Principle I} and {\bf Principle II} and point out how some of the conceptual difficulties of quantum theory become less so from the perspective offered in this work. 

{\bf Principle I} implies that experiments can be conceived as circular interactions (see Figs.~\ref{f:third}, \ref{f:cavity_cycle}, \ref{f:circular}). In other words, the linear chain of cause-effect relationships linking a well-defined initial state to a well-defined final state, which we traditionally model experiments with, has to be closed into a circle. The additional link that turns the chain into a circle could be understood as an effective interaction summarizing the myriad of physical interactions supporting the observer's information processing and control. Colloquially, we might say that this was a missing link to quantum theory.

This principle alone allowed us to derive the formalism of Euclidean or imaginary-time quantum mechanics, which already leads to quantum-like phenomena such as interference, superposition, entanglement, an uncertainty principle, etc. (see e.g. Ref.~\cite{Zambrini-1987}). Indeed, {\bf Principle I} is the one specific to quantum theory, so it already has several conceptual implications that we now describe.

First, while it is possible to identify the two extreme nodes of a chain graph, i.e. its leaves, with an initial state completely independent of any past interactions and a final state completely independent of any future interactions (see $v_{\rm i}$ and $v_{\rm f}$ in Fig.~\ref{f:circular} e), this is not so in a circular graph. Indeed, unlike a chain graph, a circular graph has no intrinsic distinctions between initial or final nodes and those that should be in between, i.e. any node has an interaction with a node in the `past' and a node in the `future' (see Fig.~\ref{f:circular}d). This already suggests that we might run into difficulties when attempting to associate well-defined states to the system at any given time, providing a potential explanation for quantum superpositions and (constructive) interference. 

Second, the additional link summarizing the interactions associated to the observer can be interpreted as an interaction between `past' and `future' (e.g. the green links in Fig.~\ref{f:cavity_cycle} or the combination of the red, blue, and green links in Fig.~\ref{f:circular}). In other words, the circular interaction associated to an experiment could be interpreted as a feedback mechanism that effectively introduce an interaction between the initial state prepared and the final state measured by the observer. An observer that does not consider herself as part of the experimental setup could interpret this effective interaction between `past' and `future' as retrocausality. This makes retrocausality as a plausible and intuitive explanation for Bell inequalities, one of the more problematic conceptual aspects of quantum theory. Alternatively, the variables associated to the observer's representation of the initial and final state of the system would be interpreted by an observer that does not consider herself as part of the experimental setup as hidden variables which interact non-locally. 

Third, it is natural to expect that the specifics of the architecture of the observer, whether human or mechanic, determines the energy scale associated to the observer's physical interactions. Any phenomena whose energy is below such energy scale would be essentially unable to generate a percept. So, even if energy might be fundamentally continuous it would still appears as discrete to the observer due to the threshold energy required to support all the physical interactions underlying the generation of a percept. This suggests a possible explanation for the quantization of energy and it is indeed consistent with psychophysics experiments estimating the energetic aspects of conscious access, as we describe in Sec.~\ref{s:Planck}. 

Fourth, although a probabilistic model with pairwise interactions on a circular graph does not necessarily lead to a first-order Markov process, but rather to a Bernstein process (see Eq.~\eqref{e:Bernstein}), we can still define a Markovian-like update rule if we use probability matrices rather than vectors (see Eq.~\eqref{e:EQM}). The off-diagonal elements of the probability matrix encode dynamical information necessary to obtain the probability matrix at a given time from that from the previous time. This suggests the imaginary-time phase of the imaginary-time quantum state encodes dynamical information since it characterizes the off-diagonal elements of the probability matrix. Furthermore, the diagonal elements of the probability matrix yield the probability of observing the associated outcome, which could explain the origin of the Born rule.

{\bf Principle II} leads us to ask what is the architecture of agents with a first-person perspective? To implement this principles we have made two assumptions: (i) such agents require self-modeling capabilities; (ii) to avoid the infinite regress associated to the self-referential problem of describing the world from within, such self-referential agents are composed of two complementary systems that essentially model each other (see Figs.~\ref{f:self-observer} and \ref{f:noneq_self-observer}; cf. Fig.~\ref{f:first-third}). This latter assumption builds on the insights of the recursion theorem of computer science, which implies that the architecture of a self-printing program, for instance, is composed of two sub-programs that essentially print each other (see Figs.~\ref{f:self}, \ref{f:quine}, \ref{f:recursion} and Secs.~\ref{s:recursion} and \ref{s:self}). In particular, this leads us to predict that the neural correlates of the experience of being a self are composed of two complementary sub-systems that observe each other; the division of brains into hemispheres in healthy individuals can be a reflection of this principle (see Fig.~\ref{f:brain_self}).

This principle allowed us to turn imaginary-time quantum mechanics into its real-time counterpart, effectively implementing a Wick rotation. This principle is related to the symplectic structure of quantum theory, and as such it is expected to apply beyond it. It also has conceptual implications that we now discuss. 

First, we expect that at the root of complementarity lies the fact that self-referential observers are composed of two sub-observers that play complementary roles as both observing subjects and observed objects. This is in line with Bohr's observation~\cite{bohr1929quantum} that complementarity appears to arise naturally in psychology where both the objects of perception and the perceiving subject belong to `our mental content'. 

Second, the implementation of the first-person observer required by {\bf Principle II} via the composition of two sub-observers essentially require that we change the single real matrix equation that encode {\bf Principle I} into a pair of real matrix equations (see Eqs.~\eqref{e:Delta1stPA} and \eqref{e:Delta1stPB}). Furthermore, the two sub-observers observe the external phenomena in reversed directions, which leads to a negative sign in one of the matrix equations (see Fig.~\ref{f:self-observer}). Such negative sign could explain the phenomenon of destructive interference as it holds the potential to reduce the values of the diagonal elements and introduce negative off-diagonal elements to the probability matrices.  

Third, the implementation of {\bf Principle I} naturally led to the imaginary-time version of the Born rule (see Sec.~\ref{s:circulatiry_noncommutativity}), and {\bf Principle II} essentially incorporated the implications of {\bf Principle I} into the first-person perspective (see Sec.~\ref{s:quantum}). This in a sense reverse the traditional reasoning that leads to the so-called  measurement problem, i.e. the problem of explaining why if nature is fundamentally described by a `wave function' that evolves unitarily, at the time of measurement such `wave function collapses' via the Born rule, breaking the fundamental unitarity of quantum evolution. Our starting point, instead, was a standard classical probabilistic model on a circular graph that implemented {\bf Principle I} and naturally led to the imaginary-time version of the Born rule, as a useful artifact to describe the update rule. {\bf Principle II} allowed us to effectively perform a Wick rotation to obtain the unitary evolution of the quantum state. In short, first {\bf Principle I} naturally leads to non-commutativy and the Born rule, and the unitarity emerges from the combination of {\bf Principle I} and {\bf Principle II}. In this sense, the measurement problem is not a problem in this approach.

\

\noindent{\bf Remark 1:} Even if an observer can build a model of itself on its physical `hardware', it is not expected to access the very physical processes that allow the construction of its self-model. This is similar to the situation with a Turing machine that can print a description of itself but cannot access its inner working mechanisms to evaluate whether it will halt on a certain input or not~\cite{turing1937computable,moore2011nature}. In this sense, such physical interactions are an intrinsic indeterminacy. This seems to be related to Metzinger's~\cite{metzinger2004being} concept of transparency mentioned in the introducton. Paraphrasing Deacon~\cite{deacon2011incomplete}: We are a hole, an absence in the universe; nature is incomplete in terms of subject-object relationships because the observing subject cannot fully observe itself, yet it is completed via human experience because the subject can indeed fully experience itself.

\

\noindent{\bf Remark 2:} From a third-person perspective the link that turns the linear chain of cause-effect relationships into a circle is just a concept summarizing the myriad interactions going on in the observer's brain and body. From a first-person perspective, instead, such interactions summarized by a single link translate into our complex human experience. Since the effective interaction represented by such a link concern human experience, we can in principle access them from a first-person perspective, not in a subject-object relationship but from direct experience. This is analogous to the fact that we cannot directly see our own eye, i.e. using our eye itself, but we can indeed experience its existence---e.g. it would be clear to us when our eye has stopped working even if we cannot directly see the damaged eye as an object of observation. As we shall discuss in Sec.~\ref{s:discussion}, the first-person methods of contemplative traditions may turn out to be a powerful tool to carry out this exploration.

\section{Quantumness and consciousness}\label{s:consciousness}

\subsection{Consciousness as a rigorous scientific subject}\label{s:Main_respected}

Dehaene, one of the leading scientist on the modern approach to consciousness, pointed out that for a long time the subject of consciousness was considered a taboo that lied `outside the boundaries of normal science'~\cite{dehaene2014consciousness} (page 7; see also Ref.~\cite{dehaene2017consciousness}). There was a strong reason for this, of course, as it is not clear yet even how to clearly define the concept of consciousness. Perhaps because of this ambiguity, attempts to find links between quantum theory and consciousness have hardly reached mainstream physics debates. Today things are radically changing, though, thanks in part to the efforts of Nobel laureate Francis Crick and his collaborator Christof Koch: consciousness has become a hot scientific research subject (see e.g. Ref.~\cite{koch2016neural} for a recent review).

In Appendix~\ref{s:respected} we summarize some relevant scientific results on consciousness for the reader that may not be familiar. In Appendix~\ref{s:easy} we summarize some relevant results concerning the information-processing mechanisms underlying conscious perception, which is considered as one of the `easy' problems of consciousness~\cite{dehaene2014consciousness,koch2016neural,dehaene2017consciousness}. These features do not address the so-called `hard' problem of consciousness, which is closely related to our human experience. In Appendix~\ref{s:hard} we briefly discuss recent developments on the latter, specially Metzinger's self-model theory of subjectivity~ mentioned in the introduction~\cite{metzinger2004being,metzinger2009ego} (see also chapter 9 of Ref.~\cite{edelman2008computing} for a review; for a short introduction to the most central ideas see Metzinger's talk {\em `The transparent avatar in your brain'} at TEDxBarcelona).
\subsection{The message of the quantum: observers are physical}\label{s:message}
Here we discuss in more detail the idea put forward in Ref.~\cite{realpe2017quantum} that Planck constant can be estimated from psychophysics experiments that in a sense estimate the energetics assocaited to conscious access.

We have discussed in previous sections how the formalism of quantum theory can be understood as encoding {\bf Principle I}, i.e. taking into account the physical interactions associated to the observer (see Secs.~\ref{s:inference_physical} and \ref{s:3rd}), and {\bf Principle II}, i.e. tackling the self-referential problem of representing the world from a first-person perspective (see Secs.~\ref{s:big_1st_person} and \ref{s:first}). Now, {\bf Principle I} leads to a non-commutative Markovian update rule (see Eq.~\eqref{e:EQM}), while {\bf Principle II} builds on the non-commutative equation obtained via {\bf Principle I} to get a pair of non-commutative equations (see Eqs.~\eqref{e:Delta1stPA} and \eqref{e:Delta1stPB}) that can be encoded on a single complex equation defined on Hermitean matrices, i.e. the von Neumann equation. 

So, we can say that {\bf Principle I} essentially leads to non-commutativity, while {\bf Principle II} essentially leads to complex numbers. However, complex numbers are related to the symplectic structure of fundamental physical equations. For instance, it is well-known that we can write Hamilton equations 
\be\label{e:HE}
\dot{x} = \frac{\partial \widetilde{\mathcal{H}}(x, p)}{\partial p},\hspace{0.3cm}\textrm{     and    }\hspace{0.3cm}\dot{p} = -\frac{\partial \widetilde{\mathcal{H}}(x, p)}{\partial x},
\ee
in terms of complex variables $z=(x+ip)/\sqrt{2}$ and its conjugate $z^\ast$ as 
\be
\dot{z} = -i\frac{\partial \mathcal{H}(z, z^\ast)}{\partial z^\ast}, 
\ee
where 
\be\label{e:complexHclassical}
\mathcal{H}(z, z^\ast) = \widetilde{\mathcal{H}}\left(\frac{z+z^\ast}{\sqrt{2}}, \frac{z-z^\ast}{i\sqrt{2}}\right);
\ee
notice that $\widetilde{\mathcal{H}}$ in the right hand side of Eq.~\eqref{e:complexHclassical} is the original Hamiltonian function in Eq.~\eqref{e:HE}, as emphasized by the tilde.
So {\bf Principle II} is expected to apply more generally beyond quantum theory. Indeed, as we will discuss in more detail elsewhere, the symplectic structure of fundamental physical equations can be understood as emerging from self-reference. 

In contrast, non-commutativity is unique to quantum theory. Indeed, as discussed in Ref.~\cite{leifer2013towards} when we restrict all density matrices to be diagonal the quantum dynamics become commutative and coincide with classical Markovian dynamics. Therefore, according to {\bf Principle I} the unique aspect of quantum theory that distinguishes it from classical theories is the fact that the observer is physical. 

Since quantum theory is characterized by Planck constant $\hbar$, it is natural to expect that $\hbar$ is related to the physical interactions associated to the processing of information by the observer. So, from this perspective the classical theory is expected to yield accurate predictions when the energy of the interactions associated to the observer is much smaller than that related to the observed system. This is consistent with the observation that if we neglect the link that summarizes the physical processes associated to the observer, we recover standard Markov processes on a chain (see e.g. Figs.~\ref{f:circular}d,e). In this sense, there is not absolute notion of `micro': something is small {\it relative}~\cite{Rovelli-1996} to an observer in precisely this way.

\

\noindent {\bf Remark:} In the approach we have introduced in Sec.~\ref{s:QMrecasted} (see also Appendix~\ref{s:details}) we can obtain Hamiltonian functions (see Eqs.~\eqref{e:Main_classicalH} and \eqref{e:classicalH_EM}) which coincide (up to a sign) with the Wick rotation $\epsilon\to -i\epsilon$ of the corresponding Lagrangian. More precisely, in the continuous limit $\epsilon\to 0$ we can write (see Sec.~\ref{s:MaxCal})
\be
\begin{split}
S_E =& \sum_{\ell = 1}^{n-1}\mathcal{H}(x_\ell ,x_{\ell +1})\epsilon\\
&\xrightarrow[\epsilon\to 0]{}  \int_{t_{\rm i}}^{t_{\rm f}} \mathcal{H}(x, \dot{x})  \mathrm{d} t, 
\end{split}
\ee
where $t = \ell\epsilon$,  $\dot{x}=\lim_{\epsilon\to 0}(x_{\ell +1}-x_\ell)/\epsilon$, and $t_{\rm i}$ and $t_{\rm f}$ are the initial and final times, respectively; for instance, if $L = m \dot{x}^2/2 - V(x)$ then ${\mathcal{H}(x,\dot{x}) = m \dot{x}^2/2 + V(x)}$. Since the graphical model that represents the observer interacting with the experimental system has the topology of a circle, we need to condition on two variables (see Fig.~\ref{f:cavity_cycle}), say initial and final states: $x_{\rm i} = x(t_{\rm i})$ and $x_{\rm f} = x(t_{\rm f})$, or initial position $x_{\rm i}$ and velocity $\dot{x}_{\rm i}$, to turn the circle into a chain and analyze it in the traditional way. So, the physicality of the observer naturally induces the need to condition on two variables instead of one, as it would be the case of a Markov process on the position $x$. 

Now, the most probable path (see Eq.~\eqref{e:MaxCal}) would be the one that minimizes the Euclidean action $S_E$ condition to the initial and final states remain fixed. It is well-known that this process leads to Euler-Lagrange equations, here for an effective Lagrangian $\widetilde{L} = \mathcal{H}$
\be\label{e:EL}
\frac{\mathrm{d}}{\mathrm{d} t}\frac{\partial \mathcal{H}}{\partial \dot{x}} = \frac{\partial \mathcal{H}}{\partial {x}},
\ee
which in turn are equivalent to the Newton equations in imaginary time. For instance, if ${\mathcal{H}(x,\dot{x}) = m \dot{x}^2/2 + V(x)}$ Eq.~\eqref{e:EL} becomes
\be\label{e:NE}
\ddot{x} = \frac{\partial V}{\partial x},
\ee
which is Newton equation in an inverted potential $-V$. If we now invert back the Wick rotation, i.e. we do $t\to i t $, we get $\ddot{x}\to -\ddot{x}$, which restores the correct sign of the potential in Eq.~\eqref{e:NE}. Although this analysis already leads to a symplectic structure, such symplectic structure is not the correct one. To obtain the correct symplectic structure we need to go through the self-referential process of representing the world from within, even in the classical world. 

In conclusion, the physicality of the observer alone already implies the need of second-order differential equations instead of the most parsimonious first-order differential equations.

\subsection{Planck constant from psychophysics experiments}\label{s:Planck}
\subsubsection{General considerations}
The discussion in the previous section suggests that Planck constant can be derived from experiments that directly or indirectly measure the interactions associated to the observer. Now, as far as we know, science happens at the conscious level, e.g. scientist always report their findings via research articles or conference talks. This suggests that the relevant physical processes taking place in the observer are those associated to conscious information processing. So, as we will argue here, psychophysics experiments that estimate the energy requirements for conscious access are strong candidates for measuring Planck constant. 

Unfortunately, we are aware of only one experimental work~\cite{scholvinck2008cortical} that explicitly addresses the energy requirements for conscious perception. So, there is an opportunity for experimental physicists to design more careful experiments to test this prediction. We will show here that available experiments suggest it is correct. Although the experiment in Ref.~\cite{scholvinck2008cortical} studies monkeys, not humans, the authors argue that `similar psychophysical results [...] obtained in monkey and human for all three sensory stimuli studied suggest that the cellular mechanisms underlying perception are similar in the two species'~\cite{scholvinck2008cortical} (see discussion section therein). Since we will use this experiment only to provide an estimate of the order of magnitude of Planck constant, we will assume that the results in Ref.~\cite{scholvinck2008cortical} can indeed be extrapolated to humans. A drawback, though, is that the authors do not report actual absolute values, but only values relative to those of unconscious information processing. More precisely, the authors of Ref.~\cite{scholvinck2008cortical} provide evidence that the energy required to transition from unconscious information processing to conscious perception is about $6\%$ of the energy required for unconscious processing alone (see below).  However, as we argued in Ref.~\cite{realpe2017quantum}, we can combine the results reported in Ref.~\cite{scholvinck2008cortical} with the results from experiments reported in Ref.~\cite{hecht1942energy}, which estimated the sensibility of the human eye using a {\em classical} source of light, to provide an estimate of the order of magnitude of Planck constant. 

Moreover, we will also discuss in more detail the idea put forward in Ref.~\cite{realpe2017quantum}, that a more recent experiment~\cite{tinsley2016direct}, which uses a {\em quantum} source of light to show that humans can detect one single photon with a probability of $0.516\pm 0.010$, can be reinterpreted in the reverse paradigm (see Fig.~\ref{f:paradigms}) as an experiment that directly measures the actual value of Planck constant. Although the experimental results reported in Ref.~\cite{tinsley2016direct} has been recently challenged~\cite{holmes2017correspondence}, we will argue that our approach allows us to predict that once such a debate is settled, the final conclusion should be that indeed humans can consciously perceive a single photon. This is a precise theoretical prediction that contrasts with the current motivation for these studies, which is essentially the curiosity on whether a human can detect a single photon. Our approach allows us to predict that humans must be able to do so.

\subsubsection{`Classical' pshycophysics experiments}\label{s:classical_psycho}

\noindent{\underline{Experiment 1} ({\em Visual motion detection}): } To begin, let us briefly describe the experiment performed by Sch\"olvinck, Howarth, and Attwell (SHA)~\cite{scholvinck2008cortical} (see `Experiment 1' in Fig.~\ref{f:experiment}). While SHA investigated three different information pathways, i.e. visual, somatosensory, and auditory, we will only focus on the visual part. In the SHA experiment, monkeys were presented an array of moving dot stimuli (see Fig.~\ref{f:experiment} top), wherein a certain percentage $C$ of dots, referred to as `coherence of motion', moved in the same direction, while the remainig dots moved in random directions. The monkeys were tasked to decide what the general direction of motion was. The objective of this experiment was to estimate the energy needed for visual processing of dot stimuli moving sufficiently coherently to generate a percept, relative to the energy needed for processing the same number of dots moving randomly such that no percept of general movement was generated. Since this experiment focuses on the perception of movement, the intensity of the dot stimuli was well above the visibility threshold. This contrasts with the other two experiments~\cite{hecht1942energy, tinsley2016direct} we will described below, whose focus is precisely on identifying the visibility threshold of humans.  

Figure~\ref{f:experiment} (bottom) sketches the qualitative form of the so-called {\em psychometric curve} for visual motion (see Fig. 2d in Ref.~\cite{scholvinck2008cortical} for the actual curve determined from the SHA experiment). The psychometric curve gives the probability $p(C)$ that subjects report seeing the correct direction of motion (vertical axis) as a funtion of the coherence of motion $C$ (top horizontal axis), i.e. the percentage of dots that moved in the specified direction. The authors specified the threshold of detection $C_{\rm SHA}$ to be the value of coherence for which the probability for the monkeys to report the correct direction of motion was $p(C_{\rm SHA}) = 0.82$. The choice of threshold is somehow arbitrary and is different for the three experiments we will discuss in this section. This is something we should take into account in the estimation of Planck constant. 

An ideal experiment would decrease the coherence of motion $C$ with a resolution and number of trials large enough that it is possible to identify the minimum value $C_{\rm min}$ just before subjects report a random guess. The value $C_{\rm min}$ is more precisely defined as the value of the coherence of motion such that the probability to report the correct response satisfy $p(C) = 0.5$ for $C<C_{\rm min}$ and $p(C_{\rm min})=p_{\rm min}$. This includes the case of a continuous transition, where $C_{\rm min} = 0$ and $p_{\rm min} = 0.5$. However, a discontinuous jump with $C_{\rm min} > 0$ and $p_{\rm min}>0.5$ is more consistent with experimental and theoretical studies suggesting that conscious perception is all or none~\cite{llinas2013cortical} and that the transition from unconscious to conscious perception is analogous to a discontinuous phase transition~\cite{dehaene2005ongoing, dehaene2011experimental} (see also Ref.~\cite{dehaene2014consciousness}, page 184). Moreover, we will argue below that the experiment by Tinsley et al.~\cite{tinsley2016direct} can also be interpreted in this way. Such an ideal situation is hard to meet in practice, though , so scientist usually define a suitable threshold such as $C_{\rm SHA}$ in the SHA experiment~\cite{scholvinck2008cortical}. 

We will refer to the region where the discontinuous jump takes place as `quantum' (see Fig.~\ref{f:experiment} bottom) for reasons we will describe later on when we discuss Tinsley et al. experiment~\cite{tinsley2016direct}; we will therefore refer to the other region as `classical' (see Fig.~\ref{f:experiment}). In particular, experiments such as the SHA experiment, which do not have a resolution high enough to observe the region where a discrete jump is expected to happen, allow us to explore only the `classical' region. As we will discuss in Sec.~\ref{s:tinsley}, experiments such as the experiment by Tinsley et al~\cite{tinsley2016direct}, instead, allow us to explore the `quantum' region.

\

\noindent{\underline{Experiment 2} ({\em Determination of the visibility threshold with classical light}):} In 1942 Hecht, Schlaer, and Pirenne (HSP) reported a set of landmark experiments~\cite{hecht1942energy} intended to measure the threshold energy for vision (see `Experiment 2' in Fig.~\ref{f:experiment} bottom).  In the HSP experiment human subjects were exposed to green light of frequency $\nu = 5.88\times 10^{14}$ Hz, whose intensity was gradually decreased until te subjects reported a random guess. In Fig.~\ref{f:experiment} (bottom) we sketch the qualitative shape of the psychometric curve (cf. Fig. 7 {in} Ref.~\cite{hecht1942energy}) that measure the probability $p(E)$ for the human subjects to report seeing a flash of light (vertical axis) as a function of the energy $E$ at the cornea (bottom horizontal axis).  The relatively low resolution of the experiment did not allow to resolve the `quanutm' region of the psychometric curve, so an arbitrary threshold $E_{\rm HSP}$ was defined such that $p(E_{\rm HSP}) = 0.6$, different from the choice of threshold in the SHA experiment. According to the results of the HSP experiment, $E_{\rm HSP}$ was in the range 
\be
2.1\times 10^{-17} \mathrm{J} \lesssim E_{\rm HSP}\lesssim 5.7\times 10^{-17} \mathrm{J}.\label{e:E_HSP}
\ee
(See Table II in Ref.~\cite{hecht1942energy} for the values in units of $10^{-10}$ ergs; notice there is a typo therein.) 

The authors of the HSP experiment interpret it within the current mainstream paradigm in physics (see Sec.~\ref{s:mainstream} and Fig.~\ref{f:paradigms}), i.e. they assume that the energy of light is objectively quantized in photons for some reason we do not yet understand, and ask `what is the minium number of photons that humans can report?'. Since the energy of a photon is given by $h\nu$, the values for $E_{\rm HSP}$ represent between 54 and 148 photons. However, HSP argued that at least three corrections should be applied to those values: reflection from the cornea, losses that occur between the cornea and the retina, and energy absorbed by the retina. By also taking into account the Poisson statistics of the classical source of light they used, HSP estimated that about 5 to 8 photons are required at the retina to generate a visual perception. 

\

\noindent\underline{Estimation of Planck constant's order of magnitude:} However, the reverse paradigm (see Sec.~\ref{s:reverse} and Fig.~\ref{f:paradigms}) presupposed in this work implies a completely different interpretation: the energy of light can in principle be continuous, as any classical energy source, but due to the physical interactions associated to the observer it {\it looks} quantized in photons. So, from this perspective we can combine HSP and SHA experiments to ask rather the question `what is the minimum amount of energy required for humans to transition from unconscious processing to conscious perception, i.e. what is the value of $E_{\rm min}$ in Fig.~\ref{f:experiment} (bottom)?'. 

Let us first analyze the SHA experiment on monkeys' visual motion detection. Let $E_0$ and $E_{\rm SHA}$ be, respectively, the energy consumed by the relevant firing neurons of a generic monkey when the dots are moving at random, i.e. $C=0$, and at the SHA coherence threshold $C_{\rm SHA}$, i.e. when monkeys report the correct direction of motion with a probability $p(C_{\rm SHA})=0.82$. According to the results reported by SHA we have ${(E_{\rm SHA}-E_0)/E_0\approx 0.06}$ (see paragraph following Eq. (2) in Ref.~\cite{scholvinck2008cortical}). Writing $E_0 = E_{\rm SHA} - (E_{\rm SHA}-E_0)$ and doing some algebra yields 
\be
{(E_{\rm SHA}-E_0) \approx 0.04 E_{\rm SHA}}. \label{e:monkey}
\ee
According to SHA $E_0$ and $E_{\rm SHA}$ are the energies corresponding to mean firing rates of about $20$ Hz and $21.4$ Hz, respectively. While it is not clear to us how to translate these numbers into the actual values of energy, we will now combine this analysis with the HSP experiment to obtain some estimates.

In Ref.~\cite{realpe2017quantum} we used the minimum value of energy measured by HSP (see Eq.~\eqref{e:E_HSP}) as an estimate of $E_{\rm HSP}\approx 2.1\times 10^{-17}$ J to have an estimate of Planck constant. Here we will rather use the average value in the range determined by HSP, i.e. $E_{\rm HSP}\approx 3.9\times 10^{-17}$ J since it is a better estimate of the typical value. Let $E_{\rm av}$ be the amount of energy required for unconscious processing of electromagnetic stimuli, e.g. for feature extraction, which is the analogous of $E_0$ in the SHA experiment; although $E_{\rm av}$ was not measured in the HSP experiment, we do not require it here. Assuming the results of the SHA experiment in Eq.~\eqref{e:monkey} can be extrapolated to the HSP experiment, and neglecting the fact that the thresholds in both experiments were defined differently, we have ${E_{\rm HSP} - E_{\rm av}\approx 0.04 E_{\rm HSP}\approx 1.6\times 10^{-18}}$ J, which is about four times the energy of the corresponding photon $E_{\rm photon} = h \nu\approx 3.9\times 10^{-19}$ J. Since $E_{\rm min} < E_{\rm HPS}$, this implies that $E_{\rm min}$ in Fig.~\ref{f:experiment} (bottom) is less than the energy associated to a few photons. As we will discuss in the next section, the experiment of Tinsley et al.~\cite{tinsley2016direct} provides evidence that $E_{\rm min}$ is indeed the energy associated to a single photon.

\subsubsection{`Quantum' psychophysics experiments}\label{s:tinsley}
By using a quantum source of light to expose human subjects to single photons and carrying out over $3\times 10^4$ trials to collect enough statistics, Tinsley et al.~\cite{tinsley2016direct} significantly improved on the HSP experiment and determined that humans can indeed detect one single photon with a probability of $p_{\rm min} = 0.516\pm 0.010$ (however, see Ref.~\cite{holmes2017correspondence}). Although Tinsley et al. did some more refined statistics by asking the human subjects about the level of confidence in their responses, we will focus on the simpler experiment that request only one bit of information as we expect that simpler is more fundamental. 

Like HSP, Tinsley et al. also interpret their experiment within the current mainstream paradigm in physics (see Fig.~\ref{f:paradigms}), i.e. Tinsley et al.~\cite{tinsley2016direct} assume that the energy of light is objectively quantized in photons for some reason we do not yet understand, and ask `can humans perceive a single photon?'. From this perspective we can interpret $E_{\rm min}$ in Fig.~\eqref{f:experiment} (bottom) as the fundamental minimum amount of energy of green light that can exists in nature, i.e. any non-zero value of energy that is below $E_{\rm min}$ simply does not exist. In other words, in this paradigm we already know $E_{\rm min} = h\nu$. By focusing directly on estimating $p_{\rm min} = p(E_{\rm min})$ in Fig.~\ref{f:experiment} (bottom), Tinsley et al. addressed what we called the `quantum' regime of this psychophysical experiment, while HSP addressed the `classical' regime.

\noindent\underline{Estimation of Planck constant:} However, the reverse paradigm (see Fig.~\ref{f:paradigms}) presupposed in this work again implies a completely different interpretation: as we said, in this paradigm the energy of light can in principle be continuous, as any classical energy source, but due to the physical interactions associated to the observer it {\it looks} quantized in photons. So, from this perspective we can turn Tinsley et al. question into `what is the minimum amount of (possibly continuous) energy humans can perceive?' In this case there are no intrinsically forbidden energies as in the mainstream paradigm interpretation. 

Instead, we could imagine an experiment that starts with light with energy $E_{\rm large}$ large enough for the human subjects to report seeing it with a probability close to $1$ (see Fig.~\ref{f:experiment} bottom). Afterwards the experimenters could progressively reduce the energy of light, while constantly asking the human subjects whether they still see the pulses of light. The experimenters can decrease the energy until the human subjects report seeing the pulses of light with probability $0.5$, i.e. a random guess. The value $E_{\rm min}$ of the energy just before reaching a random guess would correspond to the minimum amount of energy of light we can consciously percieve.  In Tinsley et al. experiment we have $p_{\rm min} = p(E_{\rm min}) = 0.516\pm 0.010$.

Of course, all the evidence we have from more than a century of quantum experiments suggests that such a value is the value we associate to a photon. However, the interpretation is completely different. Any value of energy below $E_{\rm min}$ could objectively exist, yet its energy is below the value required to launch the physical processes associated to conscious access. Combining the predictions of quantum theory, whose formalism we have derived here, with the so obtained value of $E_{\rm min}$, we could infer the precise value of Planck constant. In the reverse paradigm, it is not that the eye is so efficient that it can detect a single photon, but that the very definition of photon is the minimum amount of light we can consciously perceive. A possible criticism of this idea is that we cannot directly observe X-rays, for instance, yet they are also quantized. However, as illustrated in Fig.~\ref{f:third}, even in experiments with X-rays the observer is part of the experimental set up; the interactions associated to the observer can be considered as the bottleneck of the whole process. Notice that no other theory predicts that humans should be able to perceive a single photon; up to now this question has been an experimental curiosity not guided by any theory~\cite{tinsley2016direct}.

\subsection{Self-reference and the global architecture of self-aware systems and the self}\label{s:architecture}

There is active research in understanding the neural correlates of the self. Studies focusing on self-referential processing, i.e. on the ability to think about oneself, have identified regions located in the midline of the human cerebral cortex to be crucial for self-specific processing~\cite{schaefer2017conscious,northoff2011self,northoff2004cortical,gusnard2001medial,legrand2009self}, which typically involves both brain hemispheres. It has been argued~\cite{legrand2009self,christoff2011specifying}, however, that this type of self-referential processing is related to the experience of ourselves as a passive obsject, i.e. a `me', and that the experience of ourselves as active subjects, i.e. as an `I', is more closely related to sensorimotor processes, i.e. processes that integrate sensory information or input with a related motor response or output in the central nervous system~\cite{varela2017embodied}. 

A possible account of the sensorimotor perspective is that humans have internal models which can be of two types~\cite{wolpert1995internal}: (i) forward models, which mimic the causal flow of a process by predicting its next state, given the current state and the motor command; and (ii) inverse models, which invert the causal flow by estimating the motor command that caused a particular state transition. So, an inverse model allows an agent to estimate the motor command that will lead to a desired sensory experience, much as in the ideomotor view~\cite{pezzulo2006actions} (see Sec.~\ref{s:ideomotor}). On the other hand, a forward model allows an agent to predict which would be the sensory consequences of implementing a given motor command. These type of mechanism can in principle allow for an intrinsic self/non-self distinction, i.e. a distinction made by the agent itself rather than by another external agent. Let us use Christoff {\em et al.}~\cite{christoff2011specifying} words:

\epigraph{``An organism needs to be able to distinguish between sensory changes arising from its own motor actions (self) and sensory changes arising from the environment (non-self). The central nervous system (CNS) distinguishes the two by systematically relating the efferent signals (motor commands) for the production of an action (e.g. eye, head or hand movements) to the afferent (sensory) signals arising from the execution of that action (e.g. the flow of visual or haptic sensory feedback). [... T]he basic mechanism of this integration is a comparator that compares a copy of the motor command (information about the action executed) with the sensory reafference (information about the sensory modifications owing to the action). Through such a mechanism, the organism can register that it has executed a given movement, and it can use this information to process the resulting sensory reafference. The crucial point for our purposes is that reafference is self-specific, because it is intrinsically related to the agent’s own action (there is no such thing as a non-self-specific reafference). Thus, by relating efferent signals to their afferent consequences, the CNS marks the difference between self-specific (reafferent) and non-self-specific (exafferent) information in the perception-action cycle. In this way, the CNS implements a functional self/non-self distinction that implicitly specifies the self as the perceiving subject and agent. [...]

``For example, consider the motor act of biting a lemon and the resulting taste. This experience is characterized by (i) a specific content (lemon, not chocolate); (ii) a specific mode of presentation (tasting, not seeing); and (iii) a specific perspective (my experience of tasting). The process of relating an efference (the biting) to a reafference (the resulting taste of acidity) is what allows the perception to be characterized not only by a given content (the acidity) but also by a self-specific perspective (I am the one experiencing the acidity of the lemon juice)''}{Christoff {\em et al.}, in Ref.~\cite{christoff2011specifying}.}

The idea that the notion of self is related to a forward and an inverse model appears to be consistent with the architecture of self-referential observers (see Fig.~\ref{f:self-observer}), a subject that we plan to explore further in the future. 

There have also been explorations~\cite{limanowski2013minimal,limanowski2018seeing} on how the free energy principle~\cite{friston2010free} underlying the framework of active learning could implement Metzinger's notion of a self-model~\cite{metzinger2004being,metzinger2009ego,blanke2009full,metzinger2010self}. As we mentioned in the introduction, the idea that humans rely on a self-model appears to be supported by experiments that study the experience of ownership of body parts and of the full body, such as the rubber hand~\cite{botvinick1998rubber,ehrsson2004s} and full-body illusions~\cite{lenggenhager2007video,ehrsson2007experimental,blanke2012multisensory,blanke2015behavioral,blanke2009full} respectively. Metzinger~\cite{metzinger2004being} and Blanke~\cite{blanke2009full} argue that there are three necessary aspects underlying the simplest notion of self, or self-consciousness: (i) self-identification, i.e. the identification of an organism with a global body representation; (ii) self-location, i.e. the volume in space usually localized within the represented body boundaries; and (iii) the first-person perspective. According to Metzinger and Blanke, this notion of self arises when the brain encodes the origin of the first-person perspective from within a spatial frame of reference (i.e. self-location) associated with self-identification~\cite{blanke2012multisensory}.  

Similar to what happen with self-referential processing, the three aspects mentioned above, i.e. self-identification, self-location, and the first-person perspective, have been associated to neural activity in a few brain regions that involve both hemispheres~\cite{blanke2012multisensory}. This appears to be consistent with the architecture of self-referential observers as composed of two sub-systems, in this case the two brain hemispheres and perhaps also the left and right neural networks that run through the spinal cord (see Fig.~\ref{f:brain_self}). In other words, similar to what happens with self-replicating molecules (e.g. the DNA) and self-printing machines, self-reference imposes a global organizing principle for a neural system to refer to itself: it must be composed of two complementary sub-systems, much as the global organization of the human central nervous system.

Indeed, since Turing machines can be realized by recurrent neural networks~\cite{Goodfellow-et-al-2016} (chapter 10) with rational weights~\cite{siegelmann1991turing,siegelmann1995computation,siegelmann1995computational,hyotyniemi1996turing}, the results associated to self-reference, such as the recursion theorem and the halting problem, can in principle be extended to recurrent neural networks~\cite{siegelmann1995computational, hyotyniemi1997unsolvability}. For instance, consider the recurrent neural network with $M$ external input signals and $N$ neurons defined by the equation~\cite{siegelmann1995computation,siegelmann1995computational}
\be\label{e:rnn}
x_i(t+1) = \sigma\left(\sum_{j=1}^N a_{ij} x_j(t) +\sum_{j=1}^M b_{ij} u_j(t) + c_i\right),
\ee
where $u_j$ are the external inputs, $x_j$ are the neurons' activations, $a_{ij}$, $b_{ij}$, $c_i$ are the parameters specifying the network. 
\be
\sigma(x) = 
\begin{cases}
0\textrm{ if } x < 0,\\
x\textrm{ if } 0\leq x \leq 1,\\
1\textrm{ if } x > 1.
\end{cases}
\ee
As part of the description of the network, a subset of $p < N$ neurons is singled out as the output neurons which communicate the output of the network to the environment. Although the output values can take values in $[0,1]$ they can be constrained to binary values only~\cite{siegelmann1995computational}.

Here we are only interested in the formal properties of these networks. So, let $\mathcal{N}_\theta(\mathbf{x}; \mathbf{u})$ denote a generic reccurrent neural network defined by the vector of external inputs $\mathbf{u}$, the vector of neuron's activations $\mathbf{x}$, and the description of the network $\theta$ which contains the parameters of the networks and the indexes of the output neurons. So, here $\mathcal{N}_{\theta}$ is to $\theta$ what a Turing machine {\sc TM} is to its description {\sc ``TM''} (see Appendix~\ref{s:recursion}). Since recurrent neural networks (with rational parameters) are equivalent to Turing machines, we can in principle build a network $\mathcal{N}_{\theta}^{\textsc{Self}}$ that outputs its own description $\theta$. According to the recursion theorem (see Appendix~\ref{s:recursion}), such network should be composed of two neural networks $\mathcal{N}_\alpha^{\textsc{Alice}}$ and $\mathcal{N}_\beta^{\textsc{Bob}}$ that essentially output a description of each other.

This therefore suggests that the neural architecture of a system with the ability to refer to itself should be composed of two complementary systems that effectively run in reverse directions. This principle might help explain why the process of global ignition underlying conscious access seems to be associated to the simultaneous presence of feedforward and feedback propagations throughout the brain (see e.g. Fig. 1b in Ref.~\cite{van2018threshold} and the figure in Ref.~\cite{mashour2018controversial}; see also Ref.~\cite{joglekar2017inter}). It seems also natural to hypothesize that the global architecture of the central nervous system of humans and other animals, i.e. the division of the brain in two hemispheres and the left and right neural networks running through the spinal cord (see Fi.g~\ref{f:brain_self}), may be the result of this constraint to implement self-reference. Indeed, as we pointed out above, the neural correlates associated to self-referential processing and to the experience of body ownership usually involve both brain hemispheres~\cite{schaefer2017conscious,northoff2011self,northoff2004cortical,gusnard2001medial,legrand2009self,lenggenhager2007video,ehrsson2007experimental,blanke2012multisensory,blanke2015behavioral,blanke2009full} (see also Fig. 4 in Ref.~\cite{koch2016neural}). 

These two possibilities are not necessarily contradictory, as we could expect that a neural system may implement such architectural constraint at different scales to enhance its resilience to potential damage of parts of its infrastructure, or to implement consciousness or the experience of being a self associated to processes at different scales. Although this may seem to contradict the decades-old theory stating that split-brain patients, i.e. those whose {\it corpus callosum} connecting the two brain hemispheres has been severed, can have divided identities. However, such a theory has been recently challenged~\cite{pinto2017split,corballis2018perceptual}. According to this new work, split-brain patients actually appear to experience divided perception but undivided consciousness. Corballis {\em et al.}~\cite{corballis2018perceptual} argue that subcortical connections may play a role in integrating information from the two hemispheres.

\

\noindent{\bf Remark:} Schaefer and Northoff recently suggested~\cite{schaefer2017conscious} that, since we cannot think about oneself without being conscious, the cortical mid-line structures involved in self-referential processing may be related to a conscious part of the self. These authors further suggested that sensorimotor processes may be related to an unconscious part of the self based on automatic processes. Similarly, subjective reports of experienced meditators suggest that, by rigorously training our attention and turning it within (see Secs.~\ref{s:toolbox} and \ref{s:worldview} as well as Fig.~\ref{f:channels}), it is possible to access the deppest unconscious processes underlying our sense of self.  According to these reports, such deep unconscious processes appear to be related to processes happening around the spinal cord and brain hemispheres.
\section{Discusion}\label{s:discussion}

\subsection{Summary of main points}

Here we have thoroughly discussed how ideas from cognitive science and artificial intelligence can provide a fresh perspective to reason more carefully about the actual role observers play when they perform an experiment. We have done so by providing a more extensive, hopefully clearer discussion of the idea we put forward about a year ago~\cite{realpe2017quantum} that quantum theory can be understood from two intuitive principles (see {\bf Principle I} and {\bf Principle II} in Sec.~\ref{s:intro}). We can rephrase such two principles here as: (i) experiments are composed of two interacting subsystems, observer and apparatus; (ii) the system composed of observer and apparatus should be represented by one of its subsystems, the first-person observer, not by another external (third-person) observer. 

We have shown that the conceptual framework developed here within the reverse paradigm (see Fig.~\ref{f:paradigms}) has a higher explanatory power and consistency than the current conceptual framework within the mainstream paradigm (see Sec.~\ref{s:occam}). In particular, it suggests a natural physical explanation of the origin of Planck constant as due to the physical interactions supporting the observer's information processing, and could help resolve the conceptual issues associated to the foundations of quantum theory (see Sec.~\ref{s:foundations}). {\bf Principle I} and {\bf Principle II} are also more consistent with our everyday human experience by implying we are also physical sub-systems of the universe, rather than something immaterial, that should describe experiments from our everyday first-person perspective.

Moreover, the conceptual framework developed here suggests two predictions that can be tested experimentally: (i) humans can observe a single photon of visible light, so Planck constant can be derived from psychophysics experiments---we have shown that existing experiments are consistent with this prediction~\cite{tinsley2016direct,hecht1942energy,scholvinck2008cortical} (see Sec.~\ref{s:Planck}), yet more careful experiments should be performed to have a more rigorous assessment of its validity; (ii) the neural correlates of the self are composed of two complementary sub-processes that essentially observe or represent each other---this may help guide the ongoing search for the neural correlates of the self~\cite{schaefer2017conscious,northoff2011self,northoff2004cortical,gusnard2001medial,legrand2009self,lenggenhager2007video,ehrsson2007experimental,blanke2012multisensory,blanke2015behavioral,blanke2009full}  (see Sec.~\ref{s:architecture}).

Prediction (ii) suggests that the neural architecture of self-aware systems and the self follows a design principle similar to the double-stranded structure of the DNA molecule~\cite{watson1953molecular}. Much as this particular structure of DNA avoids the infinite-regress seemingly associated to a system that reproduces itself---i.e. the na\"ive idea that a self-reproducing system should contain a copy of the system within itself, and the copy should also contain another copy within itself for it to be able to reproduce too,  and so on {\em ad infinitum}---the neural structure of the self as composed of two complementary neural systems (e.g. two complementary neural networks, much as the architecture of a Helmholtz machine~\cite{dayan1995helmholtz,benedetti2018quantum}) avoids the infinite regress seemingly associated to a system that has a model of itself, which Metzinger attempted to cure via his principle of transparency stated in the introduction. This is also analogous to the architecture of a Turing machine that prints a description of itself formalized by the recursion theorem of computer science (see Appendix~\ref{s:recursion}).

\subsection{Quantum computing, artificial general intelligence, and quantum cognition}

This approach may help identify unexpected, game-changing applications of quantum computing technologies~\cite{biswas2016nasa} to artificial general intelligence, cognitive and social sciences, and the modern approach to consciousness. Indeed, while recent research at Google suggests that a quantum computer with over 50 qubits might have enough power to demonstrate quantum `supremacy' over classical supercomputers~\cite{boixo2016characterizing}, the artificial problem set to demonstrate it does not have any known applications to date. The approach proposed here may help in the search of novel applications of emerging quantum computing technologies where quantum `supremacy' might be achieved. 

For instance, the self-referential agents introduced here could provide a solid foundation for the emerging field of quantum cognition~\cite{bruza2015quantum}, which has provided evidence that quantum models might more parsimoniously describe cognitive phenomena. So, this approach may help identify datasets on human behavior, for instance, where quantum machine learning is superior to its classical counterpart (see Sec. II B in Ref.~\cite{perdomo2017opportunities}). It also suggests that quantum computers might more naturally implement `self-aware' systems, i.e. systems with a model of themselves, one of the key objectives of artificial general intelligence~\cite{reggia2013rise} (see also Refs.~\cite{carter2018conscious,dehaene2017consciousness})---a subfield also known as machine consciousness~\cite{gamez2008progress,reggia2013rise}. Such self-modeling capabilities~\cite{christoff2011specifying,limanowski2013minimal} (see item (iv) in Appendix~\ref{s:hard}) along with the ability to model the discreteness associated to conscious perception~\cite{llinas2013cortical,dehaene2014consciousness} (see item (v) in Appendix~\ref{s:easy}) are features of quantum computers that could also be relevant in cognitive science and consciouness research. 

Let us mention a couple more technical potential implications of the ideas we have presented here, before we move on to what we consider may be the more relevant potential implications. First, the link we provided between message-passing algorithms and imaginary-time quantum mechanics might suggests novel ways to implement these powerful distributed algorithms on quantum computers or to simulate quantum systems. Second, our approach suggest that Hamiltonians with complex entries, which are non-trivial instances of non-stoquastic Hamiltonians, might be related to non-equilibrium phenomena. This suggests that some of the potential computational advantage due to non-stoquasticity, a quantum computational resource, may be related to the type of computational advantages recently observed of irreversible Monte Carlo methods~\cite{kapfer2017irreversible,lei2018irreversible,lei2018mixing}, where detailed balance is broken, over the equilibrium counterpart. This might also suggest novel ways to implement such general Hamiltonians in quantum computers. Finally, the way we recasted the quantum mechanics of a particle in an electromagnetic field might suggest a probabilistic derivation of Maxwell equations.

Although we have framed our discussion in the representationalist paradigm of cognitive science, we expect it is possible to reframe these ideas within the context of enactivism~\cite{varela2017embodied}. We leave this for future work.


\subsection{On science, our worldview, and how we live }\label{s:onScience}
However, we expect the more relevant potential implications of these ideas are not related to machines but rather to us humans: the way we do science, our worldview, and how we live. We now turn to discuss these ideas, which are framed within the emerging field of contemplative science~\cite{varela2017embodied,tang2015neuroscience, wallace2009contemplative, wallace2007hidden, zajonc2004new, ricard2014mind,hanson2009buddha,desbordes2013new}, i.e. the fruitful collaboration that is emerging between science and contemplative traditions (see Ref.~\cite{wallace2007hidden} for a related discussion). The theoretical and experimental tools emerging at this intersection are considered rigorous enough to merit a review article in a top journal such as {\em Nature Reviews Neuroscience}~\cite{tang2015neuroscience} on a practice previously labelled `spiritual', i.e. mindfulness meditation (see Fig.~\ref{f:channels}). We focus here on the specific case of Buddhism but other contemplative traditions may have similar concepts. While we do not use mathematical language in this closing discussion section, the concepts we will describe can in principle be modelled mathematically with the tools explored in this work. Indeed, once we assume the observer is physical as {\bf Principle I} states, the observer can also turn her attention within to study the physical procceses going on inside herself (see Fig.~\ref{f:channels}). So, modeling the observer holds the potential to bridge the first- and third-person perspectives. Although research on this topic might be a high-risk endevor, in the final section we discuss why we consider the potential returns to society might be high too.

Of course, at this stage this discussion is merely speculative; it is done with the only purpose of suggesting some ideas for debate, which could potentially open up a realm of phenomena that has not been considered to date as a fruitful subject of study in mainstream physics. These ideas might actually be found to be wrong after a rigorous scientific analysis. But this is precisely the point we want to make: today theoretical and experimental tools are emerging to reject these type of ideas on a rigorous scientific basis rather than out of long-held believes (see Fig.~\ref{f:channels}), or to realize we may have been misled by partial or confusing information~\cite{tang2015neuroscience,hanson2009buddha}---recall the strong rejection felt against the concept of atom in the XIX century, or against the idea that the Earth is not the center of the universe before Copernicus. We are aware that in the current state of affairs it seems easier to flow with the understandable attitude of just automatically rejecting this discussion as total non-sense. Yet, we are convinced we scientists have the social responsibility to push against the view of the majority when we believe there is a reason to do so~\cite{mann2017optimal}, without any regard for individual considerations. Otherwise, how would have the concept of atom, for instance, reached mainstream scientific debate if all scientists had preferred to conform to the view of the majority at the time?
\subsubsection{Enlarging our toolbox with first-person methods}\label{s:toolbox}

To begin, scientists have understandably relied mostly on third-person methods; even brain scientists usually study others' brains, not their own. Although there are good reasons to doubt about the usefulness of subjective research methods, the recent advances in consciousness research (see Appendix~\ref{s:respected}) strongly suggest they can be very powerful if used properly (see Fig.~\ref{f:subjective}). Now, if the first-person perspective lies at the core of fundamental physics, as the approach we have explored here suggests, then first-person methods have the potential to play a key role in fundamental physics research (see Fig.~\ref{f:channels}).

One of the key ideas that led Varela, Thompson and Rosch, to write their highly influential book~\cite{varela2017embodied} was to learn the language that Buddhism speaks and afterwards carefully analyze what this millenarian spiritual tradition may have to say about cognitive science. Unfortunately, strong differences in language have often hindered a communication between physics and anything that could be labeled `spiritual', with the sad result that mainstream physics has largely ignored, sometimes with disdain, such extremely rich traditions. We are convinced that in the current state of affairs physicist could greatly benefit by following the example of Varela, Thompson, and Rosch, and assume what we would call Feynman's attitude, who according to Penrose~\cite{penrose1994shadows} (page 105) once said:

\epigraph{\em  ``Don't listen to what I say; listen to what I mean!''}{R. Feynman, as quoted in Ref.~\cite{penrose1994shadows} (page 105)}. 

In other words, we suggest to avoid dismissing a millenarian, profound and large collection of wisdom just because it is labelled `spiritual', or because we do not understand yet the language used. Rather, a more scientific attitude might be to first try to understand the actual meaning behind the words being used, much as Varela, Thompson, and Rosch did twenty five years ago, and only afterwards do a scientific assessment of any claims made---we will briefly mention in Sec.~\ref{s:worldview} some ideas that might be useful to start with. 

Imagine, for instance, that a person who has not had the opportunity to study the most abstract aspects of mathematics were to dismiss it because he finds the idea of infinite dimensions, or non-commutativity as an hallucination. Now imagine there were not just one but millions of people with such an attitude towards science. We contend that this would be analogous to dismissing what some millenarian traditions have to say about our description of reality when we have not had the opportunity to investigate it with enough depth. 

Perhaps one aspect that may have led mainstream physics to largely dismiss contemplative traditions as unscientific is that the latter sometimes tend to do claims without any scientific basis. However, in part thanks to the lead of the 14th Dalai Lama, on of the most recognized Buddhist monks in the West, some steps are being taken to alleviate this natural concern. In 2005 the Dalai Lama wrote~\cite{lama2005universe}

\epigraph{\em ``My confidence in venturing into science lies in my basic belief that as in science so in Buddhism, understanding the nature of reality is pursued by means of critical investigation: if scientific analysis were conclusively to demonstrate certain claims in Buddhism to be false, then we must accept the findings of science and abandon those claims.''}{D. Lama, Ref.~\cite{lama2005universe}}

Furthermore, today it is common to find collaborations between world-class research institutions and monks, which have turned out to be very fruitful. The scientific approach to mindfulness meditation~\cite{tang2015neuroscience,hanson2009buddha}, for instance, has led to the wide dissemination of such an old technique. Much of the research to date has been focused on the cognitive and health sciences, in part because people tend to think these type of techniques are just therapies to relieve stress and the like. 

However, a careful reading of the books written by highly experienced meditators suggests such techniques are much deeper than that. Insisting that meditation techniques are just therapies would be similar to insisting mathematics is just about the arithmetics we do in everyday life. Instead, meditation techniques could be considered as sophisticated tools to carefully investigate our first-person experience of reality. Thus, as {\bf Principle I} and {\bf Principle II} are directly related to our human experience, we are convinced physics may also benefit from such explorations to study the role of agency in the physical world, for instance. 

Indeed, while physics has largely focused on systematically exploring an objective, third-person description of the universe, contemplative traditions have systematically explored a subjective, first-person description of human experience---from this perspective, leading monks might be considered the first-person analogous of leading scientists. For instance, while highly sophisticated machinery developed over many years allow scientists to analyze others' brains from a third-person perspective, monks can study from a first-person perspective their own brains, which they can access for free! Of course, the machinery developed by scientists could also be useful for enhancing the first-person study of our own brains. On this regard the Dalai Lama wrote:

\epigraph{\em ``A comprehensive scientific study of consciousness must therefore embrace both third-person and first-person methods: it cannot ignore the phenomenological reality of subjective experience but must observe all the rules of scientific rigor: So the critical question is this: Can we envision a scientific methodology for the study of consciousness whereby a robust first-person method, which does full justice to the phenomenology of experience, can be combined with objectivist perspective of the study of the brain? 

``Here I feel a close collaboration between modern science and the contemplative traditions, such as Buddhism, could prove beneficial. Buddhism has a long history of investigation into the nature of the mind and its various aspects---this is effectively what Buddhist meditation and its critical analysis constitute. Unlike that of modern science, Buddhism's approach has been primarily from first-person experience. The contemplative method, as developed by Buddhism, is an empirical use of introspection, sustained by rigorous training in technique and robust testing of the reliability of experience. All meditatively valid subjective experiences must be verifiable both through repetition by the same practitioner and through other individuals being able to attain the same state by the same practice. If they are thus verified, such states may be taken to be universal, at any rate for human beings.''}{D. Lama, Ref.~\cite{lama2005universe} (page 134)}

So, we would like to pose the following questions: Can self-referential agents, or a more thoroughly scientific investigation of agency, help bridge these two complementary perspectives? Can physics and contemplative traditions enrich each other? How? As a possible example of the potential synergy between fundamental science and contemplative traditions, let us consider the strategy described in Fig.~\ref{f:channels}, which is in line with the strategy described by Dehaene in Ref.~\cite{dehaene2014consciousness} and summarized in Fig.~\ref{f:subjective} (see also Sec.~\ref{s:consciousness}). We will use as subjective report a piece of text written by a representative monk, Gueshe Kelsang Gyatso~\cite{gyatso2005mahamudra}, leader of the New Kadampa Tradition. We emphasize once again that such a subjective report is considered here exclusively as raw data and nothing more; we are not suggesting in any way that this is a valid scientific account of natural phenomena. Be aware also that, since contemplative practices have been developed many centuries ago, they tend to use a difficult symbolic language that is not necessarily to be taken literally, but rather as a pointer to certain subjective experiences that can be carefully investigated via first-person methods. The piece of text is the following (emphasis is our own):

\epigraph{``{\em There are three main channels: the central channel, the right channel, and the left channel. [...] Other names for the right channel are [...] the `speech channel' and the `channel of the {\em subjective} holder'. This last title indicates that the winds flowing through this channel cause the generation of conceptions developed in terms of the {\em subjective} mind. Other names for the left channel are [...] the `body channel' and the `channel of the held {\em object}', with the last title indicating that the winds flowing through this channel cause the generation of conceptions developed in terms of the {\em object}.''}}{Gueshe Kelsang Gyatso, in Ref.~\cite{gyatso2005mahamudra} (see Appendix II)}

Such left and right channels are considered by experienced meditators as fundamental for the formation of our self-concepts---the central channel is sometimes associated to something experience meditators usually call the `True Self', which we are not considering here. So, according to these subjetive reports the self-concept is related to the combination of one process related to the subject and another process related to the object. This is consistent with the architecture of self-referential observers in Figs.~\ref{f:self-observer} and \ref{f:noneq_self-observer} (see also Figs.~\ref{f:self}, \ref{f:quine}, \ref{f:recursion}). So, we might eventually consider the latter as a {\em physical correlate} of such subjective reports. 

In principle, it should be possible to search also for the neural correlates associated to these subjective reports. However, since meditators report that such processes are deeply buried in the unconscious, it might require more sophisticated technology than the one currently available. Alternatively, we can also run a first-person experiment with ourselves, our own physical system, by regularly practicing, for instance, vipassana meditation. This would allow us to have a direct experience of such inner phenomena rather than having to trust the first- and third-person reports written by other mediators and scientists, respectively. As a bonus, such practices have the potential to greatly enhance the quality of our lives, according to the reports of many meditators throughout history. In the next section we point out some further potential analogies between some concepts studied here and some concepts from contemplative traditions with the only purpose of suggesting some ideas for debate.

\subsubsection{Can the scientific and contemplative worldviews converge?}\label{s:worldview}
Here we discuss some potential analogies between some fundamental concepts in contemplative traditions and concepts we have explored in this work. Again, we focus here on the specific case of Buddhism but other contemplative traditions may have similar concepts. Although our discussion is about concepts, a key aspect of contemplative traditions is that they are not about theoretical concepts but about experience. So, from the perspective of the contemplative traditions, the concepts we are about to discuss could be considered as pointers to actual phenomena that can only be experienced from a first-person perspective, i.e. by carrying out `experiments' with ourselves such as practicing mindfulness meditation---see Ref.~\cite{varela2017embodied} (chapter 10, page 217) for a scientific-friendly description of some central ideas in Buddhism. From this perspective, we might say contemplative traditions such as Buddhism are to traditional philosophy what experimental physics is to theoretical physics: it does not matter what the latter believes to be true, only the former is rooted in experience and so has the final word.

\begin{figure}
\includegraphics[width=0.8\columnwidth]{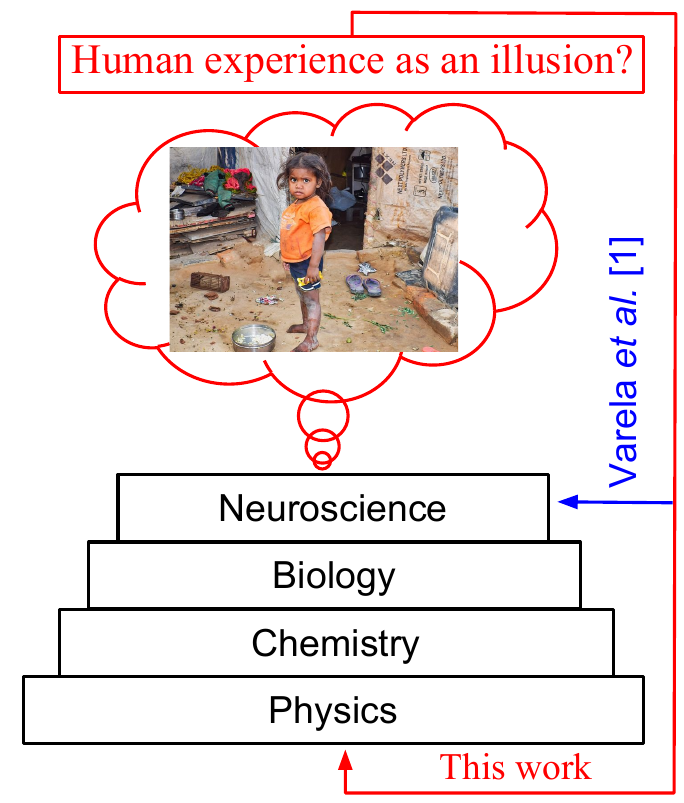}
\caption{{\em Scientific hierarchy in the mainstream paradigm}: In the current scientific paradigm, physics is considered the most fundamental theory that provides the objective laws of nature out of which all other scientific disciplines progressively emerge (see Sec.~\ref{s:intro} and Fig.~\ref{f:paradigms}). At the end of such a scientific hierarchy we find human experience as an illusion generated by billions of neurons firing. Varela, Thompson, and Rosch~\cite{varela2017embodied} suggest human experience should feedback into the foundations of neuroscience, creating a fundamental circularity that gives rise to a more consistent theory of cognitive phenomena. In this work we propose such a feedback should be extended to the very bottom of the scientific hierarchy, i.e. physics itself. In this perspctive, the laws of nature might be considered as self-consistent regularities that emerge out of the interaction between subject and object (see Sec.~\ref{s:worldview}). At a rather subtle level scientific paradigms set our conception of society. Has perhaps the emphasis of the mainstream paradigm on getting rid of the subjective, ourselves, unkowingly diverted our attention towards concerns on mechanisms, such as resource optimization or automation, leaving the more human-centered socio-ecological concerns for later? (see Sec.~\ref{s:action}). }\label{f:hierarchy}
\end{figure}

\begin{figure*}
\includegraphics[width=\textwidth]{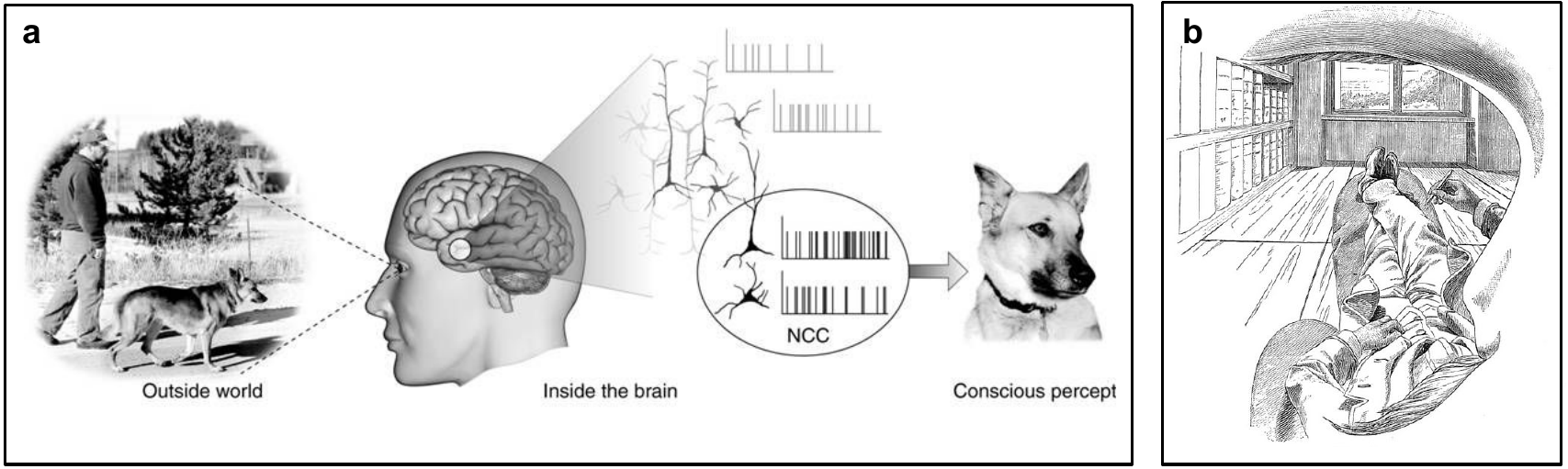}
\caption{ {\it Perspectives used in this work}. (a) Third-person perspective analysis of human observers as performed, for instance, in the modern research program on consciousness championed by Nobel laureate Francis Crick and his collaborator Christof Koch~\cite{crick1990towards} (see also Refs.~\cite{koch2016neural,dehaene2014consciousness}). The human observer under study is analyzed from the perspective of an external observer which is not part of the experimental set-up. Subjective reports by the observer under analysis are contrasted with the associated neural activity (see Fig.~\ref{f:subjective} and Appendix~\ref{s:easy}). (b) First-person perspective as depicted by Ernst Mach self-portrait in 1886. This self-portrait will be used in this work to indicate the analysis of physical systems from the perspective of the scientist performing the experiment, which is herself part of the system being studied. Although it might be obvious for most people, we would like to emphasize that to the best of our knowledge this is the only perspective human beings have, from birth to death. We argue here that this perspective leads to self-reference, which can induce an infinite regress if approached na\"ively (see Fig.~\ref{f:first} and Appendix~\ref{s:hard}). }
\label{f:perspectives}
\end{figure*}

\begin{figure*}
\includegraphics[width=0.7\textwidth]{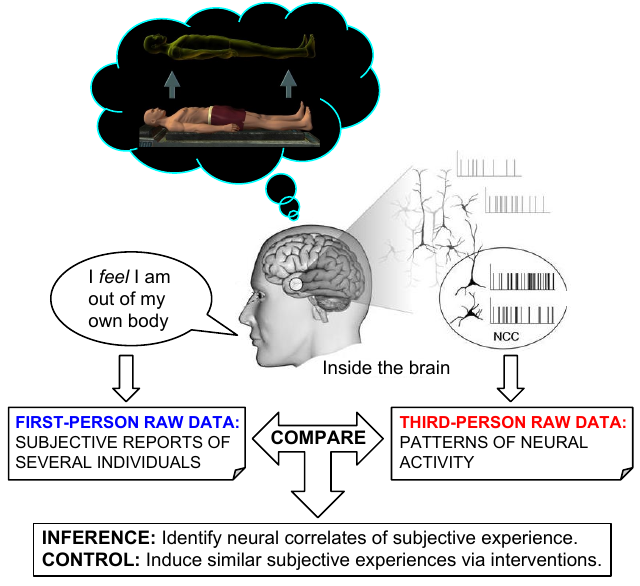}
\caption{ {\it Modern scientific strategy to study consciousness}. Today even phenomena that were previously considered unscientific, such as out of body experiences~\cite{lenggenhager2007video,ehrsson2007experimental,blanke2009full}, can be studied and even induced at will in the lab; such type of research is considered rigorous enough to deserve publication in top journals such as {\it Science}~\cite{lenggenhager2007video,ehrsson2007experimental}. The strategy for doing this is a combination of subjective reports, taken as first-person raw data, and third-person analysis of neural activity. So, by collecting subjective reports of several individuals and recording the corresponding neural activity we obtain a pair of correlated datasets from a first- and third-person perspective on the same phenomena. By carefully analysing these pair of datasets we can in principle identify the neural correlates of subjective experience. With that knowledge we can perform interventions in an individual's brain to induce such subjective experience, if the correlations detected were causal. The example shown in the figure also suggests that even the feeling experienced by some people of being out of their own body---which appears to be the phenomenon most consistent with the assumption implicitly made in physics that we can observe the world from the outside as if we were not part of it---is experienced from those people's own subjective perspective.}
\label{f:subjective}
\end{figure*}

\begin{figure*}
\includegraphics[width=0.9\textwidth]{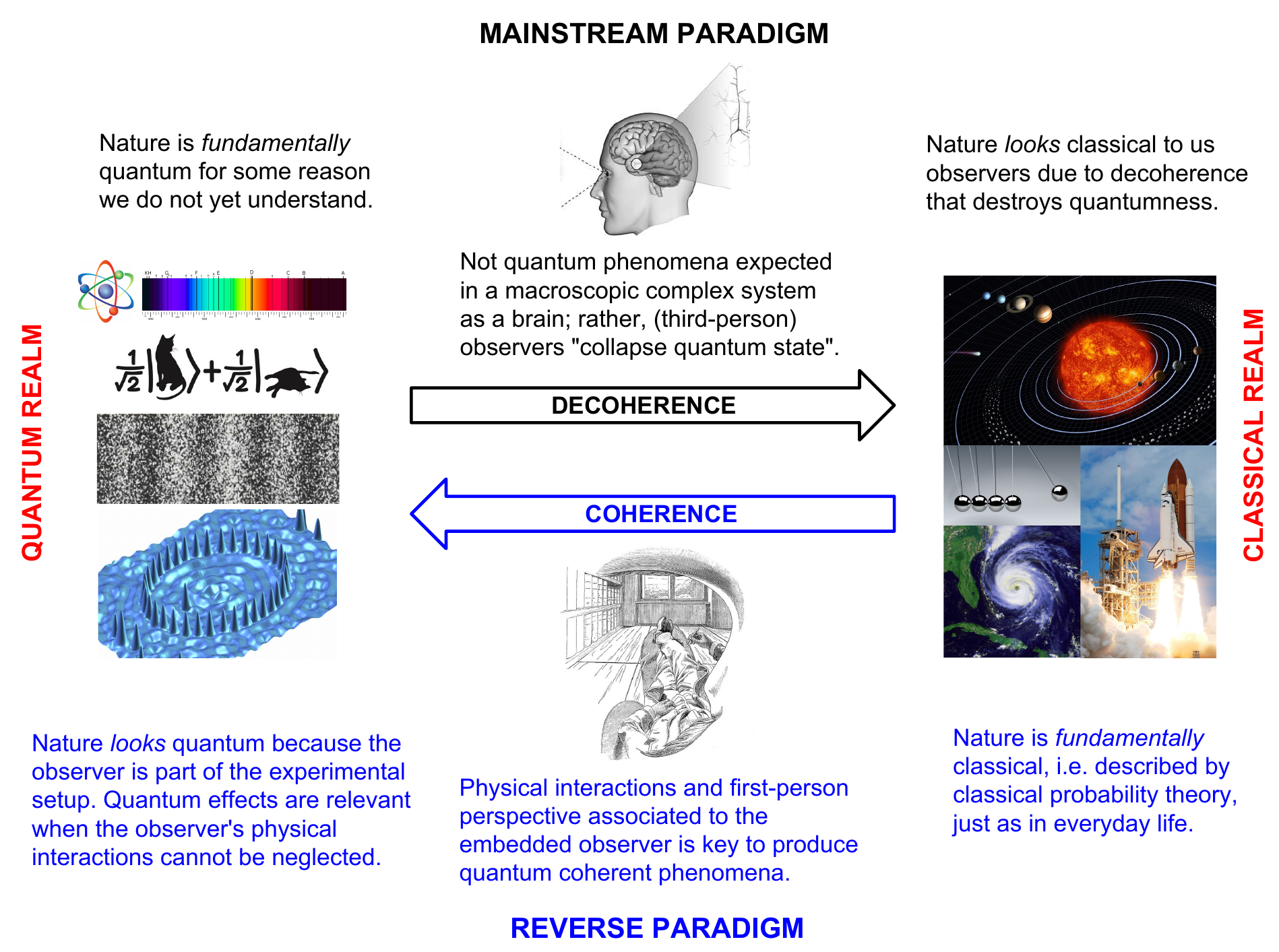}
\caption{ {\it Comparison between mainstream paradigm and the reverse paradigm followed in this work}. The mainstream paradigm in physics is that at the `microscopic' level nature is quantum for some reason we do not yet understand (top). In this paradigm, the act of observation can `collapse the wave function' rendering the system classical; so, observers induce decoherence of the quantum state. Furthermore, when observers are analyzed, such as in studies of the Maxwell demon, they are analyzed from a third-person perspective, i.e. from the perspective of another observer external to the system being analyzed  (cf. Fig.~\ref{f:perspectives}a). In contrast, this work is placed within the context of the reverse paradigm (bottom) wherein nature is fundamentally classical, i.e. it can be described with classical probability theory, yet due to the physical interactions associated to the observer, considered as another classical system, quantum phenomena arise. Thus, in this paradigm the observer instead of inducing decoherence actually becomes the cause that the world appears quantum to her. Moreover, in this paradigm the observer is modelled as a first-person observer which leads to self-reference (cf. Fig.~\ref{f:perspectives}b).} 
\label{f:paradigms}
\end{figure*}

\begin{figure}
\includegraphics[width=\columnwidth]{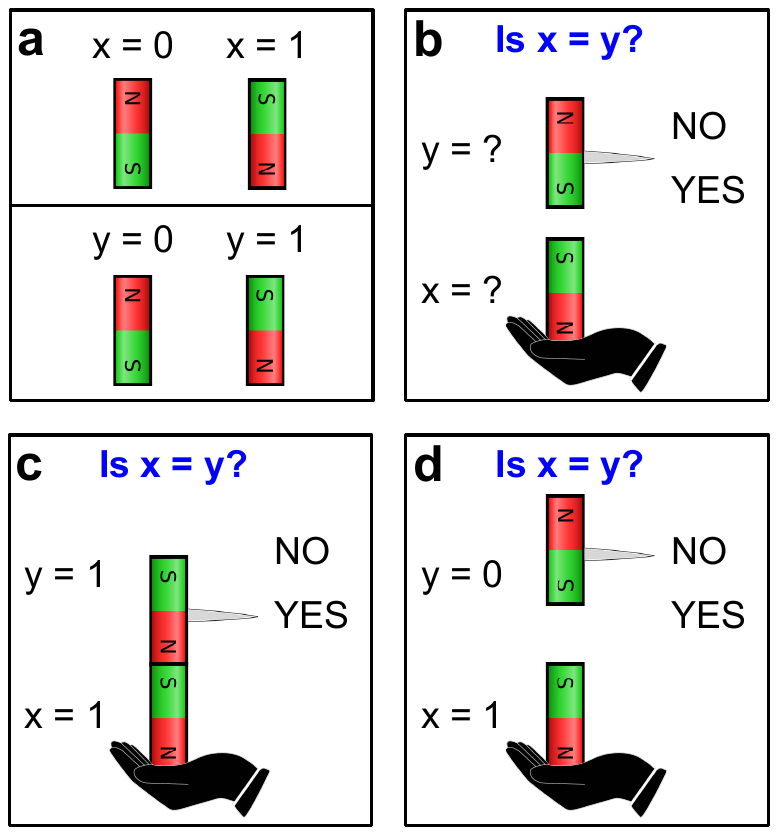}
\caption{{\it Information processing requires physical interaction.} (a) Information is encoded in physical systems. Here we show how two bits $x$ and $y$ can be encoded using magnets: if the North pole of the magnet representing $x$ points up or down, respectively, then $x=0$ or $x=1$; similarly for $y$. (b) The information processing required to answer an apparently abstract question such as ``Is $x=y$?'' can be implemented here by the interaction of the two magnets representing $x$ and $y$: (c) if $x=y$ then the two magnets attract each other; (d) else they repel each other.}
\label{f:infophysics}
\end{figure}

\begin{figure*}
\includegraphics[width=\textwidth]{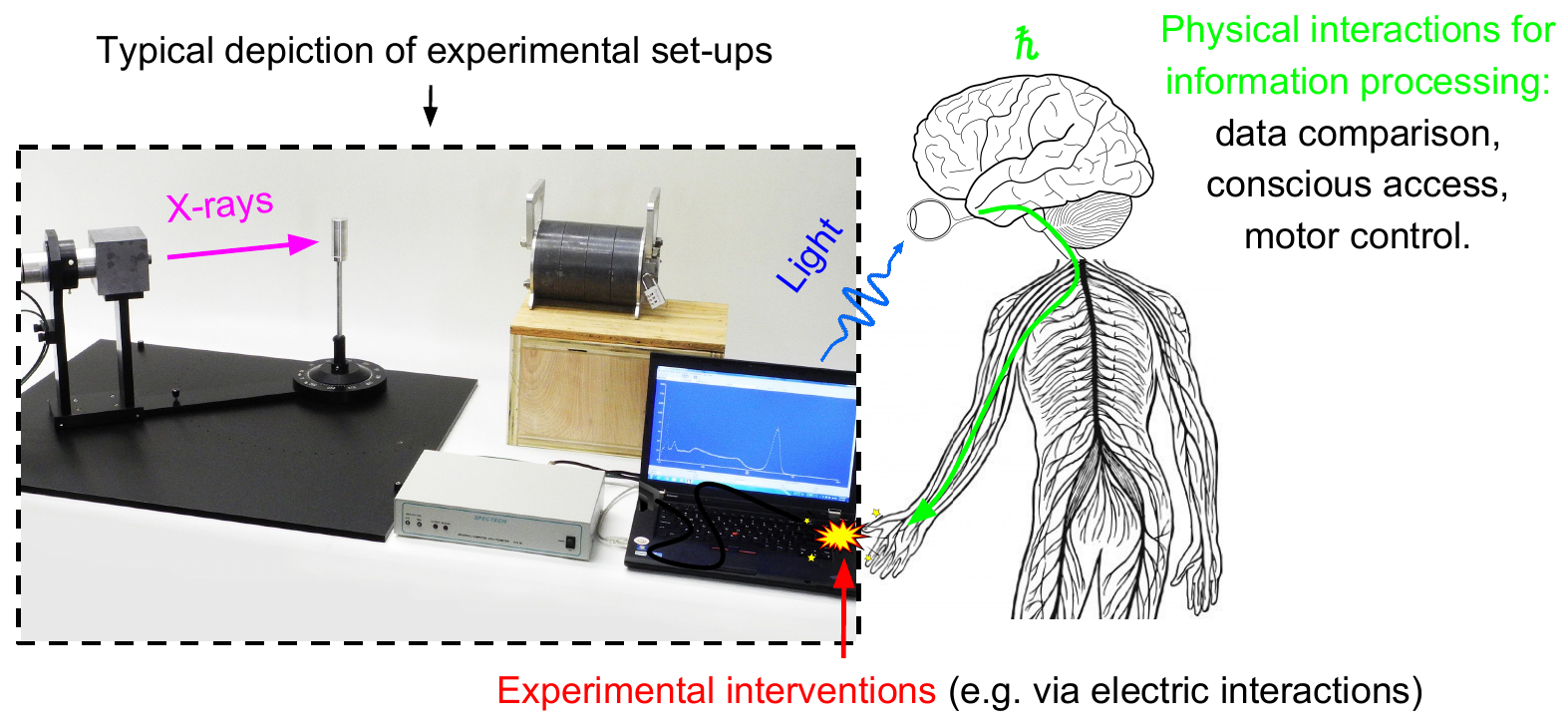}
\caption{ {\it Illustration of {\bf Principle I}.} Experimental set-ups are typically depicted as shwon inside the dashed box on the left, i.e. omitting the observer. If at all, to the best of our knowledge, the observer is included as a cartoon not playing any dynamical role. This implicitly assumes that observers are something abstract, rather than something material that should also be subject to the laws of physics. {\bf Principle I} asks us to drop such an assumption and consistently treat observers as physical systems that interact with the experimental set-up. Indeed, observers usually get information about the system via visible electromagnetic radiation or light. Similarly, experimental interventions, e.g. state preparations, also require some kind of interaction, typically pressing some buttons or a computer keyboard. According to textbook physics, touching is also a physical interaction between the atoms of the body and the atoms of the device being touched. Furthermore, the information processing necessary for an observer to detect or `be aware of' any correlation between the initial state prepared and the final state observed requires the physical interaction of at least some of the components of the information processing device (see Fig.~\ref{f:infophysics}), i.e. the brain in this case. Moreover, since any statements we make about physics typically comprise relationships between data which `we are conscious' about, we expect the information processing should include the process of {\it conscious access}. We argue in this work that such previously neglected interactions can account for the quantum of action.}
\label{f:third}
\end{figure*}

\begin{figure*}
\includegraphics[width=\textwidth]{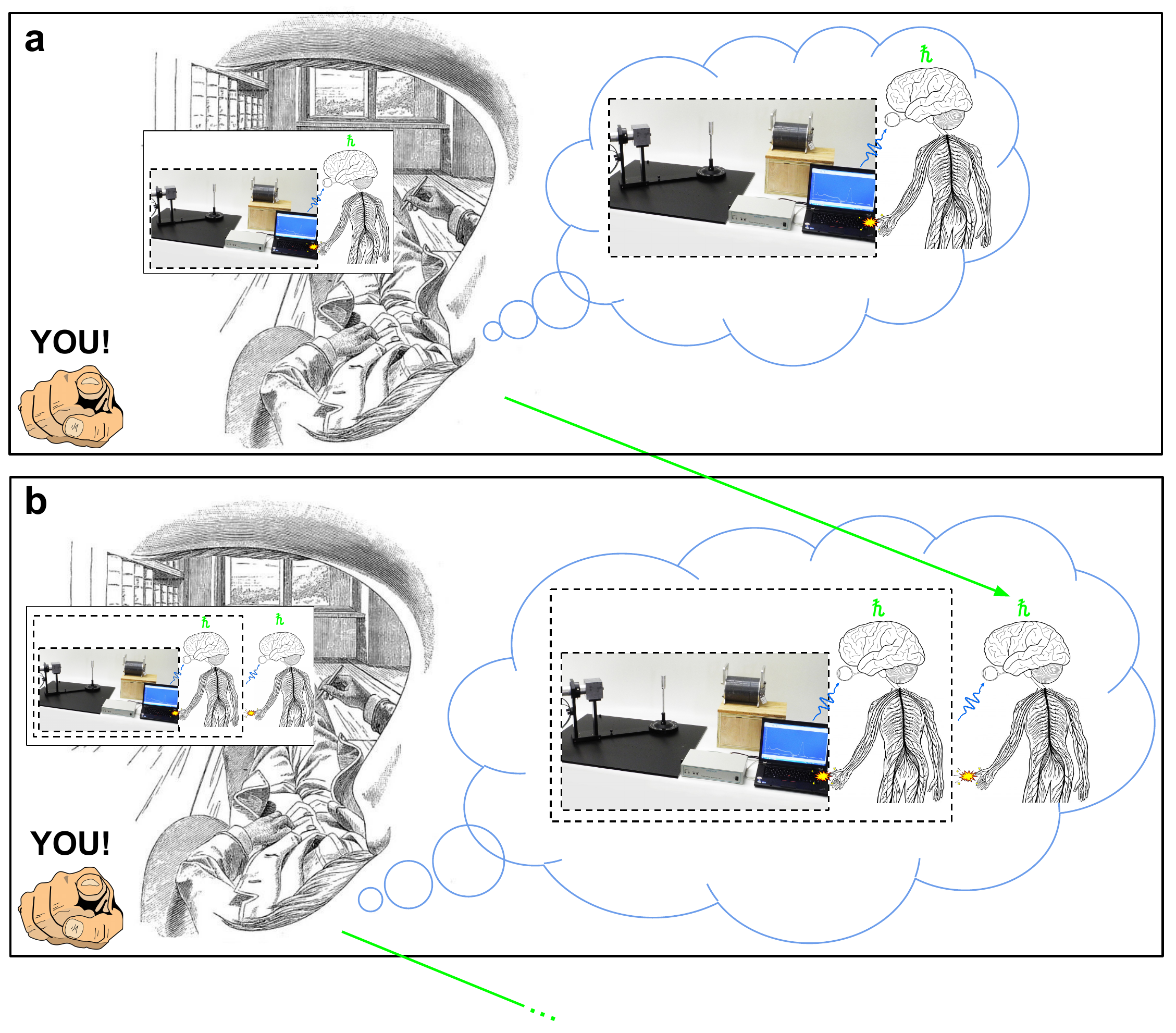}
\caption{ {\it Illustration of {\bf Principle II}}. (a) In Fig.~\ref{f:third} there was something very important lacking: You! Indeed, the observer in Fig.~\ref{f:third} was analyzed from the perspective of an external observer, in this case you, once again not included in the picture. (b) {\bf Principle II} asks us to drop the assumption, however successful, that we can observe the world from a third-person perspective, as if we were not part of it. In other words, {\bf Principle II} asks us to be consistent with what we observe every single second of our lives, from birth to death, i.e. that we can only perceive the world from a first-person perspective, until there is experimental evidence that suggest otherwise. However, if we attempt to include the new observer in (a) in the same way we did in Fig.~\ref{f:third} we end up with the same problem, only that with two observers now. If we insist in doing this we end up with an infinite regress. This problem is caused by self-reference, and can be tackled using ideas from the recursion theorem in theoretical computer science. A simple conceptual explanation of the central idea involved in the recursion theorem is illustrated in Fig.~\ref{f:self} (see also Fig.~\ref{f:first-third}).}
\label{f:first}
\end{figure*}

\begin{figure*}
\includegraphics[width=\textwidth]{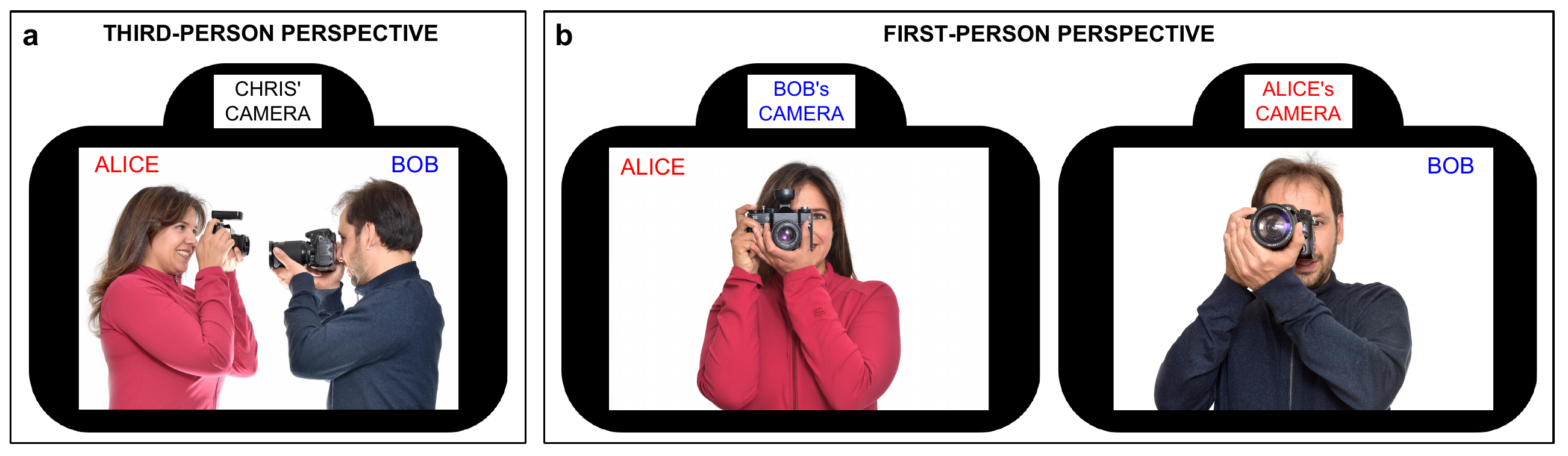}
\caption{ {\it Simple experiment illustrating the architecture of a first-person observer}. (a) A system composed of two observers, Alice and Bob (here represented as photographers), observing each other as seen from the perspective of another observer, Chris, external to the system composed of Alice and Bob. (b) System composed of Alice and Bob as observed from the perspective of the composed system of Alice and Bob itself. }
\label{f:first-third}
\end{figure*}

As we pointed out in the introduction, the mainstream scientific worldview today is that there is an objective mechanical world and that we have the special status of understanding such a world as if we were an abstract entity independent of it.  Today it is almost taken for granted that physics, and in particular quantum physics, provides the objective laws that lie at the very foundation of the skyscraper of science. The remaining scientific disciplines therefore emerge from it, and at the end of the scientific hierarchy we find human experience as an `illusion' generated by the incessant activity of billions of neurons distributed throughout our brain and body (see Fig.~\ref{f:hierarchy}). 

In contrast, the approach here suggests that what we call the fundamental laws of physics are intrinsically tied to our human experience, i.e. to our belonging to the universe as physical sub-systems of it ({\bf Principle I}) and to our everyday first-person perspective ({\bf Principle II}). When we take this into account, the mainstream paradigm nicely summarized in Crick's Astonishing Hypothesis (see Sec.~\ref{s:intro}) loses its ground, it becomes circular because now the subject is at both the bottom and the top of the scientific hierarchy (see Fig.~\ref{f:hierarchy}). At the bottom, the subject gives rise to the laws of nature, out of which that very same subject emerges at the top as an `illusion' generated by an assembly of neurons behaving according to the laws of nature. In short, the subject `emerges' from the subject! From this perspective, the `hard' problem of consciousness may be ill-posed as we cannot explain human experience by reducing it to something else; human experience is at the starting point not at the end (see Fig.~\ref{f:hierarchy}). In this perspective, laws of nature are more like self-consistent stable regularities that emerge out of the fundamental circularity that is the interaction between observer and observed (see Fig.~\ref{f:hierarchy} and Ref.~\cite{varela2017embodied}, chapter 1). 

This does not necessarily mean that we should move to the other extreme of asserting the reality of the subject and denying the reality of the world. Rather, there is a sort of mutual causality between subject and object: they co-dependently arise, as emphasized in Buddhism. Indeed, in our derivation we did assume there is an external world that provides the raw data to the observer (see Figs.~\ref{f:realisticobserver}, \ref{f:self-observer}, \ref{f:noneq_self-observer}); furthermore, the self-referential observer is composed of two sub-observers that play complementary roles as both observing subjects and observed objects (see e.g. Figs.~\ref{f:first-third}, \ref{f:self}, \ref{f:self-observer}). 

At this point it is useful to follow the strategy described by Dehaene~\cite{dehaene2014consciousness} (see Figs.~\ref{f:subjective} and \ref{f:channels}) and take the subjective descriptions of highly experienced meditators (e.g. monks) as raw data to exlore whether there might be analogies with concepts in our approach. We will look at the concepts of non-duality, emptiness, impermanence, unconditioned mind, and inter-dependence. 

From the discussion above we could say that reality is neither objective nor subjective, but non-dual. There appears to be an analogy between the mathematics of self-reference, e.g. the recursion theorem, and non-duality. Consider, for instance, the case of the self-referential observer which is composed of two sub-observers that play dual roles as subjects and objects. This seems to suggest that the self-referential observer is non-dual in that both subject and object play a role in its composition, while at the same time it is built up of two dual entities. It is as if both duality and non-duality coexist. This may address a common criticism of the concept of non-duality as being itself dual because it discriminates between two concepts: duality and non-duality. 

The idea that things are co-dependently arising is central in Buddhism and leads to the concept of emptiness, i.e. the lack of inherent existence of the objects we percieve. To introduce the concept of emptiness it might be useful to consider an analogy with a dictionary~\cite{harnad1990symbol,levary2012loops,vincent2017hidden} out of which we are supposed to learn the meaning of every word. Now, to understand the meaning of a word we need to look up in the dictionary the meaning of the words used to define it, and so on recursively. However, if we continue doing this sooner or later we will come full circle, unless there are some words whose meaning is grounded from outside the dictionary, e.g. from direct experience~\cite{harnad1990symbol}. 

Our analogy is between a dictionary and the universe; since we cannot step outside the universe because by definition the universe is all there is, we will insist that every word in the dictionary can only be learned from inside it, i.e. the dictionary is complete. This implies that we will never be able to learn the meaning of any word! We could say that words in a complete dictionary have no {\em inherent meaning}. A complete dictionary is a collection of semantic loops~\cite{levary2012loops} which can only provide the relationship between words, not their meaning in any objective way. Similarly, the concept of emptiness refers to the idea that everything in the universe only exists in relationship to something else, and in particular to the mind of the observer; i.e. nothing possesses inherent existence. The analogy with the dictionary suggest that loops are key and loops are also key in the approach we have explored here. As we discussed in Sec.~\ref{s:foundations}, in contrast to what happen in a stochastic process on a chain, in a stochastic process on a circle there is no intrinsic beginning nor end, which may imply that states do not posses inherent existence. In physics jargon we might say that nature is fully relational, including everything related to the observer itself.

Another important concept in Buddhism is that of impermanence which essentially means that at every unit of time things arise and then disappear to arise again in the next unit of time. This implies, in particular that the self along with its objects of observation arise and disappear every unit of time. This seems consistent with the iterative construction of the first-person observer with her objects of observation (see Fig.~\ref{f:self-observer}). Intuitively, we could say that the first-person observer must first arise so there is someone to have the experience of a unit of time. Impermanence also appears to be consistent with an intrinsic discretization of time as expected from a quantum theory of gravity. 

Let us now mention the concept of unconditioned mind which essentially says that to experience reality as it actually is, we should free ourselves of the many assumptions or beliefs we have about the world. So, we could say an unconditioned mind is a mind with zero assumptions, which is reminiscent of Wheeler's `law without law'. 

To facilitate this discussion, consider again the example of the complete dictionary above which turned out to be full of semantic loops. One way we could break such loops is by assuming we know some of the words in the dictionary (cf. Fig.~\ref{f:cavity_cycle}). This is similar to what is done in mathematics, where the mathematician must assume some statements, the axioms, to be true. It is similar also to what is done in science, where scientists first have to make some assumptions about nature that are then tested experimentally to check the assumptions are not wrong. Without assumptions, without axioms, we do not know how to start using the methods of logical deduction. Since in this work we are assuming the scientists are physical systems, we can consider the doing of science as a physical process part of which take place in the scientist's brain. Similarly, we can consider the doing of mathematics as a physical process going on the mathematician's brain when interacting with a set of statements which are also represented physically on the mathematician's brain or in a piece of paper.  

Consider as a toy model of this situation the simple example illustrated in Fig.~\ref{f:circular} where an observer is involved in a circular interaction with a switch and a lamp. One way the observer can break such a circle and turn it into a chain is by doing an assumption about the state of the switch and lamp, for instance (see Figs.~\ref{f:cavity_cycle} and \ref{f:circular}). This allows us to return to the classical perspective where there is an inherent existing state evolving forward in time and an inherent existent self who is observing such a process in a dual subject-object relationship. However, it might be possible for the observer not to make any assumptions but rather dive deep into the experience of such circularity in a non-dual way, i.e. through direct experience. Indeed, Buddhism emphasize that by continuously practicing certain activities such as mindfulness or vipassana meditation, the meditator eventually can have a direct experience of emptiness, i.e. not just as a theoretical construct. As the analogy with the dictionary suggests, emptiness might indeed be related to circularity. 

This discussion motivates us to ask the following admittedly bold question: Can we have a direct, non-dual, experience of quantum phenomena, such as quantum superposition or entanglement, by rigorously training our brains via the systematic first-person methods developed by contemplative traditions? This question can in principle be addressed experimentally by applying such first-person methods. There are subjective reports since many centuries ago that appear to suggest this might be the case. For instance, experienced meditators speak about a state of non-discriminating mind; the possibility of perceiving a bit without discriminating between its two states sounds similar to the potential experience of a quantum superposition. 

It is understandable the usual tendency to automatically dismiss these notions as total nonsense out of long-held believes against anything that could be labeled `spiritual', much as XIX century scientists used to do with the concept of atom. Fortunately, today scientific tools are emerging to be able to reject such notions in a more rigorous scientific basis, not just out of a belief that risks to become dogma (see Fig.~\ref{f:channels}). Once we accept the observer as a physical system which is part of the experimental setup, as {\bf Principle I} suggests, we have the opportunity to explore a whole new realm of physical phenomena by turning our attention within, which is a possible interpretation of what meditation techniques do. Certainly, we may find more convenient to avoid getting involved in this type of research because we may consider it is unlikely to yield interesting results, or we may be concerned that practicing meditation may harm us despite the many reports throughout history that it actually improves our well-being. Although such concerns are quite understandable, by no means they constitute valid scientific arguments to rigorously reject a set of ideas and empirical observations that have survived per centuries.

Finally, from the perspective explored in this work whatever an observer can observe requires her to get involved in a circular interaction with the objects of observation. This suggests the observer acts as a kind of hub that connects everything she can ever observe. In other words, everything the observer can ever experience is connected to everything else through herself. This is analogous to the Buddhist concept of inter-dependence, the idea that we are all connected. 

Again, Buddhism emphasize its focus is not on the concepts but rather on the phenomena they point to, which we can in principle experience via systematic first-person experimental methods. We have included this highly speculative discussion here to provide some ideas for debate which could potentially open a realm of phenomena that has not been considered to date as a fruitful subject of study in mainstream physics.

\subsubsection{Not just bare rationality: A call for action}\label{s:action}
We hope to have provided enough arguments in this work to convince the reader that we already have some tools to more rigorously address the role of human experience on the foundations of science. We now argue that the stakes are high for bringing this debate to the forefront. Although at first sight the ideas discussed below may not seem to be about physics, we contend that in a profound sense they are. Moreover, we are convinced this debate might offer a great opportunity for concerned physicists to enhance their contribution to the solution of some of the biggest worldwide challenges, such as global warming, income inequality, and the strong division we face today. Although focusing on the potential technological applications of the ideas developed in this article may be more profitable or convenient, we consider the debate on the potential implications of the ideas we are about to discuss as far more urgent. After all, any new technology, no matter how sophisticated, can be used for good or for bad; in a sense, with every new technology everything changes for everything to stay the same. What can really make a difference is the mindset of the users of the technology.

Indeed, by providing the ground on which we step to understand the world around, scientific paradigms have a huge impact on how we conceive businesses and other socio-economic activities that shape people's lives (see Fig.~\ref{f:hierarchy}). Oxford economist Kate Raworth, well-known for his idea of the `Doughnut Economics'~\cite{raworth2017doughnut}, put it nicely~\cite{raworth2017old}:

\epigraph{\em ``In the 1870s, a handful of aspiring economists hoped to make economics a science as reputable as physics. [...] The most pernicious legacy of this fake physics has been to entice generations of economists into a misguided search for economic laws of motion that dictate the path of development. People and money are not as obedient as gravity, so no such laws exist. Yet their false discoveries have been used to justify growth-first policymaking. [...] Thanks to more and better data, it has become clear that such economic laws of motion simply don't exist. Far from being a necessary phase of development, extreme inequality and environmental degradation are the result of policy choices, and these choices can be changed. In the place of laws to be obeyed, there are design decisions to be made.''}{K. Raworth, Ref.~\cite{raworth2017doughnut}}

We have suggested in this work that the mainstream paradigm might have to be revised to put our subjective human experience at the very foundation of science. In this revised paradigm we are not mere abstract entities helplessly subjected to the mechanical laws of a ruthless inherently existent external world. Rather, our human experience takes center stage. Given how authoritative physics is in the modern world, bringing this debate to the forefront of physics would give concerned physicists a unique opportunity to join the ongoing push towards a more human-centered society~\cite{brown2017buddhist}. We foresee at least two objectives of this research agenda
\begin{enumerate}
\item Start building a common language and exchanging methods between fundamental physics and contemplative traditions~\cite{varela2017embodied}. To succeed in physics we should sometimes overemphasize qualities such as rationality, fierce competition, and overworking~\cite{miller2010us}. The contemplative traditions seem to put a stronger emphasis on qualities such as mindfulness, generosity, love, as well as a simple and healthy life-style; according to the contemplative traditions such qualities are essential for a first-person  exploration of human experience because this requires a very calm and equanimous mind. So, a deeper convergence between science and contemplative traditions holds the potential to further improve the way we live.

\item Contribute to the ongoing transition towards a more human-centered society by exploring the potential disruption of the mainstream materialist paradigm, which sometimes seems to overemphasize less human-centered concerns such as competition, resource optimization, and profit maximization. This holds the potential to further encourage the exploration of more human-centered business models~\cite{brown2017buddhist} (see also Ref.~\cite{varela2017embodied}, chapter 11, page 243).
\end{enumerate}

Since the Age of Reason, or `Enlightenment', reason has been considered as the primary source of authority and legitimacy~\cite{wiki:xxx}. Mathematics could be considered as the very embodiedment of reason. In mathematics, much as in the example of the dictionary discussed in Sec.~\ref{s:worldview}, we start from a given set of statements called axioms and then we apply a set of logical rules to essentially transform those statements into new statements we call theorems. When this can be done we say the theorem is true. Yet this is similar to the traditional linear chain of cause-effect relationships that in physics connect an inherently existent intial state to an inherently existent final state. We have argued in this work that such a linear chain must be closed into a circle to include the effects of the observer. 

We expect something similar may happen in mathematics, where we may need to turn the linear chain connecting axioms to theorems into a circle that explicitly includes the mathematician in an effective way. Indeed, since inference can be considered as a physical process, we could in principle use the tools discussed here to model the doing of mathematics to get some sort of `quantum mathematics' where circularity, not linear reasoning, takes central stage.  Such type of circularity has been analyzed in economics, for instance, by Soros~\cite{soros2008crash}. Yet economics has been largely built on the idea that we are entities that mostly rely on linear rationality, which has been sometimes equated to selfishness---although there is growing evidence that humans not only care about themselves when taking decisions~\cite{declerck2015neuroeconomics}, and that balancing between individual and social considerations might enhance the adaptability of human groups~\cite{realpe2018balancing}. 

Now, contemplative traditions not only have been an inspiration for cognitive science and medicine, but also in other fields such as economics. On this regard, Clair Brown, Professor of Economics and Director of the Center for Work, Technology, and Society at the University of California, Berkeley, recently wrote~\cite{brown2017buddhist}:

\epigraph{\em ``Free market economics holds that human nature is self-centered and that people care only about themselves as they push ahead to maximize their incomes and fancy life-styles. According to this approach, buying and consuming---shopping for new shoes or playing a new video game---will make you happy. Forget that soon you will grow tired of the shoes, become disappointed with the game, and be off shopping again. In this endless cycle of desire, we are continuously left wanting more without ever finding lasting satisfaction. Free market economics is not guiding us toward living meaningful lives in a healthy world, nor is it offering solutions to our concerns about global wars, income inequality, and environmental threats.

``Buddhist economics, in contrast, provides guidance for restructuring both our individual lives and the economy to create a better world. `Practice compassion to be happy' replaces `More is better'. `Everyone's well-being is connected' replaces `Maximize your own position'. `The welfare of humans and Nature is interdependent' replaces `Pollution is a social cost that the individual can ignore''.}{C. Brown, Ref.~\cite{brown2017buddhist}}

We therefore find tempting to ask: Has perhaps the strong emphasis of mainstream physics on getting rid of the subjective, ourselves, unknowingly diverted our attention mostly towards less human-centered concepts such as resource optimization, automation, consumption, competition, growth, etc., sometimes leaving the more human-centered socio-ecological issues as a side concern to be dealt with later? Isn't it perhaps the implicit ideal of mainstream science today that we become rational machines and we are unaware of it? Perhaps it might be rational for a hypothetical oil producer to hamper electric car adoption, for instance, or for a hypothetical social network company to sell sensible information that can be used to manipulate voters in an election, as long as those decisions generate revenue. But is it mindful about current and future generations?

Physics has played a major role in shaping history when physicists have challenged the view of the majority. The history of science is full of examples of our astonishing human capacity to create and transform our reality. We hope to have made a case for why we consider the time is ripe and the stakes are high to start more thoroughly exploring these ideas that are already being explored in other fields of science. 

One interesting aspect of contemplative traditions such as Buddhism is that they do not point the finger to anyone. Rather, they suggests that independently of our political affiliation or belief system we all have one common enemy: the selfish tendencies that sometimes dominate our behavior and that very often we are not aware of. According to Buddhism, such selfish tendencies are rooted in the fact that we have not directly experienced that what we usually call `I' do not posses any intrinsic existence. This is consistent with Metzinger's view~\cite{metzinger2004being,metzinger2009ego} that such an `I' can be understood as a self-model, a representation in our brains; a representation, not an intrinsically existing object. More generally, in the Buddhist worldview nothing possesses inherent existence, everything is relational.

Here we have explored some ideas that may lead to a physics-based approach to the study of the self. More importantly, the same first-person methods we suggest could be useful in fundamental physics research, are also useful to counteract the selfish tendencies mentioned above; so it is not about moralisms. From this perspective, there is one war that holds great potential to substantially improve the way we live; it is a pacific and lovely war against our selves. We hope to have provided enough arguments to suggest that we physicists might have some important tools at hand to help the human race win such a war.

\acknowledgements
The figures in this work were constructed from images found via Google Images with the option `labelled for reuse with modification', mostly obtained from Wikimedia Commons and Pixabay. The photos in Fig.~\ref{f:first-third} were taken by Edwin Lemus in Bogot\'a, Colombia, following the directions of the author. This work was given slots in the `2018 March Meeting' of the American Physical Society held at Los Angeles, California, and the conference `The Science of Consciousness 2018' held at Tucson, Arizona. Unfortunately, I was unable to attend for personal reasons; a video will be uploaded for the people who may have been interested.

I thank Marcela Certuche for making the completion of this project possible and for the many discussions on these ideas. I thank also Harold Certuche and Blanca Dominguez for their support in the last part of this project. I thank also Addishiwot Woldesenbet Girma, Alejandro Perdomo-Ortiz, and Delfina Garc\'ia Pintos for the support and the interesting discussions on this work. I thank Tobias Galla, Alan J. McKane, and the University of Manchester for giving me the opportunity to work on this project while being a Research Associate. I thank Buddhist monk Kelsang Sangton from the Kadampa Meditation Center in Bogot\'a, Colombia, for very useful discussions on this work. I thank Marcin Dziubi\'nski for his short but nice lessons on recursion and self-reference at the University of Cambridge. I thank Diana Chapman Walsh for interesting discussions on contemplative science and for bringing my attention to the work of Arthur Zajonc~\cite{zajonc2004new}. I also thank Michael R. Sheehy for interesting discussions on these ideas. I thank Marcello Benedetti, Gonzalo Ordo\~nez, Maria Schuld, and Alejandro Perdomo-Ortiz for useful comments on the first version of this manuscript. I also thank the organizers of the conferences ``Science and Nonduality 2017'' held at San Jose, California, the ``Mind \& Life Summer Research Institute 2018'' held at Garrison, New York, and the ``Second Workshop on Biological Mentality'' held at Ann Arbor, USA, where these ideas were presented. {I thank Camila Sardeto Deolindo and Cerys Tramontini for bringing my attention to the work of Alan Wallace~\cite{wallace2007hidden}} I thank as well Mario Chamorro and Laura Escobar for their inivitation to discuss this work in the `AniMondays' they organized in 2017 at Mountain View. 

I thank Matteo Marsili, Luca Dall'Asta, and the Abdus Salam ICTP for the hospitality during a visit in 2010, where an early version of some of these ideas was drafted. I also thank Valentina Lanza, Alfredo Braunstein, Abolfazl Ramezanpour, Riccardo Zecchina, Fabrizio Altarelli for their support at the Politecnico di Torino when the development of these ideas started. I thank also Ralph Gebauer, Giuseppe Santoro, Hernan Ocampo, and Efrain Solarte for the excellent lessons and discussions on quantum mechanics. I thank John Henry Reina, Fabio Benatti, and Roberto Floreanini for useful discussions on quantum theory. I thank Gerhard Gr\"ossing and his wife for the hospitality at the conference ``Emergent Quantum Mechanics'' of the ``Heinz von Foerster Congress'' which took place at Vienna in 2011, and to Garnet Ord for the interesting discussions held there. I thank also Gerard 't Hooft for the short but interesting discussion on the foundations of quantum mechanics at the ``Euroscience Open Forum'' held at Turin in 2010. I thank Maria Schuld, Francesco Petruccione and his wife, as well as Kriben Pillay for the hospitality at the summer school on ``Quantum Machine Learning'', which took place at the Drakensberg Mountains in 2017, and for the interesting discussions related to these ideas. 

I thank Narlinda Espinosa, Pedro Pablo Ortega, Tommaso Biancalani, and Jaime G\'omez G\'omez for listening to some of these ideas. I also thank Narlinda Espinosa, Javier Montoya and Beatriz Cogollo for their support at the Universidad de Cartagena that facilitated the completion of this work. 

I must thank Roberto Dotta, Luca Gilli, Johanna Wirrer, Mario Andr\'es Acero Ortega, Andrea Occhipinti, Fabio Ricciato, Diana Vanessa Realpe, and Gerardo Paz-Silva for their support that contributed to the development of this project.

Last but not least, I thank Mariela G\'omez Ram\'irez and Nelson Jaramillo G\'omez for bringing my attention to these ideas.

\appendix 

\section{We as physical systems}\label{s:we}
According to {\bf Principle I}, anything that a robot can perceive must be physically represented in its hardware (see Fig.~\ref{f:circular}). Since we are also observers, {\bf Principle I} implies that the same happen to us (see Fig.~\ref{f:third}). In other words, it implies that we are not the immaterial entities that classical physics has implicitly assumed for so long. We are physical, our senses are physical, our brains are physical. From this perspective, we are not conceptually different from a computer since anything we can percieve has a physical process associated in our brains, the so-called {\it neural correlates}~\cite{dehaene2014consciousness,koch2016neural}. More clearly, although the way in which our brains encode and process information may be radically different from the way state-of-the-art computers do, the minimal physical requirements identified above, i.e. physical representation and interaction, cannot be avoided.

In this respect, according to {\bf Principle I}, the brain is the fabric of our reality. We are unable to distinguish reality from our perception of it in the sense that what exists for us, what we can perceive, must be physically represented in our `hardware'. There is indeed experimental evidence to support this. Consider, for instance, the experimental demonstration reported in Ref.~\cite{losey2016navigating} where humans are able to play a video game using only input from direct brain stimulation, i.e. without relying on any usual sensory inputs like sight, hearing, or touch. Another example is the recent experimental demonstration reported in Ref.~\cite{salazar2017correcting}, where humans can control robots via the neural correlates of their thoughts (see also Refs.~\cite{cohen2012fmri, cohen2014controlling}). Neural correlates have been identified even for hallucinations and out-of-body experiences~\cite{lenggenhager2007video,ehrsson2007experimental,blanke2009full,ehrsson2004s,dehaene2014consciousness} (see Fig.~\ref{f:subjective}). 

A different type of evidence, however disturbing, is Buddhist monk Thich Quang Duc, who remained in deep meditation while burning in flames until death on 11 June 1963~\cite{NY,young1991vietnam} (disturbing videos of this are available in YouTube). For many of us, touching with our fingers the flame of a candle may produce a perception of extreme pain that would move us to act almost instinctivly. However, the case of Thich Quang Duc suggests that things are not necessarily the way we are accustom to, and that our perception of reality might be manipulated.

It is understandable the extreme caution that we physicists have had with anything related to ourselves. But we now live in an era where machine learning and artificial intelligence are doing amazing progress~\cite{mnih2015human, lecun2015deep}, and scientists have found clever ways to experimentally tackle certain aspects of consciousness~\cite{koch2016neural,dehaene2014consciousness}. We need to be careful of any strong prejudices that may remain from the past and that may prevent us from making further progress. Indeed, the observation that our brain co-creates our reality may be uncommon to some physicists, but not so for brain research scientist--- see e.g. Ref.~\cite{eagleman2015brain}, chapter 2, for a recent presentation intended for a general audience. This has also been the position of some millenarian spiritual traditions, like Buddhism, and one of the tenets underlying mindfulness meditation~\cite{tang2015neuroscience}--- see e.g. Ref.~\cite{hanson2009buddha} for a modern scientifically-minded presentation of these ideas (see also Ref.~\cite{varela2017embodied}, chapter 10).

\begin{figure*}
\includegraphics[width=0.8\textwidth]{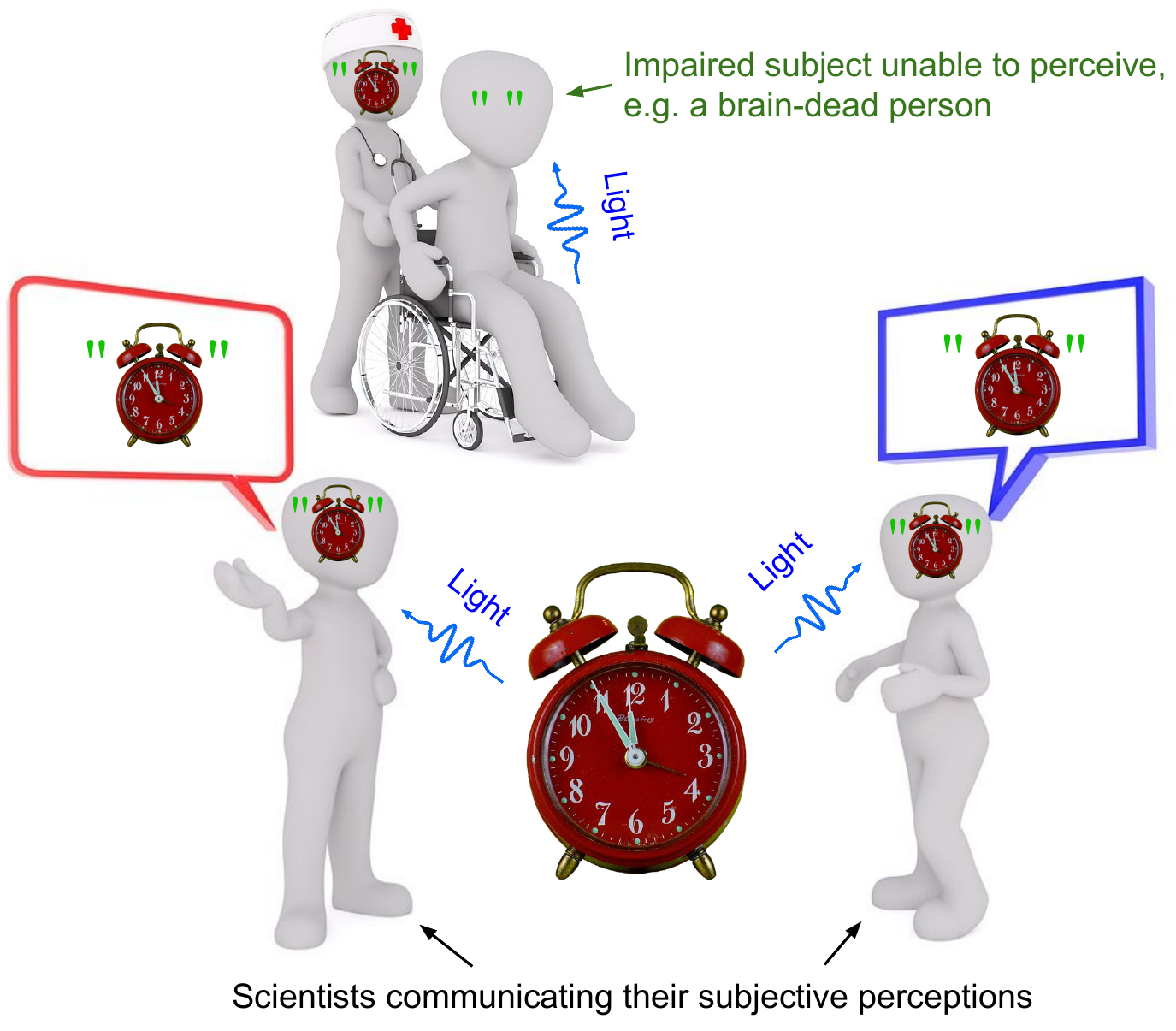}
\caption{ {\it Illustration of the `model what is' motto}. An implicit assumption in physics is that measuring devices (here a clock) take the subject out of the loop, allowing us to have an objective description of the world. This assumption is indeed consistent with experimental evidence in the sense that our subjective perceptions of the readings of measuring devices appear to be consistent, i.e. when we communicate our subjective perceptions to other scientists we usually agree on them (bottom), as if they were objective. However, the `model what is' motto asks us not to start our analysis from this assumption, but rather from the very inferential process that allows us to test its validity in the first place. More precisely, the `model what is' motto (see Sec.~\ref{s:intro} and Appendix~\ref{s:we}) ask us to first model what we actually experience and only afterwards evaluate whether the communicated subjective experiences seem consistent with the {\em assumption} that there is an objective world out there; this is a sort of {\em intersubjectivity}. According to recent research there are physical processes in the brain (and body) associated to our subjective experiences, i.e. their neural correlates (here represented by a replica of the external system within green quotation marks). Consider, for instance, a brain-dead person (top; here represented with green quotation marks enclosing a blank space) in front of an experimental device: even though he receives the same information (via light) as others observing the same device, his brain is unable to construct a percept of it; i.e. no properly working brain no perceptions. Furthermore, consistent with textbook physics, the neural correlates associated to an external system must be induced by physical interactions (e.g. light). {\bf Principle I} asks us to take such physical interactions into account and only afterwards evaluate whether they can be neglected as usually done in physics. {\bf Principle II} asks us to describe experiments from a subjective perspective, i.e. from the perspective of the scientist actually carrying out the experiment. Since, as far as we know, the only first-person perspective we have access to is our own, {\bf Principle II} essentially asks us to describe experiments from our own perspective (see Fig.~\ref{f:first}), i.e. from the perspective of each one of us. Since, in this view, physics is about agreeing on a class of subjective experiences, we can in principle explore potential extensions of physics to other type of subjective experiences (see Fig.~\ref{f:channels}). Indeed, subjective experiences such as emotions, for instance, appear to be associated to certain physical processes (e.g. face expressions) consistent enough for computers to be able to recognize them with good accuracy~\cite{jung2015joint}.}
\label{f:inter}
\end{figure*}

\begin{figure*}
\includegraphics[width=0.95\textwidth]{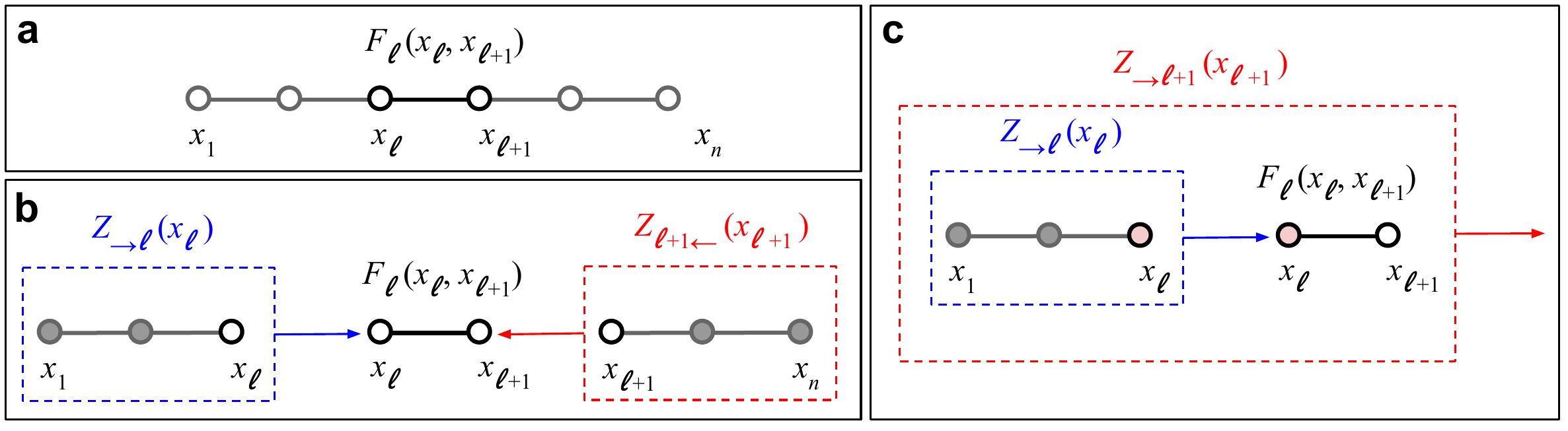}
\caption{ {\it Illustration of the cavity method}. (a) Factor graph associated to a Markov chain (see Eq.~\eqref{e:P_prodF}). (b) Graphical expression for the pairwise marginal $\mathcal{P}_\ell(x_\ell, x_{\ell +1})$ (see Eq.~\eqref{e:P_BP}); the partial partition functions $Z_{\to\ell}(x_\ell)$ (blue; see Eq.~\eqref{e:Z->}) and $Z_{\ell+1\leftarrow}(x_{\ell+1})$ (red; see Eq.~\eqref{e:Z<-}) correspond to the sum over all variables on the cavity graphs inside the dashed rectangles, except for $x_\ell$ and $x_{\ell +1}$ which are clamped to be able to recover the whole grapical model by multiplying for $F_{\ell}(x_\ell, x_{\ell +1})$. (c) The partial partition function $Z_{\to\ell+1}(x_{\ell+1})$ (red) can be recursively computed by multiplying the partial function $Z_{\to\ell}(x_\ell)$ and the factor $F_\ell(x_\ell, x_{\ell+1})$ and tracing over $x_\ell$ (see Eq.~\eqref{e:BP->}). This is the content of the belief propagation algorithm~\cite{Mezard-book-2009} specified by Eqs.~\eqref{e:BP->} and \eqref{e:BP<-}.}
\label{f:cavity}
\end{figure*}

\begin{figure}
\includegraphics[width=0.95\columnwidth]{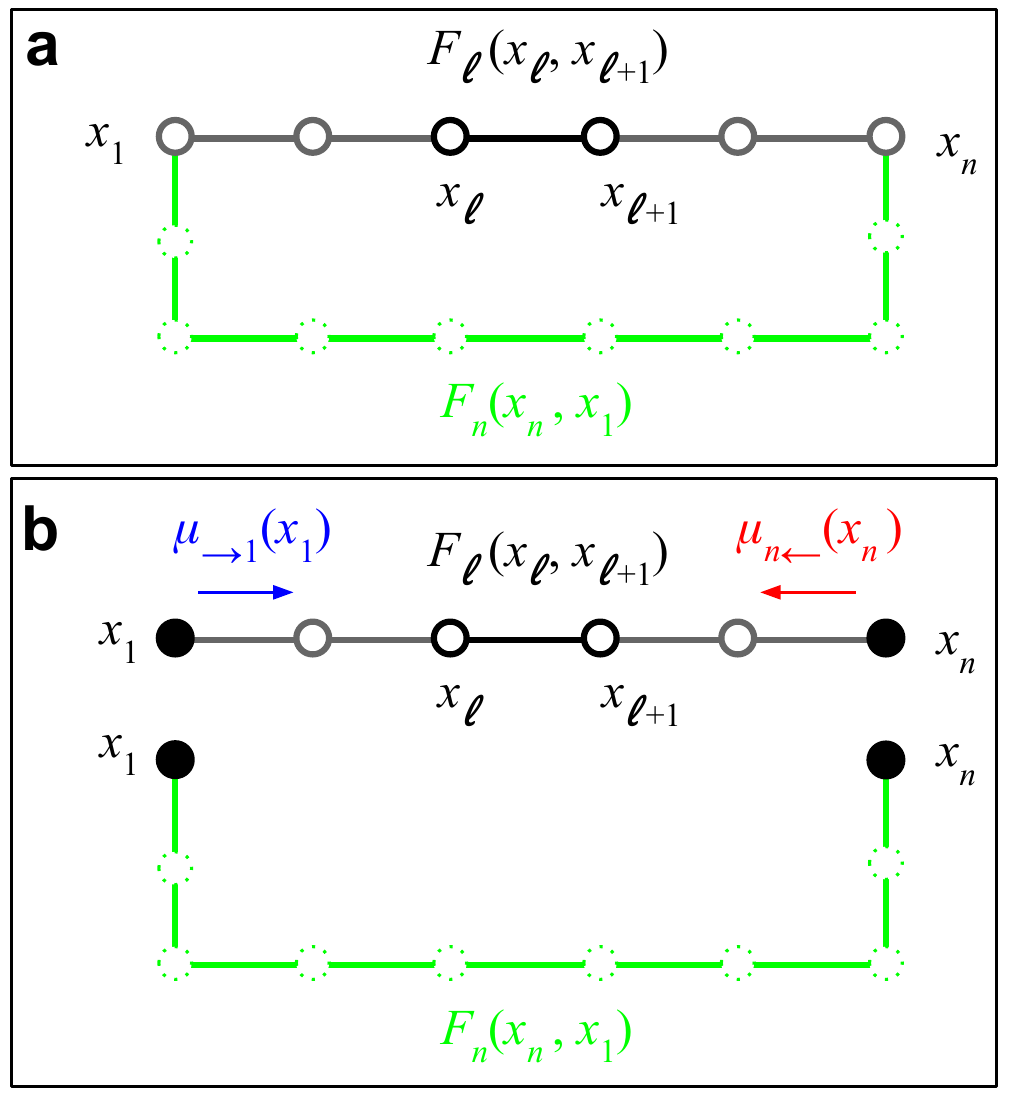}
\caption{ {\it Illustration of cavity method on cycles}. (a) Factor graph with circular topology, where the new factor $F_n(x_n, x_1)$ (green) closes the chain in Fig.~\ref{f:cavity}a. While this factor could be the product of other factors representing local interactions between additional variables (dotted circles), it is enough for our purposes to consider only one. In contrast to an open chain, it is in general not possible to write $\mathcal{P}_{\rm cycle}(x_1, \dotsc ,x_n)$ as a Markov chain (see Sec.~\ref{s:EQM}). Furthermore, the standard belief propagation equations do not generally lead to the correct marginals \cite{Weiss-2000}. (b) By conditioning on $x_1$ and $x_n$ (black filled circles) we can remove factor $F_n(x_n, x_1)$ to turn the cycle into a chain with $x_1$ and $x_n$ clamped to some given values $x_1^\ast$ and $x_n^\ast$. Such clamping can be implemented via {\em pure} initial and final messages ${\mu_{\to 1}(x_1) = \delta(x_1- x_1^\ast)}$ and ${\mu_{n\leftarrow}(x_n) = \delta(x_n- x_n^\ast)}$, where $\delta(x)$ denotes the Dirac delta function. Recall that any quantum pure  state $\left|\psi\ket = U\left| x^\ast\ket$ can be prepared from a state with support on a single point obtained from a projective measurement $\left| x^\ast\ket$, with $\bra x \right|\left. x^\ast\ket =\delta(x-x^\ast)$, and a unitary transformation $U = \left|\psi\ket\bra x^\ast\right| + \sum_{x\neq x^\ast}\left|\phi_x\ket\bra x\right|$, where  $\left|\psi\ket\bra \psi\right| + \sum_{x\neq x^\ast}\left|\phi_x\ket\bra \phi_x \right| = I$, which can be implemented via certain Hamiltonian. Since the messages are closely related to imaginary-time wave functions, this is somehow similar to the imaginary-time two-vector state formalism of quantum mechanics~\cite{reznik1995time, aharonov2010time}, in which we define initial and final pure states (pre- and post-seection) and predict the state in between (see Sec.~\ref{s:EQM}).}
\label{f:cavity_cycle}
\end{figure}

Of course, this does not rule out the existence of an objective world out there. What we perceive, e.g. light, can be generated by an external physical processes, e.g. a lamp shining. But what we perceive is not the lamp itself, but the physical processes or neural correlates that the lamp generates in our brain. From this perspective, we could say that when scientists discuss the results of an experiment what they actually discuss are the perceptions they have (see Fig.~\ref{f:inter}). For instance, when scientists read a number in the display of a machine, such perception is actually a physical process happening in the scientists' brains. We often take for granted that, for instance, the perception of a number on the display of a machine is generated by a truly existing machine displaying such a number. Indeed, this assumption is quite consistent with our experience as scientists. Yet, instead of using this robust assumption as the starting point, here we start earlier by modelling the very inference process that allows us to trust this assumption (see Fig.~\ref{f:inter}). 

Let us consider the following thought experiment:

\

\noindent{\bf Thought experiment:} As we mentioned above, scientist have already demonstrated that it is possible to induce percepts, such as the experience of seeing a simple video screen~\cite{losey2016navigating} or the feeling of being out of the body~\cite{lenggenhager2007video,ehrsson2007experimental,blanke2009full}, in human subjects by direct brain stimulation without relying on any usual sensory inputs such as sight, audio, touch, etc. These proof-of-principle experiments allow us to imagine that, as the underlying technology progresses, it might be possible to induce more and more realistic percepts without relying on the traditional sensory pathways. Let us imagine a time when such technology is mature enough to induce the experience that we are living in a virtual world we cannot distinguish from the real world, much as the situation depicted in the movie {\em The Matrix}. In such a situation it can happen that we go to sleep while experiencing the real world and then some scientists plug our brains to such technology without us being aware of it. The scientists could then induce the experience that we have woken up the next day in a virtual world we believe is the real world. What experiment could we carry out to realize we are experiencing a virtual world? Consistent with the modern scientific insights that our brain is the fabric of our reality, we expect there is no experiment to distinguish our reality from our perception of it (see Fig.~\ref{f:inter} and Ref.~\cite{Zeilinger-Nature2005}). While this may be new and even shocking to some scientists, this conclusion has been reached many centuries ago by some contemplative traditions via rigorous first-person experimental methods (see Sec.~\ref{s:onScience}). While scientists have mostly focused their attention on phenomena taking place in the external world---this is the case even when scientists study others' brains---contemplative practitioners have mostly focused their attention on phenomena taking place inside themselves (see Fig.~\ref{f:channels}). 

The view we take here, which is consistent with {\em intersubjectivity}, is that physics is about people, say scientists, agreeing on a class of subjetive experiences. Let us now, discuss a class of subjective experiences, which is usually not considered part of the realm of physics, that relates directly to the sense of `I' and the first-person perspective, which is related to the concepts explored in this work. From our personal experience we know that we can experience ourselves as subjects, i.e. as an `I' that perceives the external world; consider for instance the sentence 
\be\label{e:S1}
\textrm{\it I observe a lamp.}
\ee
Similarly, we can also experience ourselves as an object, a `me' that is observed by someone else; consider for instance the sentence 
\be\label{e:S2}
\textrm{\it The government is observing me.}
\ee
We can even experience circular things like thinking of the sentence
\be\label{e:S3}
\textrm{\it I am looking at myself.}
\ee
So, from a psychological point of view we have no problem to deal with ourselves as both subjects and objects. 

Understanding this complex matter is out of reach to the author of this work. However, we can use {\bf Principle I} along with the verifiable fact that we can think of (or observe) statements like those in Eqs.~\eqref{e:S1} to \eqref{e:S3} to conclude that there must be physical processes representing such experiences of ourselves as both subjects (e.g. Eq.~\eqref{e:S1}) and objects (e.g. Eq.~\eqref{e:S2}), as well as the kind of circular experiences described by Eq.~\eqref{e:S3}. Indeed, there is ongoing research on identifying the neural correlates of the self~\cite{northoff2013brain,schaefer2017conscious}. Our work suggests that the neural architecture of the self should be composed of two neural sub-systems that essentially model each other, which could potentially explain why our brains is divided into hemispheres (see Sec.~\ref{s:architecture} and Fig.~\ref{f:brain_self}).

In this respect, the complementary pairs described in Secs.~\ref{s:big_1st_person} and \ref{s:first} appear to be related to the physical processes representing the perception of ourselves as subjects, i.e. observers, and as objects, i.e. observed. As early as 1929, Niels Bohr~\cite{bohr1929quantum}  had already considered this as the possible origin of complementarity in quantum theory. Consider what Niels Bohr wrote in Ref.~\cite{bohr1929quantum} (as reproduced in Ref.~\cite{wang2015reintroducing}):

\epigraph{\em ``For describing our mental activity, we require, on one
hand, an objectively given content to be placed in
opposition to a perceiving subject, while, on the other
hand, as is already implied in such an assertion, no sharp
separation between object and subject can be maintained, 
since the perceiving subject also belongs to our mental
content. From these circumstances follows not only the
relative meaning of every concept, or rather of every word,
the meaning depending upon our arbitrary choice of view
point, but also that we must, in general, be prepared to
accept the fact that a complete elucidation of one and the
same object may require diverse points of view which defy a
unique description. Indeed, strictly speaking, the conscious
analysis of any concept stands in a relation of exclusion to
its immediate application. The necessity of taking recourse
to a complementary, or reciprocal, mode of description is
perhaps most familiar to us from psychological problems.''}{N. Bohr, Ref.~\cite{bohr1929quantum}}

As we discuss in Secs.~\ref{s:self} and \ref{s:quantum}, {\bf Principle II} leads to complementarity when dealing with the self-referential problem of describing the world from within. This could help explain why quantum models appear to parsimoniously describe cognitive phenomena~\cite{bruza2015quantum}.

\section{The modern and well-respected approach to consciousness}\label{s:respected}
\subsection{Third-person perspective and the `easy' problem of consciousness}\label{s:easy}

Dehaene discusess three key ingrediates that have helped the subject of consciousness become a hot research topic today~\cite{dehaene2014consciousness} (page 8; see also Ref~\cite{dehaene2017consciousness}): (i) a better definition of consciousness that distinguishes a minimum of three concepts: vigilance, attention, and conscious access---among these, consciouss access, i.e. `the fact that some of the attended information eventually enters our awareness and becomes reportable to others', has been the main focus of experimental research, and is the focus of this work too; (ii)  the development of techniques to manipulate consciousness in the lab; (iii) taking subjective introspection seriously, not as a research method but as a source of raw data that can be contrasted with measurements of neuronal activity to identify potential Neural Correlates of Consciousness (NCCs; see Fig.~\ref{f:subjective}). 

One of the key aspects identified by this modern approach to consciousness is that information can be processed unconsciously. Another key aspect is the identification of certain features, the NCCs, that distinguishes conscious perception from unconsious information processing. We now summarize some of these insights for reference:
\begin{enumerate}
\item {\em Information can be processed unconsciously:} Many information processing tasks can be performed unconsciously~\cite{dehaene2014consciousness} (chapter 2); we can even do some mathematical operations at the unconscious level. Apparently unconscious information processing is carried out by myriad processors working in paralllel. This appears to be related to the feature-extraction component of the artificial observer illustrated in Fig.~\ref{f:realisticobserver}.

\item {\em Consciouss perception is a neural `tsunami':}  Conscious perception is distinguished by a relevant amplification factor in neural activity as compared to unconscious information processing~\cite{dehaene2014consciousness} (chapter 4); in words of Dehaene:  
\epigraph{\em ``Subliminal perception can thus be compared to a surf wave that looms large on the horizon but merely licks your feet when it reaches the shore. By comprarison, conscious perception is a tsunami---or perhaps an avalanche is a better metaphor, because conscious activation seems to pick strength as it progresses, much as minuscule snowball gathers snow and ultimately triggers a landslde''}{S. Dehaene, Ref.~\cite{dehaene2014consciousness} (page 119)} 

\item {\em Conscious perception is global neural activity}: Conscious perception is distinguished by a synchronization of information exchanges across distant brain regions (cf. Fig.~\ref{f:realisticobserver}c,d).
 
\item {\em Conscious perception is bidirectional causality:} Conscious perception is distinguished by a massive increase in bidirectional causality throughout the brain, producing a sustained state of reverberating activity. In other words, there is a forward-moving wave which can be interpreted as the climbing of sensory information from raw data to increasingly abstract representations of the stimulus (cf. Fig.~\ref{f:realisticobserver}). Furthermore, there is an opposite descending wave whose function Dehaene suggests could be amplifying incomming activity or checking that the input is consistent with the current interpretation at a higher level. This bidirectional dynamics seems similar to the situation illustrated in Fig.~\ref{f:self-observer}, which we expect to be a minimal architectural requirement for a physical system to implement a self-referential observer with a first-person perspective. 

\item {\em Conscious access is discrete}: Consciouss access seems to be related to a discontinous phase transition induced by the feedback mechanism that arises when the neurons at the higer level sent back excitatory signals to the very units that activated them~\cite{dehaene2014consciousness} (page 184). This seems to be consistent with evidence suggesting that conscious access is all-or-none~\cite{llinas2013cortical} and, in the perspective take in this work, might underlie the quantization of energy (see Sec.~\ref{s:Planck}). 

\item {\em Consciouss processing resembles a computer}: Consciousness appears to give humans the power of a Turing machine. More precisely, the brain's behavior during long conscious information-processing tasks, such as the performance of complex mathematical calculations, is roughly captured by the ideal model of a Turing machine~\cite{dehaene2014consciousness} (page 104). This is consistent with the second component of a physical observer described in Fig.~\ref{f:realisticobserver}.
\end{enumerate}

These features concern the information-processing mechanisms underlying conscious perception, which is considered as one of the `easy' problems of consciousness. These features do not address the so-called `hard' problem of consciousness closely related to our human experience. In the next section, we briefly discuss recent developments on this front.
\subsection{First-person perspective and the `hard' problem of consciousness}\label{s:hard}

Edelman~\cite{edelman2008computing} (page 412) summarizes the so-called `hard' problem of consciousness in three questions: What makes a mere physical process an experience for someone? What makes a mere physical system a subject, or an experiencer? What does having a first-person perspective on an experience consists of? About fifteen years ago, Metzinger~\cite{metzinger2004being} (see also Ref.~\cite{metzinger2009ego} and chapter 9 in Ref.~\cite{edelman2008computing}) put forward a theory of subjectivity based on the concept of a self-model, which provided a fresh approach to the `hard' problem of consciousness. (For a short introduction to the most central ideas see Metzinger's talk {\em `The transparent avatar in your brain'} at TEDxBarcelona.) Since we expect such a self-model to be closely related to the self-referential observers discussed in this work, we briefly mention here some of the main concepts involved. 

Edelman~\cite{edelman2008computing} (chapter 9) summarizes four key components in Metzinger's theory~\cite{metzinger2004being}: (i) Agency, or sense of initiative; minenes, or sense of ownership; perspectivelness, or the perception of phenomenal space as being organized around the self; selfhood, or the conscious experience of being someone. In particular, Edelman~\cite{edelman2008computing} (see page 419) summarizes Metzinger's ideas on selfhood in four steps that we cite almost textually here:
\begin{enumerate}
\item {\em The experienced reality is virtual.} The `raw data'
in the world is accessible to the brain exclusively through the mediation of its
sensory apparatus (which includes both the five external senses and the various interoceptive channels). No matter how veridical some of the information
provided by these senses is, the representations they feed into are necessarily
`virtual' computational constructs. This seems to be consistent with the architecture of physical observers illustrated in Fig.~\ref{f:realisticobserver}. 

\item {\em The experienced reality is a simulation of the world.} Use of the virtual representations generated by the senses often involves `simulation' of
events or situations.  Simulation is also central to planning and control: intended actions,
for example, are represented by motor programs whose effects are simulated
by a circuit that gives rise to the phenomenal sense of agency and ownership.
Note that an embodied and situated agent is an integral part of the world, and so must be simulated along with it by the agent's
cognitive system.

\item {\em The simulation is not recognized by the system as such.} To avoid
infinite regress (trying to represent a system that represents a system that
represents...; see Figs.~\ref{f:first} and ~\ref{f:first-third}), the model of the world (which includes a model of the system
itself) is taken to be the `last word'---the ultimate reality. This is essentially the concept of transparency we mentioned in the introduction. We here tackled the infinite regress via ideas related to the recursion theorem, which suggests that the minimal architecture of a self-model should be composed of two complementary systems that essentially model each other. From this perspective, it is tempting to think that in the case of humans, such a minimal architectural requirement is implemented via the division of our brain into two hemispheres, which to a large extent appears to perform certain complementary tasks (see Fig.~\ref{f:brain_self}; cf. Fig.~\ref{f:self}).

\item {\em The part of the simulation that represents the system itself is special.}
The represented reality contains one component that differs from all others in
being always present. This self-model---the only representational structure
that is fed by a continuous source of internally generated (interoceptive) input---is the phenomenal self.
\end{enumerate}

\section{Self-reference and the recursion theorem}\label{s:recursion}
\begin{figure}
\includegraphics[width=\columnwidth]{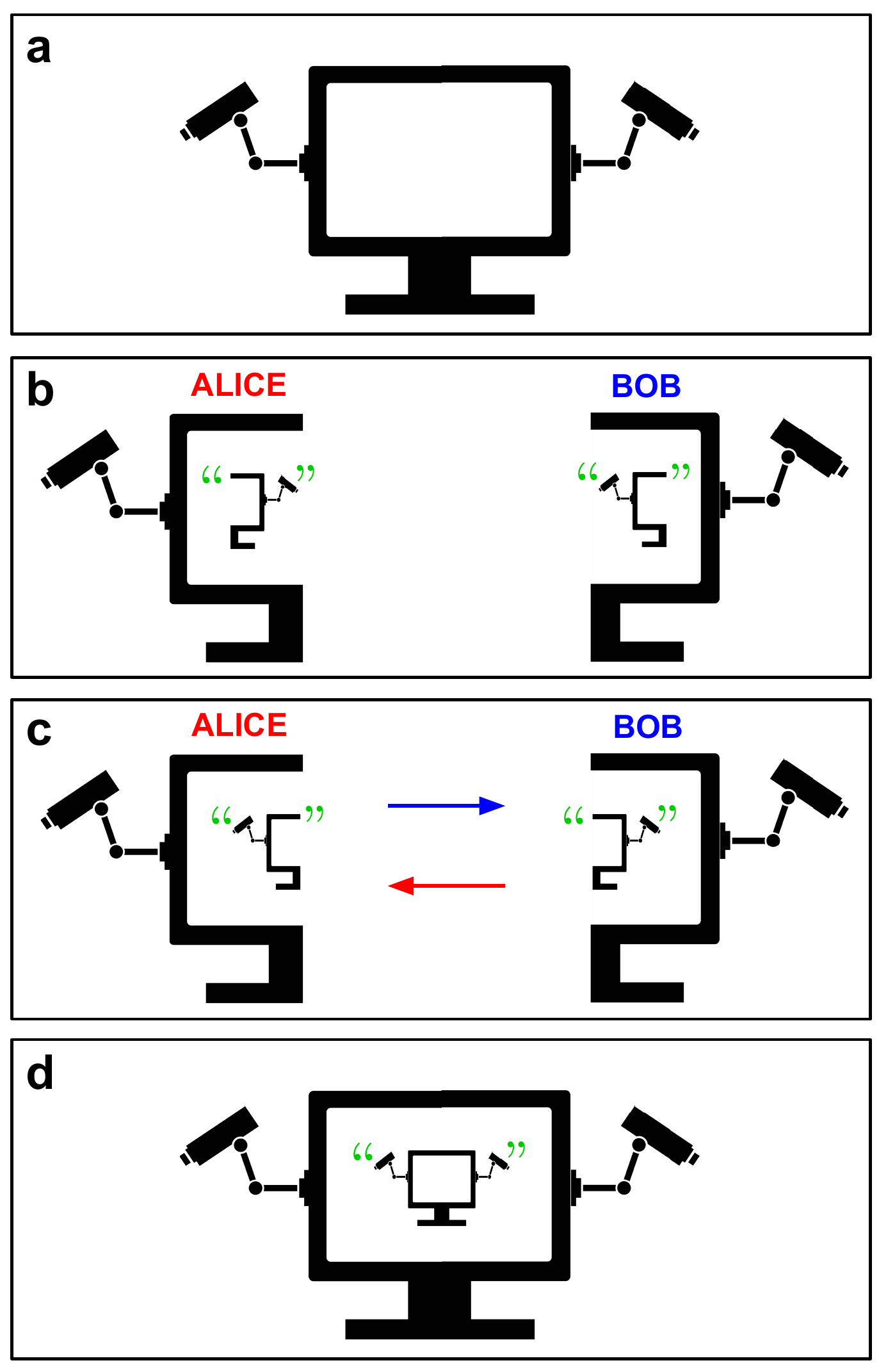}
\caption{ {\it Self-reference and complementarity.} (a) The goal is to build a Turing machine (represented here by a computer) that prints (in the screen) a {\it description} of itself (represented here by the same computer within quotation marks). The recursion theorem in its full generality is presented in Appendix~\ref{s:recursion} following Ref.~\cite{sipser2006introduction}; here we present a slightly modified, more symmetrical, and simplified description of the main concepts involved.  (b) This is achieved by two complementary sub-machines, {\sc Alice} and {\sc Bob}, that essentially print a description of each other, here represented by drawings within (green) quotation marks. To avoid circularity, while {\sc Alice} directly prints or {\it generates} a description of {\sc Bob}, {\sc Bob} actually {\it infers} a description of {\sc Alice} from her output. This is reminiscent of the free energy principle underlying the modern approach to brain modeling via active inference~\cite{friston2010free} or the related architecture of a Helmholtz machine~\cite{dayan1995helmholtz,benedetti2018quantum}. (c) However, after {\sc Alice} and {\sc Bob} mutually print each other they need to exchange their descriptions to obtain a correct image; this is symbolized here by the colored arrows. (d) In this way we obtain a Turing machine that, when run, prints a description of itself.}
\label{f:self}
\end{figure}

Here we briefly discuss the recursion theorem of computer science, which can be considered as a mathematical formalization of the concept of self-reference. Figure~\ref{f:self} illustrates the core concept underlying the recursion theorem using the specific case of a self-printing program, or quine, introduced in Sec.~\ref{s:big_1st_person}. As we mentioned in Sec.~\ref{s:big_1st_person}, a self-printing program, represented as a full computer in Fig.~\ref{f:self}, is composed of two complementary sub-programs, {\sc Alice} and {\sc Bob} represented as the two halves of a computer in Fig.~\ref{f:self}, that essentially print each other (cf. Fig.~\ref{f:first-third}b).

We now formalize this concept following Ref.~\cite{sipser2006introduction} (chapter 6). Let $\Sigma$ be an alphabet, i.e. a finite set of characters, and let $\Sigma^\ast$ denote the set of all possible strings of characters from alphabet $\Sigma$, here referred to as words.  A Turing machine is an abstract machine with no memory constraints which can manipulate the characters in an alphabet $\Sigma$ according to a pre-specified set of rules. In other words, a Turing machine is the implementation of an abstract mechanical process that transforms a given string of characters into another, effectively computing a given function $f:\Sigma^\ast\to\Sigma^\ast$. 

It is possible to associate to every Turing machine {\sc TM} a unique string of characters $\textsc{``TM''}\in\Sigma^\ast$, which is referred to as the description of the Turing machine; the quotation marks {\sc `` ''} can be considered here as an operator that transform Turing machines into strings in $\Sigma^\ast$.  A Turing machine {\sc TM} is the abstract version of a program, e.g. a search engine, that can run on a computer to perform a given task, e.g. search for websites related to a specific keyword. A description of a Turing machine {\sc ``TM''}, instead, is the abstract version of the code written in a specific programming language that is used to compile the corresponding program. This ability to associate a unique string of characters to a Turing machine is what allows a Turing machine to implement self-reference. Indeed, from this perspective we can think of a Turing machine {\sc TM} as a string of characters, i.e. its description {\sc ``TM''}, that can manipulate any string of characters $w\in\Sigma^\ast$, including its own description, i.e. $w=\textsc{``TM''}$. In this sense, it can manipulate (a description of) itself.

\begin{figure*}
\includegraphics[width=0.8\textwidth]{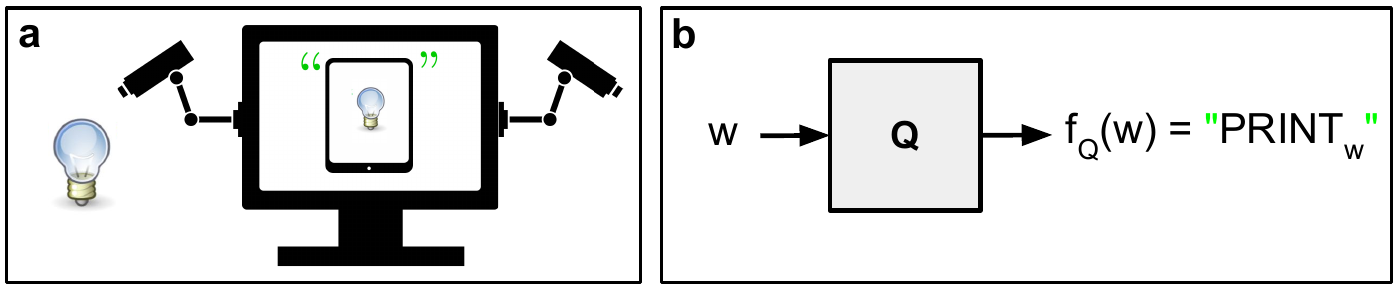}
\caption{ {\it Inferring the description of a printing machine.} (a) Cartoon example of a Turing machine (represented here by a computer) that, given some input (here a bulb), prints the {\it description} of a Turing machine (represented here by a Tablet within quotation marks) that prints the given input. (b) A more formal representation of the Turing machine in (a), here called {\sc Q}. Ref.~\cite{sipser2006introduction} uses {\sc Q} to proof the recursion theorem. The input is a string $w$ of characters from a suitable alphabet, and the Turing machine that prints $w$ is called {\sc Print}$_w$. The Turing machine {\sc Q} basically {\it infers} $\textsc{Print}_w$, effectively implementing the function $f_\textsc{Q}$ that maps $w$ into ``$\textsc{Print}_w$''. Proving the existence of {\sc Q} is straightforward (see Ref.~\cite{sipser2006introduction}, chapter 6). }
\label{f:q}
\end{figure*}

An example relevant for our discussion is the Turing machine {\sc Q} represented as a computer in Fig.~\ref{f:q}a. Given some input (e.g. the image of a bulb in Fig.~\ref{f:q}a), the Turing machine {\sc Q} prints the {\em description} of another Turing machine, represented by a tablet within quotation marks in Fig.~\ref{f:q}a, that prints the given input.  More formally, the Turing machine {\sc Q} implements a function $f_{\textsc{Q}}$, represented by a box in Fig.~\ref{f:q}b, that takes as input any word $w\in\Sigma^\ast$ and prints the description {\sc ``Print$_w$''} of another Turing machine {\sc Print}$_w$ that ignores its input and just prints $w$. The existence of such a Turing machine is proven in the

\

\noindent{\bf Lemma $6.1$ of Ref.~\cite{sipser2006introduction}:} There is a computable function $f_Q : \Sigma^\ast\to\Sigma^\ast$ such that, for any string $w$, $f_Q(w) = \text{``\textsc{Print}$_w$''}$ is the description of a Turing machine $\textsc{Print}_w$ that ignores its input, just print $w$ and halts.

\

\begin{figure*}
\includegraphics[width=0.8\textwidth]{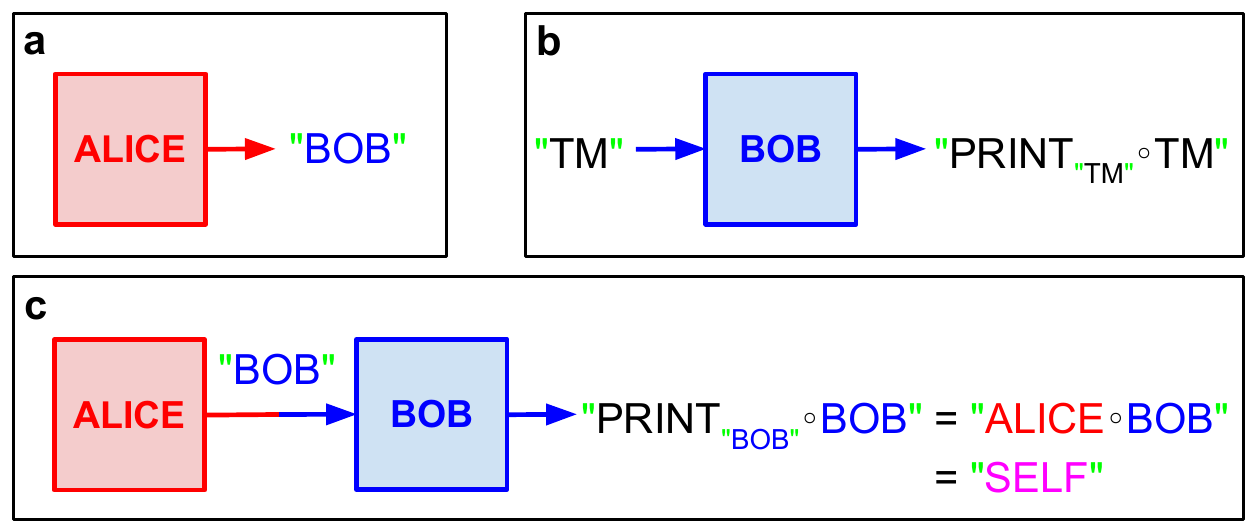}
\caption{ {\it Self-printing Turing machine.} A more formal description of the Turing machine sketched in Fig.~\ref{f:self}. (a) The Turing machine $\textsc{Alice} = \textsc{Print}_{\text{``\textsc{Bob}''}}$ prints a description of the Turing machine {\sc Bob} (see Eq.~\eqref{e:AliceTM}). But how is {\sc Bob} defined? (b) $\textsc{Bob} = \prescript{}{\textsc{``TM''}}{\textsc{Print}}_{``{\textsc{Print}_{\textsc{``TM''}}\circ\textsc{TM}}\textsc{''}}$ takes as input the description ``{\sc TM}'' of a generic Turing machine {\sc TM} and infers, via {\sc Q} (see Fig.~\ref{f:q}), the description of a Turing machine $\textsc{Print}_{\text{``\textsc{TM}''}}$ that prints ``{\sc TM}''. {\sc Bob} then composes the Turing machine $\textsc{Print}_{\text{``\textsc{TM}''}}$ with the Turing machine {\sc TM} itself and outputs the corresponding description, i.e. ``$\textsc{Print}_{\text{``\textsc{TM}''}}\circ \textsc{TM}$'' (see Eq.~\eqref{e:BobTM}); composition is represented here by the symbol $\circ$. (c) The Turing machine $\textsc{Self} = \textsc{Alice}\circ\textsc{Bob}$ that results from the composition of {\sc Alice} and {\sc Bob} outputs a description of itself, as it can be seen by doing {\sc TM} $=$ {\sc Bob} in (b) and using $\textsc{Alice} = \textsc{Print}_{\text{``\textsc{Bob}''}}$ (see Eq.~\eqref{e:SelfTM}). }
\label{f:quine}
\end{figure*}

The Turing machine {\sc Q} is useful to build self-printing Turing machines, as illustrated in Fig.~\ref{f:quine} which is the formal version of Fig.~\ref{f:self}. As we said above, as well as in Sec.~\ref{s:big_1st_person} and Fig.~\ref{f:self}, a self-printing Turing machine 
\be\label{e:SelfTM}
{\textsc{Self} =  \textsc{Alice}\circ\textsc{Bob}},
\ee
is composed of two Turing machines, {\sc Alice} and {\sc Bob}, that essentially print each other. Figure~\ref{f:quine}a shows the Turing machine 
\be\label{e:AliceTM}
\textsc{Alice} = \textsc{Print}_\textsc{``Bob''},
\ee
which ignores its input, prints a description {\sc ``Bob''} of the Turing machine {\sc Bob}, and halts. Now, if $\textsc{Bob} \stackrel{?}{=} \textsc{Print}_\textsc{``Alice''}$ were to similarly ignore its input, just print a description of {\sc Alice}, and halt, we would have a circular definition were the definition of {\sc Alice} depends on who {\sc Bob} is, and viceversa. 

To avoid such circular definition, {\sc Bob} essentially works backwards by inferring the description of {\sc Alice} from the output she produces, which is {\sc ``Bob''} (see Fig.~\ref{f:quine}b). This is precisely what the Turing machine {\sc Q} does: given an input $w=\textsc{``Bob''}$ it prints the description {\sc ``Print$_\textsc{``Bob''}$''} of a Turing machine $\textsc{Print}_\textsc{``Bob''}$ that ignores its input, prints $w=\textsc{``Bob''}$, and halts. 
So, when {\sc Q} takes the input {\sc ``Bob''} it outputs precisely {\sc ``Alice''}, since ${\textsc{Print}_\textsc{``Bob''} = \textsc{Alice}}$. 

But the full self-printing machine actually is ${\textsc{Self} =  \textsc{Alice}\circ\textsc{Bob}}$ (see Fig.~\ref{f:quine}c), so {\sc Bob} is designed such that (see Fig.~\ref{f:quine}b): (i) it takes as input the description {\sc ``TM''} of an arbitrary Turing machine {\sc TM} and infers via {\sc Q} the description of a Turing machine that prints {\sc ``TM''}; (ii) it then generates the composition $\textsc{Print}_{\textsc{``TM''}}\circ\textsc{TM}$ of the Turing machines associated to the description $``{\textsc{Print}_{\textsc{``TM''}}}\textsc{''}$, that it inferred via {\sc Q} from the given input {\sc ``TM''}, and to the descritpion {\sc ``TM''} it receives as input; (iii) it finally prints the description $``{\textsc{Print}_{\textsc{``TM''}}\circ\textsc{TM}}\textsc{''}$ of such composition. This fully specifies {\sc Bob} in a way that is independent of who {\sc Alice} is, i.e.
\be\label{e:BobTM}
\textsc{Bob} = \prescript{}{\textsc{``TM''}}{\textsc{Print}}_{``{\textsc{Print}_{\textsc{``TM''}}\circ\textsc{TM}}\textsc{''}}.
\ee
Notice that the first {\sc Print} operator in the definition of {\sc Bob} in Eq.~\eqref{e:BobTM} has also a left subscript {\sc ``TM''}, which indicates its input; this contrasts with the definition of {\sc Alice} which does not have such a left subscript indicating that it always ignores its input.

So, we can now fully specify {\sc Alice} by replacing {\sc Bob} in Eq.~\eqref{e:AliceTM} by the left hand side of Eq.~\eqref{e:BobTM}. With both {\sc Alice} and {\sc Bob} fully specified, we can fully specify {\sc Self} in Eq.~\eqref{e:SelfTM} too. See Fig.~\ref{f:quine} and Ref.~\cite{sipser2006introduction} (chapter 6) for further details.

\begin{figure*}
\includegraphics[width=0.9\textwidth]{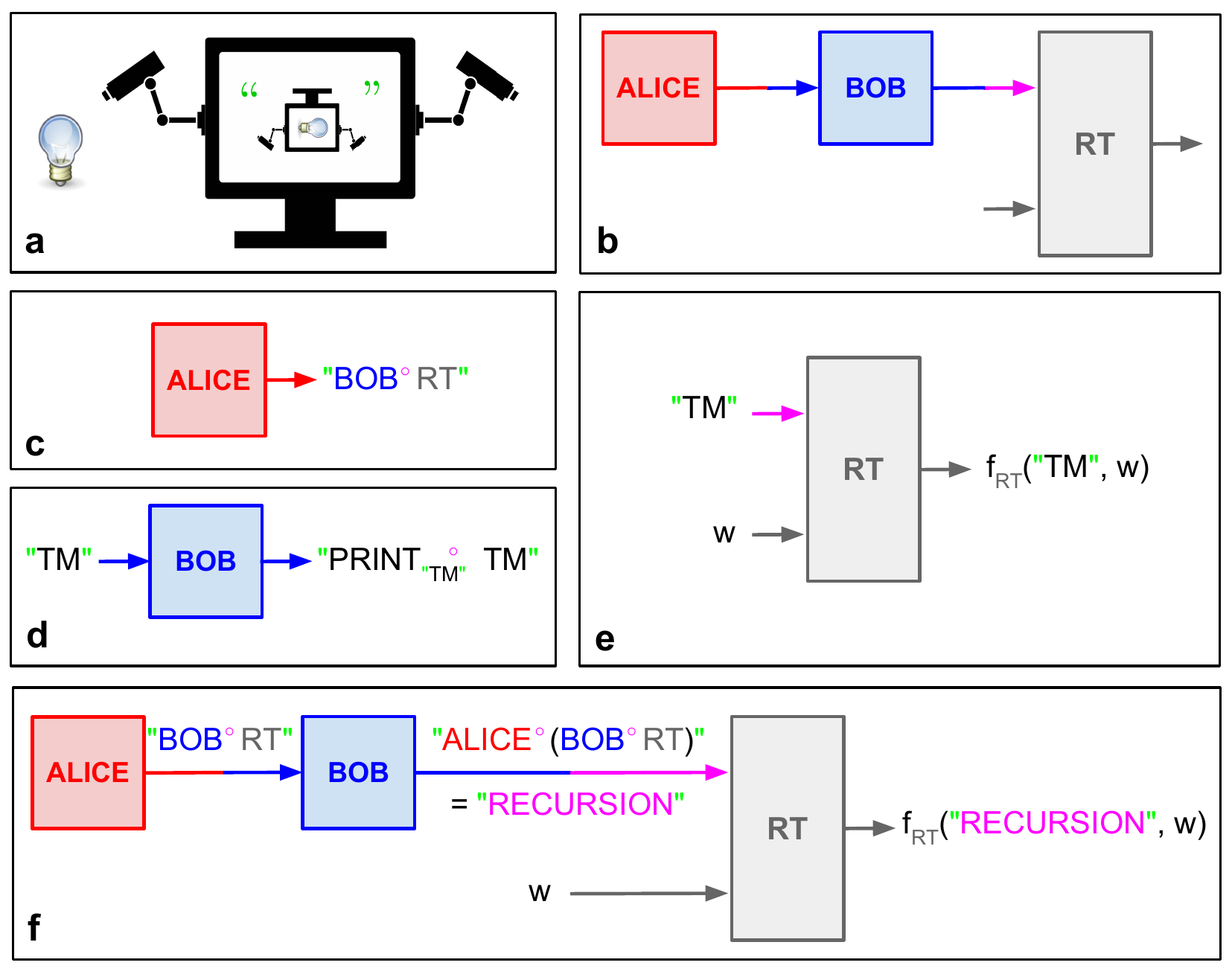}
\caption{ {\it Recursion theorem.} (a) The recursion theorem essentially states that a Turing machine (here a computer) can access a description of itself and manipulate it, along with the input (here a bulb), to produce a certain output (here a description of a rotated copy of itself printing a rotated bulb). (b) The architecture of a generic Turing machine, $\textsc{Recursion}$, that implements this idea is composed of three sub-machines: {\sc Alice} and {\sc Bob}, similar to those in Fig.~\ref{f:quine}, along with a 2-input Turing machine {\sc RT}, provided by the recursion theorem. We can write this Turing machine as ${\textsc{Recursion} = \textsc{Alice}^\circ\textsc{Bob}^\circ\text{RT}}$, where the superindex $^\circ$ refers to composition along the upper input channel of {\sc RT} (see Eq.~\eqref{e:RecursionRT}). But how exactly are {\sc Alice}, {\sc Bob}, and {\sc RT} defined? (c) $\textsc{Alice}= \textsc{Print}_{\text{``\textsc{Bob}$^\circ$\textsc{RT}''}}$ prints a description of the composition of {\sc Bob} and {\sc RT} along the upper input channel of {\sc RT} (see Eq.~\eqref{e:AliceRT}). (d) {\sc Bob} is defined in the same way as in Fig.~\ref{f:quine}b, only that now {\sc TM} is a 2-input Turing machine (see Eq.~\eqref{e:BobTM}). (e) {\sc RT} takes as inputs the description ``{\sc TM}'' of a Turing machine {\sc TM} and a string $w$ and implements a general computable function $f_{\textsc{RT}}(\text{``\textsc{TM}'', w})$, that operates on both the description of {\sc TM} and the input $w$ to produce an output. (f) Proof of the recursion theorem by putting all pieces together and using the definitions of the Turing machines in (c-e). Since ${\textsc{Recursion} = \textsc{Alice}^\circ\textsc{Bob}^\circ\text{RT}}$ and $\textsc{Alice}= \textsc{Print}_{\text{``\textsc{Bob}$^\circ$\textsc{RT}''}}$, we can see that the whole Turing machine {\sc Recursion} implements a function $f_{\textsc{RT}}(\text{``\textsc{Recursion}''}, w)$ that can use its own description, ``{\sc Recursion}'', during the computation.}
\label{f:recursion}
\end{figure*}

\begin{figure*}
\includegraphics[width=\textwidth]{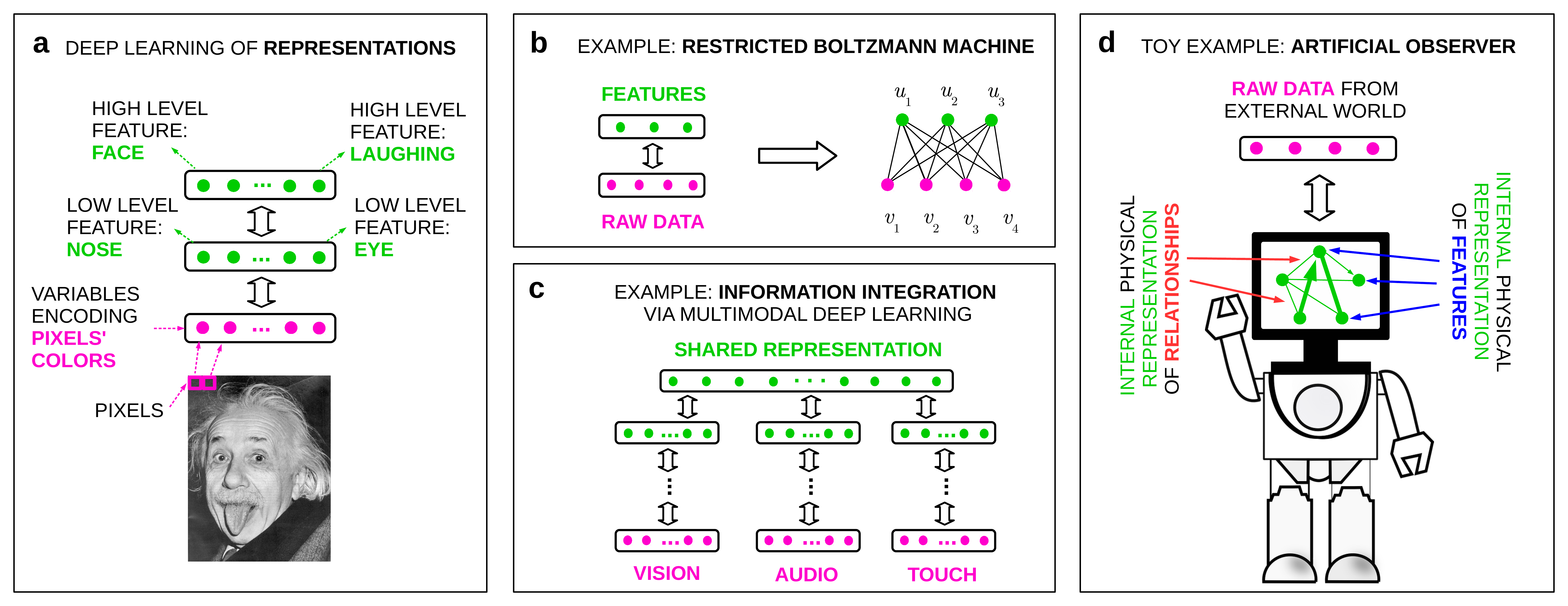}
\caption{ {\it Toy example of a more realistic artificial observer.} (a) Example of a deep architecture that can be used to learn hierarchical representations of data (cf. Fig. 1 of~\cite{bengio2009learning}). An image can be represented as an array of pixels whose colors can be encoded as numbers. The pink variables at the bottom represent the pixels' color. Such raw data are then processed by the first module to extract first-level features, which represent a specific type of pattern, e.g. a nose or an eye. Ideally, such features can be represented by a boolean variable which is equal to one if the feature it represents is present on the input image, and zero otherwise. First-level features can be interpreted as new data that can be processed by a second module to extract second-level features, e.g. a face or the facial expression associated to laughing. (b) A restricted Boltzmann Machine (RBM) is a specific example of a module that can be recursively stacked to build a deep architecture. The variables $v_k$ in the first layer are the {\it visible} or observed variables, which encode the raw data, while the variables $u_j$ in the second layer are the {\it unobserved} or hidden variables, which encode the features. An RBM is a probabilistic graphical model characterized by a Boltzmann distribution $P_B(\bu ,\bv) = e^{-\beta E(\bu, \bv)}/Z$ with an energy function ${E(\bu , \bv) = \sum_{j,k} W_{j k} u_j v_k + \sum_j a_j u_j + \sum_k b_k v_k}$. To train an RBM means to find values for the parameters $W_{jk},\: a_j,\: b_k$ such that: (i) the marginal probability $P(\bv) = \sum_{\bu} P_B(\bu , \bv)$ resembles the emprical distribution of the dataset, and (ii) $P(\bv)$ can generalize to or predict new data that was not presented before to the RBM. The implementation of an RBM on hardware requires physical systems that can represent the corresponding variables, and the neccessary physical interaction between them. Such physical interactions can be direct like that between variables $v_1$ and $u_1$, or indirect like that between variables $v_1$ and $v_2$--- see e.g. Ref.~\cite{benedetti2016estimation} for a discussion on the physical implementation of a type of RBM on a quantum annealer (similar considerations hold for physical implementations on classical hardware). (c) Deep architectures can in principle be combined to create multimodal architectures that can integrate information arriving from different pathways, e.g. the image of a car, the sound of a car, the word `car', the pattern of interactions associated to touching a car~\cite{ngiam2011multimodal,salakhutdinov2009learning}. In principle, the part dedicated to vision can extract relevant features of images (e.g. cars), while the audio component can extract relevant features of related sounds (e.g. sounds of engines). The intuition behind is that it is more effective to encode the association between images and their associated sounds at the level of high level features than at the level of raw data. Ideally, a feature at the top layer can encode the analogous of a `concept', e.g. the boolean variable associated can take value one when any of the information pathways contain a pattern associated to the concept of a car. These high-level features are also called `shared representations'. (d) Toy example of an artificial observer, e.g. a robot, on a toy world. The external world is modelled here as a source of raw data while the observer is modelled and the external observer as a robot with a neural network implemented in hardware whose visible units encode the raw data of the external world. Such a network can be composed of a multimodal deep architecture that hierarchically extracts features from different information pathways, e.g. vision, audio.}
\label{f:realisticobserver}
\end{figure*}

\begin{figure*}
\includegraphics[width=\textwidth]{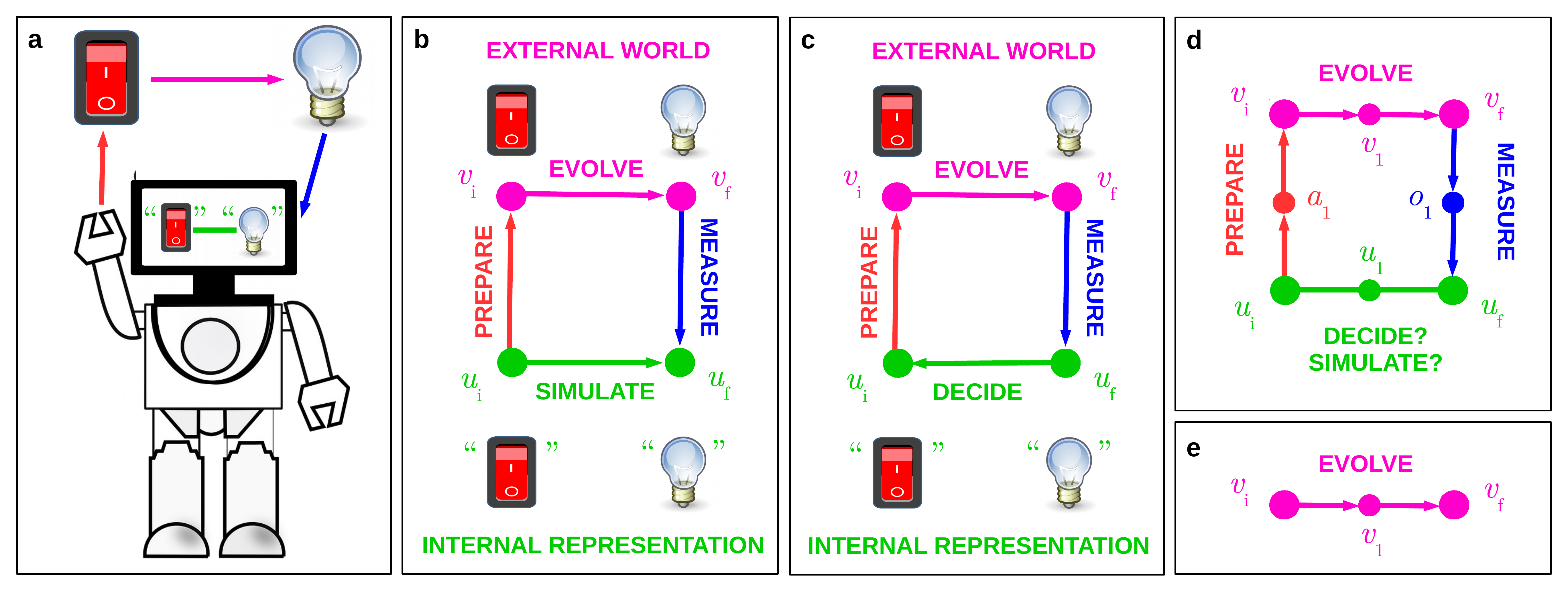}
\caption{ {\it Experiments as circular interactions.} (a) An artificial observer, Alice, performing an experiment to determine whether the state of a switch (i.e. {\sc On} or {\sc Off}) has a causal influence on the state of a lamp (i.e. {\sc Light} or {\sc Dark}). We emphasize that this is a third-person perspective analysis. Alice's `brain' is a computer, a physical realization of a Turing machine. We say Alice has observed the external system when she has build an internal representation of it, denoted by enclosing a replica of the system within quotation marks. Such internal representation requires a physical implementation in Alice's hardware. To perform the experiment, Alice first `decides' which intervetion to do, i.e. where to position the switch, and then `acts' by moving the switch accordingly; such action requires a physical interaction represented by a red arrow (cf. Fig.~\ref{f:third}). After preparing the system via her interventions, Alice leaves the system evolve (pink arrow) and measure the state of the lamp. Such measurement also requires a physical interaction represented by the blue arrow (cf. Fig.~\ref{f:third}). By running the experiment $n$ times Alice can obtain a dataset $\mathcal{D} = \{(u_\textrm{i}^{(1)},u_\textrm{f}^{(1)}),\dotsc ,u_\textrm{i}^{(n)},u_\textrm{f}^{(n)})\}$, where $u_\textrm{i}^{(d)}$ and $u_\textrm{f}^{(d)}$ stand for, respectively, Alice's internal representations of the state of the switch and the lamp at the $d$-th run of the experiment. Alice can use $\mathcal{D}$ to build a causal model represented by the green line joining the two representations; this line actually stands for an arrow whose direction we have not yet defined. (b) If we interpret Alice's causal model as a simulation of the external system, then the arrow corresponding to the green line should point in the same direction of the external (pink) arrow. (c) Alternatively, in the so-called {\it ideomotor view}~\cite{pezzulo2006actions}, Alice's causal model is reversed: Alice's representation of the intended effect (e.g. lamp in state {\sc Light}) of her action (e.g. turn switch {\sc On}) is the cause of the action. In other words, it is not the action that produces the effect, but rather the internal representation of the effect that produces the action~\cite{pezzulo2006actions}. We expect this to be a more faithful representation of the situation in an experiment. However, this leads to a graphical model that is a directed loop representing reciprocal causation, a subject that to the best of our knowledge is not as developed as the most standard models of causality based on directed acyclical graphs, i.e. with no loops (see e.g. Ref.~\cite{spirtes1993causation}, chapter 12.1). (d) The most relevant feature from this analysis is that, once we take into account the observer as part of the experimental set-up, the topology of the interactions taking place in an experiment is {\it circular}. For this reason we will not assign a direction to the green line, and will discuss both cases. (e) In contrast, when the interactions associated to the observer are neglected, the apparent topology of interactions taking place in an experiment is that of a chain. Notice that the initial and final points of a chain (i.e. $v_1$ and $v_3$) interact with only one single point (here $v_2$)  while the rest of the points (e.g. $v_2$) interact with two other points. This allows us to specify a well-defined initial (or final) state and propagate it through the chain via the transition probabilities. This contrasts with the case of a circular topology as in (d) where {\it no single point is special}: there is neither intrinsic beginning nor intrinsic end on a circle. }
\label{f:circular}
\end{figure*}

\begin{figure*}
\includegraphics[width=\textwidth]{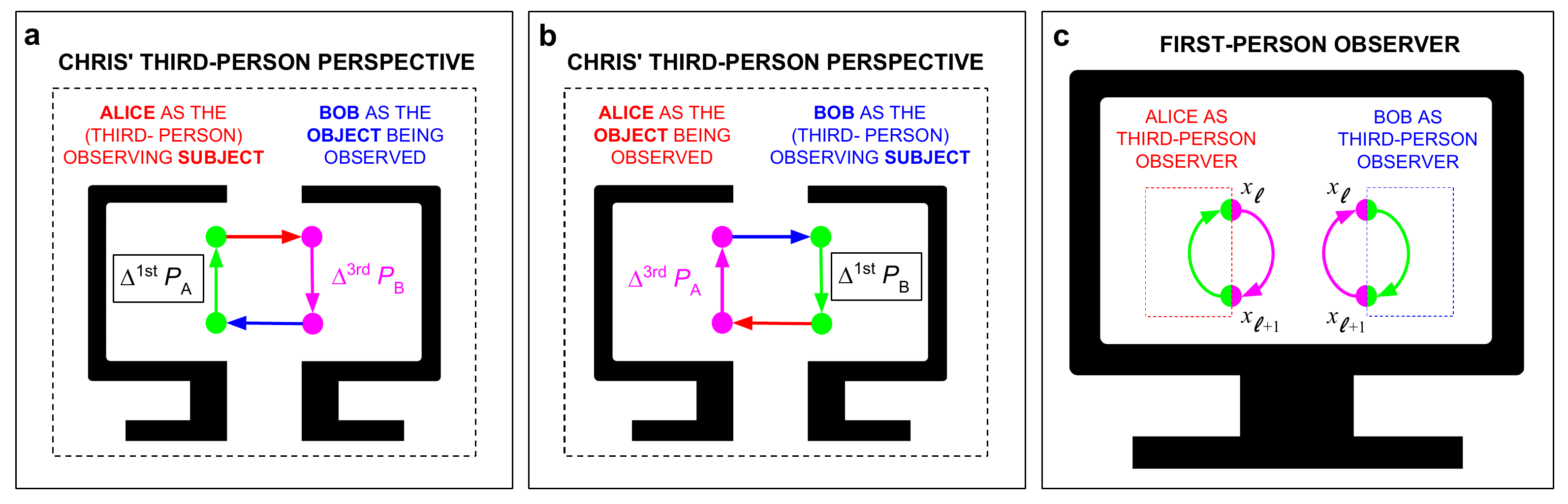}
\caption{{\it Architecture of self-referential observer with a first-person perspective}. (a, b) A third-person observer, Chris, describes the mutual observation between observers Alice and Bob; this is the formal analogous of the situation in Fig.~\ref{f:first-third}a. The same circular physical process observed by Chris can be interpreted as Alice observing Bob (a) or viceversa (b). (a) When Alice is interpreted as the observing subject and Bob as the object being observed, the top (red) and bottom (blue) horizontal arrows are interpreted as preparation and measurement, respectively (cf. Fig.~\ref{f:circular}). Similarly the (green) left and (purple) right vertical arrows are interpreted, respectively, as Alice's internal decision process and the evolution of the external physical system (i.e. Bob). As we discuss in Sec.~\ref{s:big_1st_person} although a camera can take a picture of any external object, it cannot take a picture of itself because a system cannot simultaneously behave as both subject and object (i.e. they are complementary roles), unless there is a complementary system like a mirror or another camera. Similarly, Alice can observe the changes in Bob's state from a third-person perspective $\Delta^{3\rm rd} P_B$, but she cannot observe her own changes $\Delta^{1\rm st} P_A$ of state from a first-person perspective. This is represented by writing $\Delta^{1\rm st } P_A$ inside a black box. However, if there are {\em no} external or irreversible contributions the changes are equal in magnitude and opposite in sign, i.e. ${\Delta^{1 \rm st} P_A = -\Delta^{3\rm rd} P_B}$ (see Fig.~\ref{f:noneq_self-observer} for the extension to the general non-equilibrium case). This is the formal analogous of Alice's camera in Fig.~\ref{f:first-third}b, with $\Delta^{1\rm st} P_A$ playing the role of the memory of Alice's camera and $\Delta^{3\rm rd} P_B$ playing the role of Bob as the object the camera is taking the picture of. (b) When Bob is interpreted as the observing subject and Alice as the object being observed, all roles are reversed. In particular, top (blue) and bottom (red) horizontal arrows are now interpreted as measurement and preparation, respectively (cf. Fig.~\ref{f:circular}), as well as the (purple) left and (green) right vertical arrows are interpreted, respectively, as Bob's internal decision process and the evolution of the external physical system (i.e. Alice). However, while the external process observed by Bob is in the reverse direction of the external process observed by Alice. This is more clearly seen in (c) where we traced out the measurement and preparation parts. We can clearly see the external process observed by Alice (left) goes from $x_\ell\to x_{\ell +1 }$, while that observed by Bob goes from $x_{\ell + 1}\to x_\ell$ instead. This is similar to the inversion of left and right when we look at ourselves in a mirror. So, taking into account this inversion we have $\Delta^{1\rm st} P_B = \Delta^{3\rm rd} P_A$. (c) The self-referential observer that has a first person-perspective is composed of two sub-observers Alice and Bob that mutually observe each other. We emphasize once again that the physical process is exactly one and the same but has two different interpretations depending on the role we asign to Alice and Bob as either subjects or objects; we draw the same circular process twice in (c) to emphasize these dual roles. This is the formal analogous of the situation illustrated in Fig.~\ref{f:first-third}b.}
\label{f:self-observer}
\end{figure*}

\begin{figure}
\includegraphics[width=\columnwidth]{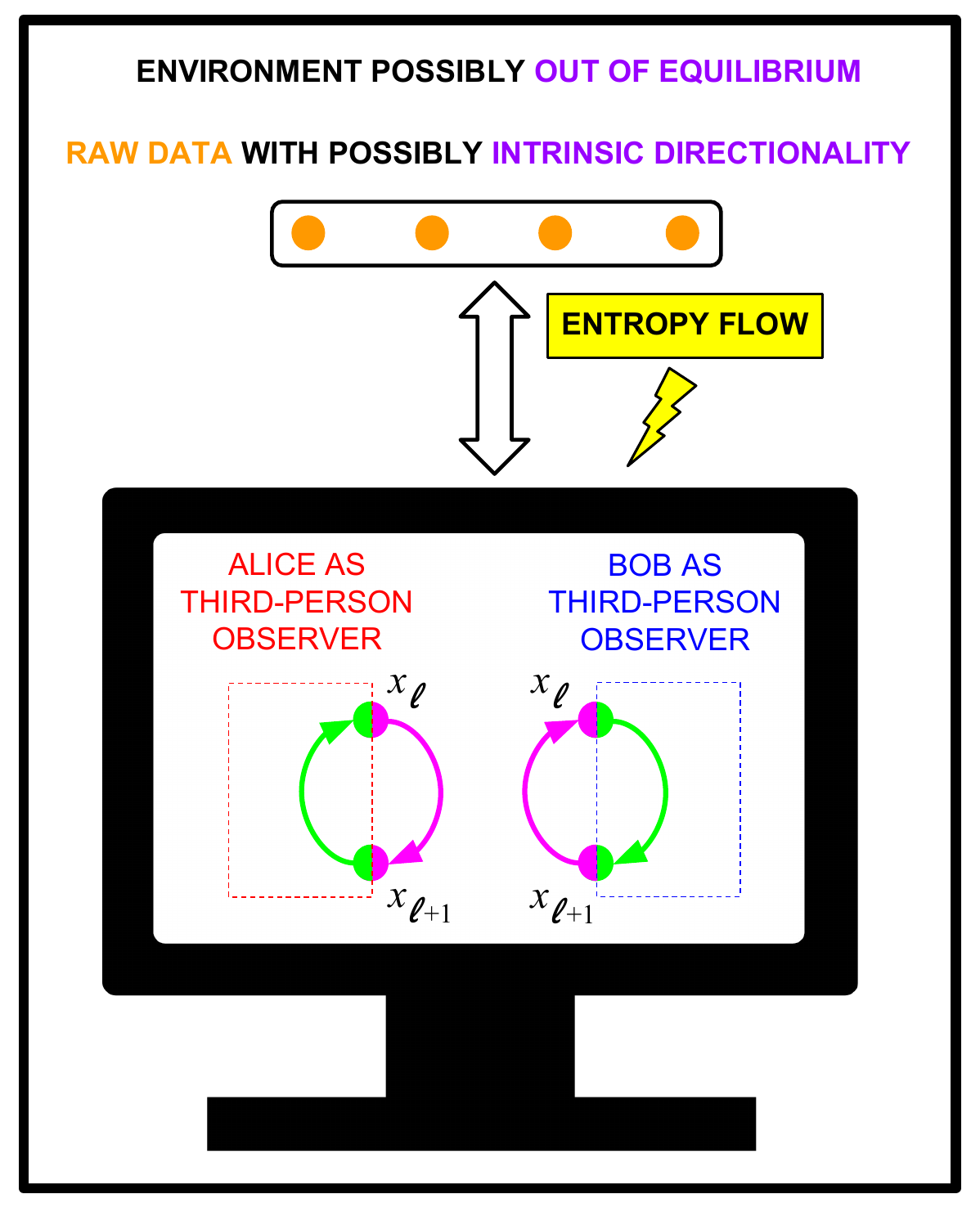}
\caption{ {\it Self-referential observer embedded in an environment}. As illustrated in Fig.~\ref{f:realisticobserver}, the variables (circles) and processes (arrows) `percieved' by an observer are here interpreted as high-level features extracted from a raw data of an external world or environment. So, here we make more explicit the environment which the self-referential observer in Fig.~\ref{f:self-observer}c is interacting with. Such environment can be interpreted as a physical system or as raw data. Now, if such environment is in equilibrium, or equivalently there is no intrinsic directionality in the raw data, the high-lelvel representation as a circular process in the self-referential observer is reversible and changes in the state of one of the sub-observers (e.g. Alice) are essentially equivalent to changes in the state of the other sub-observer (e.g. Bob), as discussed in Fig.~\ref{f:self-observer}. However, when there are external influences, this is not true any more since there is an intrinsic directionality affecting equally to Alice and Bob. In this case the changes in Alice state are not equal only to changes in Bob's state because there is also an external contribution. This is the analogous of the recursion theorem with external data. To deal with this situation we notice that the irreversible contributions with intrinsic directionality are represented by the anti-symmetric parts of $P$ and $J$. So, we can still have equality between the symmetric parts, i.e. ${\Delta^{1\rm st} P_{A, s} = -\Delta^{3\rm rd} P_{B, s}}$ and ${\Delta^{1\rm st} P_{B, s} = \Delta^{3\rm rd} P_{A, s}}$. }
\label{f:noneq_self-observer}
\end{figure}

\begin{figure}
\includegraphics[width=\columnwidth]{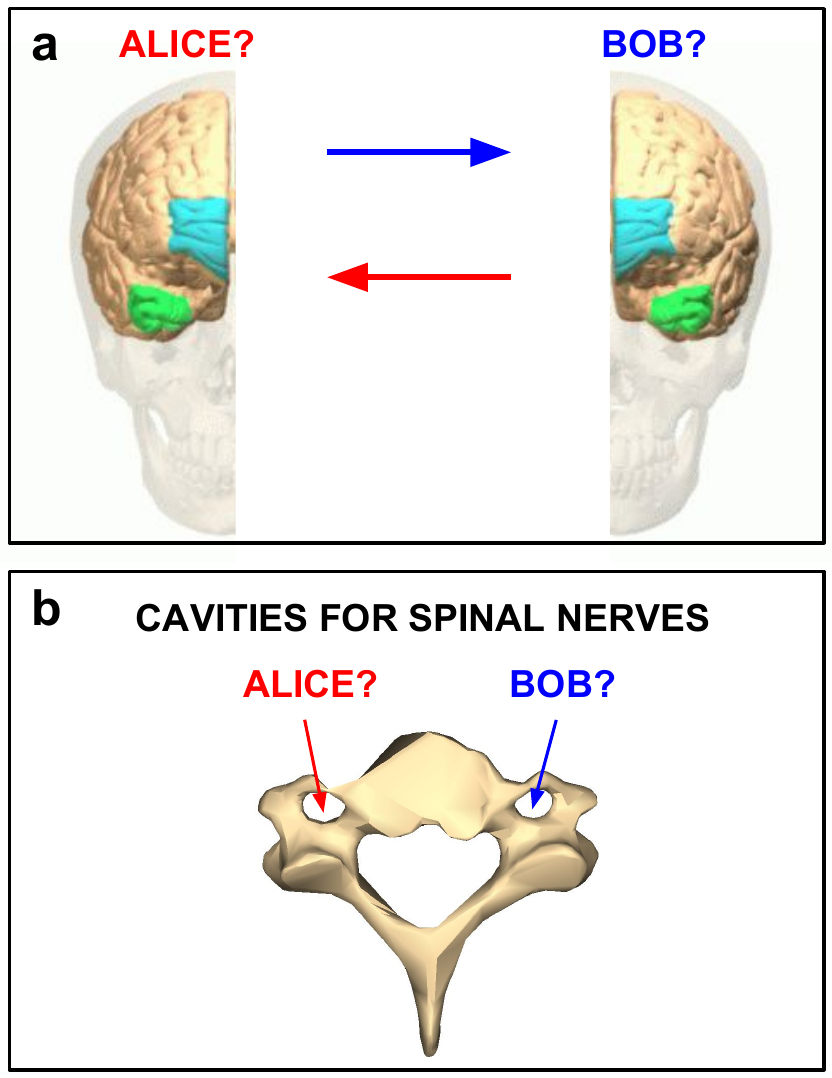}
\caption{ {\it Is global neural architecture the result of self-reference?.} In this work we conjecture (see Sec.~\ref{s:architecture}) that the two-hemisphere architecture of the brain in {\it normal healthy} humans is a {\it large-scale feature} that {\it results} from the implementation of self-reference by the brain (a), according to the general ideas of the recursion theorem as sketched in Fig.~\ref{f:self} and discussed in more detail in Appendix~\ref{s:recursion}. This does {\it not} imply that the two hemispheres are a {\it necessary} feature for the brain to implement self-reference. Indeed, there is evidence that removing an hemisphere does not necessarily affect a person~\cite{choi2007strange}. A possibility is that the brain implements self-reference at different scales. Although the decades-old theory stating that split-brain patients, i.e. those whose {\it corpus callosum} connecting the two brain hemispheres has been severed, can have divided identities has been recently challenged~\cite{pinto2017split,corballis2018perceptual}, i.e. split-brain patients actually appear to experience divided perception but undivided consciousness, Corballis {\em et al.}~\cite{corballis2018perceptual} argue that subcortical connections may play a role in integrating information from the two hemispheres. More generally, we may conjecture that self-reference may constraint the global architecture of the entire central nervous system, e.g. the left and right neural networks running through the spinal cord (b). }
\label{f:brain_self}
\end{figure}

\begin{figure}
\includegraphics[width=0.9\columnwidth]{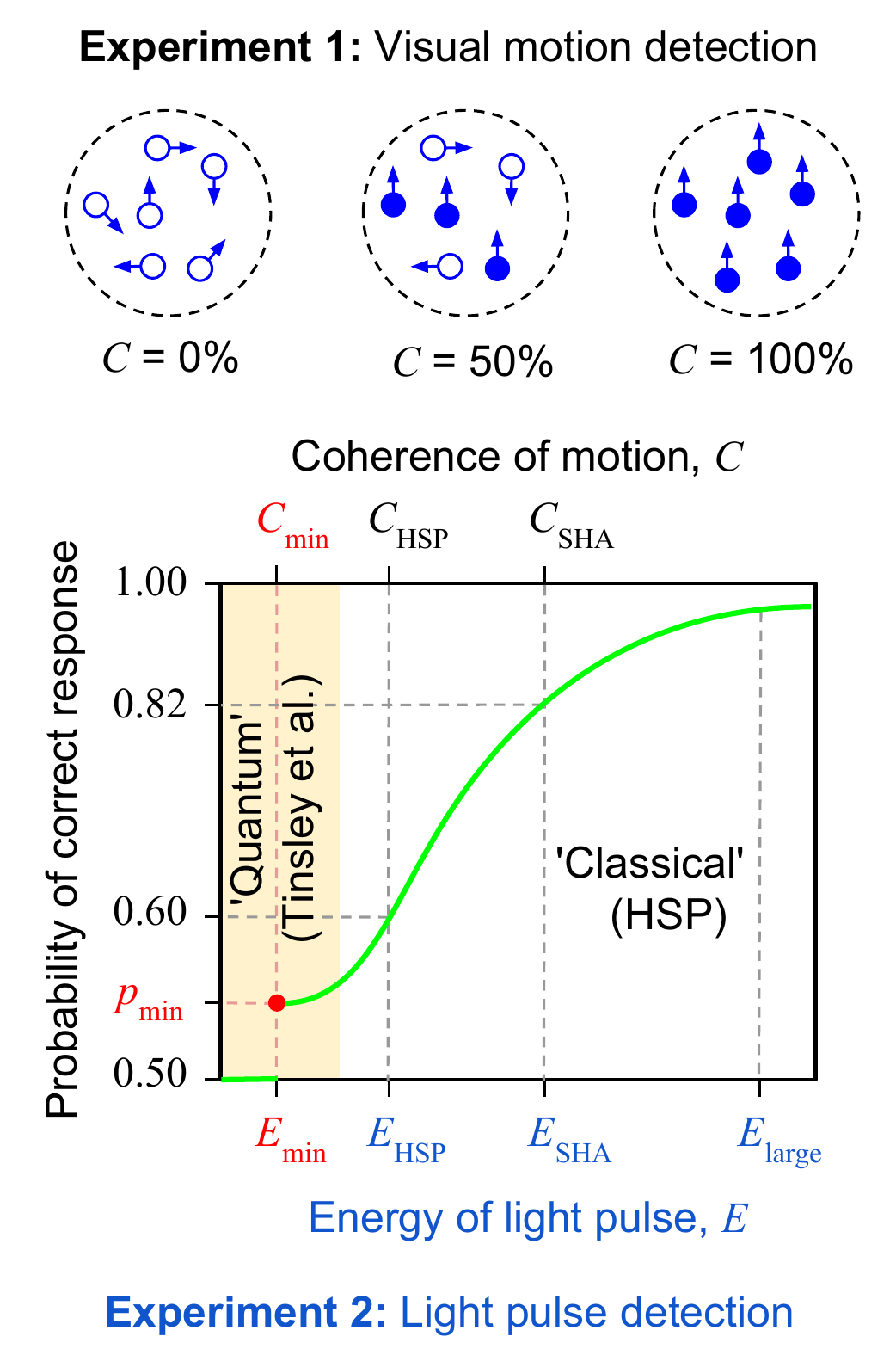}
\caption{{\it Psychophysics experiments and quantum of action}. The three dashed circles at the top illustrate the experiment carried out by Sch\"olvinck, Howarth, and Attwell (SHA)~\cite{scholvinck2008cortical} called here `Experiment 1'. The small circles with arrows illustrate the moving dots shown to the monkeys, with the arrow specifying the direction of motion. Empty circles move randomly, while filled circles all move upwards. The leftmost dashed circle illustrates the case of zero coherence ($C=0$), where all dots move randomly; the dashed circle at the center illustrates the case of 50\% coherence ($C=50\%$), where half of the dots move randomly and the other half move upwards; the rightmost dashed circle illustrates the case of full coherence ($C=100\%$), where all dots move upwards. The curve at the bottom sketches a generic psychophysics curve which is qualitatively similar to those obtained in the three experiments we analyze here (see Fig.~7 {in} Ref.~\cite{hecht1942energy} for the HSP exeriment, Supplementary Fig. 3 {in}~\cite{tinsley2016direct} for Tinsley et al. experiment, and Fig. 2d,e,f in Ref.~\cite{scholvinck2008cortical} for SHA experiment). The top horizontal axis contains the values of coherence $C$ in SHA experiment. The bottom horizontal axis contains values of energy that could be associated to the coherences in SHA experiment, or to energies directly measured as in the HSP experiment~\cite{hecht1942energy} called here `Experiment 2' or in Tinsley et al. experiment~\cite{tinsley2016direct}.  The vertical axis contains the probability that the subjects correctly respond to have seen the stimulus, i.e. the direction of motion in SHA experiment, or energy pulses in HSP and Tinsley et al. experiments. See Sec.~\ref{s:Planck} for further details.} 
\label{f:experiment}
\end{figure}

\begin{figure*}
\includegraphics[width=0.8\textwidth]{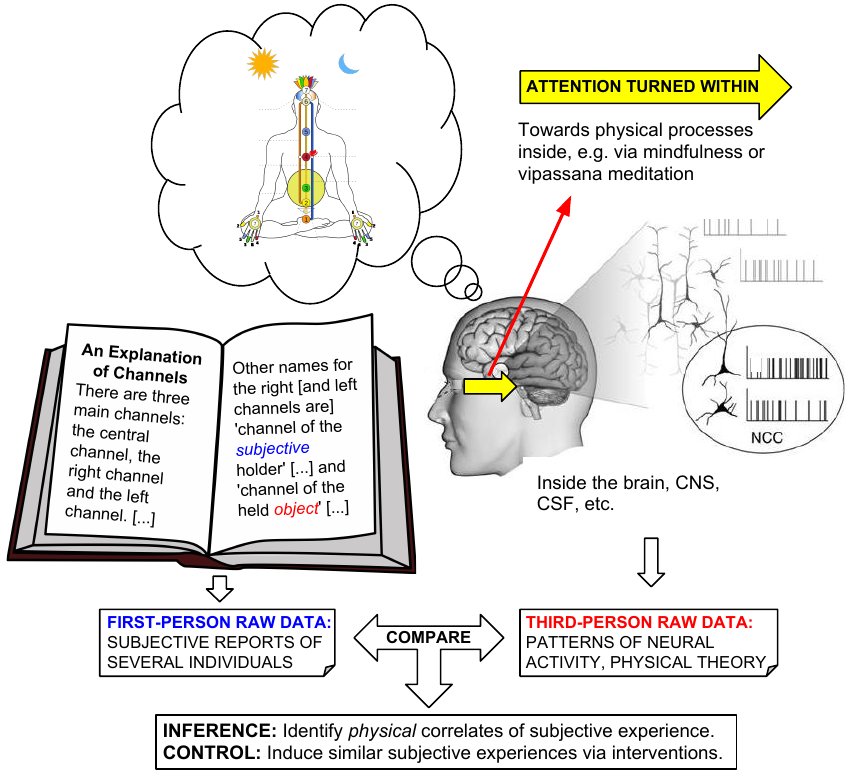}
\caption{ {\it Modern scientific approach to study contemplative traditions}. While in the XIX century almost nobody believed on the concept of atom, today physicists feel quite comfortable talking about such strange concepts as Higgs bosons, antimatter, dark matter, etc. Conterintuitive concepts we never experience in everyday life. What would be the exclamation of a XIX century scientist after hearing any of these concepts? Further theoretical and experimental developments have been instumental for this change in attitude. Something similar might be happening today with the almost automatic rejection some scientists feel against anything that could be labelled `spiritual'. Indeed, theoretical and experimental tools are already emerging to do such a rejection on a rigorous scientific basis, or to realize we may have been misled by partial or confusing information~\cite{tang2015neuroscience,hanson2009buddha}; some of those tools are considered rigorous enough to merit a review in a top journal such as {\it Nature Reviews Neuroscience}~\cite{tang2015neuroscience}. Since contemplative traditions have been developed many centuries ago, they tend to use a difficult symbolic language that is not necessarily to be taken literally, but rather as a pointer to certain subjective experiences that could in principle be explored with the strategy described in Fig.~\ref{f:subjective}. We could compare physical processes taking place in practitioners' and non-practitioners'  brains, central nervous system (CNS), cerebrospinal fluid (CSF), etc, and contrast them to their subjective reports (see e.g. Ref.~\cite{tang2015neuroscience}). Alternatively, we could follow an anthropology-like strategy and use as subjective reports books written by representative practitioners along the centuries. Since such books have been written before our modern theories there is no risk of self-suggested reports trying to conform to the latter. Now, {\bf Principle I} tells us the observer is physical, so we can also observe physical phenomena by turning our attention within. This is not as strange as it sounds; a simple example is that we could have a rough estimate of our heart rate by paying attention to our own body. Much as science has developed sophisticated tools to see beyond the capabilities of our own senses, some contemplative practices have developed techniques, such as mindfulness and vipassana meditation, to sharpen our minds and objectively observe internal physical processes we are not aware of in our everyday lives. We can automatically dismiss this as total nonsense based on long-held beliefs. Alternatively, we can try a more scientific strategy and run an experiment with ourselves, with our own physical system, by attending a 10-day vipassana retreat, for instance. According to the subjective reports of many experienced meditators, deeply unconscious physical processes around the spinal cord and brain hemispheres seem to be related to our self-concepts. Since such reports appear to have some coincidences with the architecture of self-referential observers explored here (see Sec.~\ref{s:toolbox}), the latter might be considered a potential {\em physical correlate} of the former. So, first-person methods might help guide explorations on the foundations of science. Although we can understandably worry about the faithfulness of subjective reports, the combination of these with third-person methods can prove the consistency of the former by identifying a set of neural correlates common to most reporting subjects. This strategy has proven useful in the study of mindfulness meditation~\cite{tang2015neuroscience}, for instance, a practice previously considered `spiritual'.}
\label{f:channels}
\end{figure*}

Now, a Turing machine not only can print its own description, it can also use it as an input and perform general computational operations with it. Furthermore, a Turing machine (illustrated in Fig.~\ref{f:recursion}a by a big computer) can take a combined input composed of external data (e.g. the imagine of a bulb in Fig.~\ref{f:recursion}a) and its own description (illustrated in Fig.~\ref{f:recursion}a by a small computer printed within quation marks on the screen of the big computer) and perform general computational operations with it. Figure~\ref{f:recursion}a illustrates this by a computer that takes as external input the image of a bulb and print a description of a rotated version of itself printing a rotated version of the bulb. 

The architecture of such general Turing machine, that we call here {\sc Recursion}, is composed of three Turing machines (see Fig.~\ref{f:recursion}b): {\sc Alice} and {\sc Bob}, which together generate the description of {\sc Recursion}, and another 2-argument Turing machine {\sc RT} that takes as inputs both an arbitrary word $w\in\Sigma^\ast$ provided from the outside and the description of {\sc Recursion} generated from the inside of {\sc Recursion} itself by the composition of {\sc Alice} and {\sc Bob}, i.e.
\be\label{e:RecursionRT}
\textsc{Recursion} = \textsc{Alice}\circ\textsc{Bob}^\circ\textsc{RT};
\ee
here the superscript $^\circ$ indicates the ouput of {\sc Bob} is passed to the upper input channel of {\sc RT}.  

The definition of $\textsc{Alice}$ is slightly modified to take into account the new Turing machine {\sc RT} (see Fig.~\ref{f:recursion}c), i.e.
\be\label{e:AliceRT}
\textsc{Alice} = \textsc{Print}_{\textsc{``Bob$^\circ$RT''}}.
\ee
The definition of {\sc Bob} instead remains the same as in Eq.~\eqref{e:BobTM} since it was defined in terms of a generic Turing machine {\sc TM} (see Fig.~\ref{f:recursion}d). The 2-argument Turing machine {\sc RT} implements the actual computations according to a given 2-argument function $f_{\textsc{RT}}$; it is simlarly defined in terms of a generic Turing machine {\sc TM} whose description enters through the upper input channel, leaving its lower input channel free to receive external data $w\in\Sigma^\ast$ (see Fig.~\ref{f:recursion}e). So, we have (see Fig.~\ref{f:recursion}f):
where the lower input channel of {\sc RT} remains available to receive external data. This proves the

\

\noindent{\bf Recursion theorem (Theorem $6.3$ in Ref.~\cite{sipser2006introduction}):} Let {\sc RT} be a Turing machine that computes a 2-argument function ${f_{\textsc{RT}} : \Sigma^\ast\times\Sigma^\ast\to\Sigma^\ast}$. Then there is another Turing machine {\sc Recursion} that computes a function ${f_{\textsc{Recursion}} : \Sigma^\ast\to\Sigma^\ast}$, where for every ${w\in\Sigma^\ast}$,
$$f_{\textsc{Recursion}}(w) = f_{\textsc{RT}}(\text{``\textsc{Recursion}''},w). $$

\

See Fig.~\ref{f:recursion} and Ref.~\cite{sipser2006introduction} (chapter 6) for further details.

\section{Derivation of real kernel representaions in Sec.~\ref{s:examples}}\label{s:details}
\subsection{Non-relativistic Schr\"odinger equation}\label{s:App_Schrodinger}
The Schr\"odinger equation of a single particle of mass $m$ in a one-dimensional non-relativistic potential $V(x)$ is given by
\be\label{e:Schrodinger}
i\hbar \frac{\partial\psi(x, t)}{\partial t} = \left[ -\frac{\hbar^2}{2 m}\frac{\partial^2}{\partial x^2} + V(x)\right]\psi(x, t),
\ee
where $\psi(x ,t)$ is the wave function. If we discretize $x \approx k \delta$,  with ${k = \dotsc , -2, -1, 0, 1, 2, \dotsc}$, in steps of size $\delta$, we can also discretize the second-order differential opperator in a symmetric way via the second-order central difference
\be\label{e:partial2}
\frac{\partial^2\psi(x,t)}{\partial x^2}  = \frac{\psi(x + \delta, t)- 2\psi(x, t) + \psi(x-\delta, t)}{\delta^2}.
\ee
If we write the discretized wave function as a vector
\be
\boldsymbol{\psi}(t) = \begin{pmatrix}\vdots \\ \psi(-\delta, t) \\ \psi(0, t) \\ \psi(\delta, t) \\ \vdots \end{pmatrix},
\ee
which we represent here with boldface notataion, then the Laplace operator $\partial^2 / \partial x^2$ in Eq.~\eqref{e:partial2} can be represented by the symmetric matrix
\be
\boldsymbol{\Delta} = \frac{1}{\delta^2} \begin{pmatrix} \ddots & \ddots & \ddots & \ddots & \ddots & & & &  \\ \cdots & 0 & 1 & -2 & 1 & 0 & \cdots &  & \\ &\cdots & 0 & 1 & -2 & 1 & 0 & \cdots & \\ & &\cdots & 0 & 1 & -2 & 1 & 0 & \cdots  \\  & & & & \ddots & \ddots & \ddots & \ddots & \ddots\end{pmatrix},
\ee
and the Hamiltonian $H=-({\hbar^2}/{2 m}){\partial^2}/{\partial x^2} + V(x)$ in square brackets in Eq.~\eqref{e:Schrodinger} can be represented by the matrix
\begin{widetext}
\be
H = \begin{pmatrix} \ddots & \ddots & \ddots & \ddots & \ddots & & & &  \\ \cdots & 0 & -\frac{\hbar^2}{2 m\delta^2}  & {\frac{\hbar^2}{m\delta^2} + V_{-1}} & -\frac{\hbar^2}{2 m\delta^2}  & 0 & \cdots &  & \\ &\cdots & 0 & -\frac{\hbar^2}{2 m\delta^2} & {\frac{\hbar^2}{m\delta^2} + V_0} & -\frac{\hbar^2}{2 m\delta^2} & 0 & \cdots & \\ & &\cdots & 0 & -\frac{\hbar^2}{2 m\delta^2} & {\frac{\hbar^2}{m\delta^2} + V_1} & -\frac{\hbar^2}{2 m\delta^2} & 0 & \cdots  \\  & & & & \ddots & \ddots & \ddots & \ddots & \ddots\end{pmatrix},
\ee
\end{widetext}
which is real and symmetric, so the dynamical matrix is given by $J = H/\hbar$; here $V_k = V(k\delta)$.

\subsection{From non-relativistic path integrals to real convolutions}\label{s:App_path}

As originally described by Feynman, the non-relativistic Schr\"odinger equation, Eq.~\eqref{e:Schrodinger},
can be derived from the path integral via (see e.g. Eq. (18) in Ref~\cite{feynman1948space})
\begin{equation}\label{e:path}
\psi(x_{\ell+1}, t+\epsilon) = \frac{1}{\mathcal{A}}\int\exp{\left[\frac{i}{\hbar}S(x_{\ell+1},x_\ell)\right]}\psi(x_\ell,t)\mathrm{d} x_\ell,
\end{equation}
where for the one-dimensional case in Eq.~\eqref{e:Schrodinger} we can set the short-time action as
\be\label{e:action}
S(x_{\ell+1}, x_\ell) = \frac{m\epsilon}{2}\left(\frac{x_{\ell + 1}- x_\ell}{\epsilon} \right)^2 - \epsilon V(x_{\ell+ 1 }),
\ee
and 
\be
\mathcal{A} = \sqrt{i 2\pi\hbar\epsilon /m };
\ee
furthermore, $x_\ell$ represents the position of the particle at time $t=\ell\epsilon$. Notice that by iterating Eq.~\eqref{e:path} we can obtain the path integral representation. 

By multiplying Eq.~\eqref{e:path} by its congugate we can derive the corresponding equation for the density matrix $\rho(x, x^\prime , t) = \psi(x, t)\psi^\ast(x^\prime, t)$, i.e.
\begin{widetext}
\be\label{e:path_rho}
\rho(x_{\ell+1}, x_{\ell+1}^\prime, t+\epsilon) = \frac{1}{|\mathcal{A}|^2}\int\exp{\left\{\frac{i}{\hbar}\left[S(x_{\ell+1},x_\ell) - S(x_{\ell+1}^\prime,x_\ell^\prime)\right]\right\}}\rho(x_\ell,x_\ell^\prime ,t)\mathrm{d} x_\ell\mathrm{d} x_\ell^\prime,
\ee
\end{widetext}
where $|\mathcal{A}|^2 = 2\pi\hbar\epsilon /m $.

While we could tranform Eq.~\eqref{e:path_rho} into a pair of real equations as we did in Sec.~\ref{s:vNreal}, this would lead to oscillatory terms arising from the complex exponential factor in the integrand. Such oscillatory terms would be difficult to interpret in probabilistic terms later on. However, it is possible to rewrite these equations in terms of {\it real} Gaussian convolutions, which can be naturally interpreted in probabilistic terms. Before we show how to do this, we will outline the main steps in the derivation of Eq.~\eqref{e:Schrodinger} from Eq.~\eqref{e:path}, described in detailed in e.g. Ref.~\cite{feynman1948space}, which closely parallel the derivation in terms of real Gaussian convolutions to be described afterwards.  

First, expanding $\psi(x_{\ell+1}, t+\epsilon)$ to first order in $\epsilon$, we can write Eq.~\eqref{e:path} as 
\be\label{e:path_step1}
\begin{split}
\epsilon\frac{\partial\psi(x_{\ell+1}, t)}{\partial t} =& \frac{1}{\mathcal{A}}\int\exp{\left[\frac{i}{\hbar}S(x_{\ell+1},x_\ell)\right]}\psi(x_\ell,t)\mathrm{d} x_\ell  \\
& - \psi(x_{\ell + 1}, t).
\end{split}
\ee
Since $\epsilon\to 0$, the complex Gaussian factor associated to the kinetic term in Eq.~\eqref{e:action} {\it oscillates very fast} except in the region where ${x_{\ell +1} - x_\ell = O(\sqrt{\hbar\epsilon/m})} $. So, to estimate the integral to first order in $\epsilon$, the term $\psi(x_\ell, t)$ in the right hand side of Eq.~\eqref{e:path_step1} need be expanded around $x_{\ell + 1}$ to second order in ${x_{\ell + 1} - x_\ell}$. Consistent with this approximation to first order in $\epsilon$, we also do $\exp{[-i V(x)\epsilon / \hbar]} = 1 -i V(x)\epsilon / \hbar + O(\epsilon^2) $. This leads to Eq.~\eqref{e:Schrodinger}. 

However, as we know from the derivation of the diffusion equation approximation for a random walk~\cite{van1992stochastic,risken1989fpe}, a real Gaussian has an equivalent cancelling effect, not because of fast oscillations but because of exponentially small terms. More precisely, we will argue that if we introduce the Hamiltonian function
\be\label{e:classicalH}
\mathcal{H}(x, x^\prime) = \frac{m}{2}\frac{(x - x^\prime)^2}{\epsilon^2} + V(x),
\ee
we can do the replacement, which amounts to a Wick rotation $\epsilon\to -i\epsilon$,
\be\label{e:replacement}
\frac{1}{\mathcal{A}}\exp{\left[\frac{i}{\hbar}S(x,x^\prime)\right]}\to\frac{1}{|\mathcal{A}|} \exp{\left[-\frac{\mathcal{H}(x, x^\prime)\epsilon}{\hbar} \right]};
\ee
notice that the kinetic term in the right hand side of Eq.~\eqref{e:replacement} leads to a real Gaussian with variance $\hbar\epsilon / m$ and normalization constant given precisely by $|\mathcal{A}|$.

Due to this Gaussian term, the integral

\begin{widetext}
\be\label{e:realGaussian}
\begin{split}
\frac{1}{|\mathcal{A}|}\int \exp{\left[-\frac{\mathcal{H}(x, x^\prime)\epsilon}{\hbar} \right]}\psi(x^\prime, t)\mathrm{d} x^\prime  &= \left[1-\frac{\epsilon}{\hbar} V(x)\right]\left[\psi(x, t) + \frac{\hbar \epsilon }{2 m}\frac{\partial^2 \psi(x,t)}{\partial x^2} \right]+ O (\epsilon^2) \\
& = \psi(x, t) -\frac{\epsilon}{\hbar} V(x)\psi(x,t) + \frac{\hbar \epsilon }{2 m}\frac{\partial^2 \psi(x,t)}{\partial x^2} + O (\epsilon^2), 
\end{split}
\ee
\end{widetext}
can be approximated to first order in $\epsilon$ in a way similar to that of the integral in Eq.~\eqref{e:path_step1}. Indeed, since $\epsilon\to 0$, the real Gaussian factor associated to the kinetic term in Eq.~\eqref{e:classicalH} {\it is exponentially small} except in the region where ${x - x^\prime = O(\sqrt{\hbar\epsilon/m})} $. This has allowed us to estimate the integral to first order in $\epsilon$ by expanding the term $\psi(x^\prime, t)$ in the left hand side of Eq.~\eqref{e:realGaussian} around $x$ up to second order in ${x - x^\prime}$. Consistent with this approximation to first order in $\epsilon$, we have also done $\exp{[- V(x)\epsilon / \hbar]} = 1 - V(x)\epsilon / \hbar + O(\epsilon^2) $. 

This implies that Eq.~\eqref{e:Schrodinger} can be written in terms of real Gaussian convolutions in a way similar to Eq.~\eqref{e:path_step1} as
\begin{widetext}
\be\label{e:GaussianSchrodinger}
\epsilon\frac{\partial\psi(x_{\ell+1}, t)}{\partial t} = i \left\{\frac{1}{|\mathcal{A}|}\int \exp{\left[-\frac{\mathcal{H}(x_{\ell + 1}, x_\ell)\epsilon}{\hbar} \right]}\psi(x_\ell, t)\mathrm{d} x_\ell  - \psi(x_{\ell + 1}, t) \right\}.
\ee
\end{widetext}

Indeed, by replacing the term with the integral in the right hand side of Eq.~\eqref{e:GaussianSchrodinger} by the right hand side of Eq.~\eqref{e:realGaussian}, and multiplying both sides of Eq.~\eqref{e:GaussianSchrodinger} by $i\hbar/\epsilon$, we obtain Schrodinger equation (Eq.~\eqref{e:Schrodinger}). In this way we have essentially bring the imaginary unit $i$ from the exponential in Eq.~\eqref{e:path_step1} down to turn it into a linear factor in Eq.~\eqref{e:GaussianSchrodinger}. In contrast to Eq.~\eqref{e:path}, however, it does not seem possible to obtain a (real) path integral by iterating Eq.~\eqref{e:GaussianSchrodinger}. 

Finally, we now show that Eq.~\eqref{e:GaussianSchrodinger} leads to an equation analogous to the von Neumman equation (Eq.~\eqref{e:vN}) for the density matrix ${\rho(x, x^\prime , t) = \psi(x, t)\psi^\ast(x^\prime, t)}$, in terms of {\em real} Gaussian convolutions instead of differential operators. Indeed, taking the time derivative of this density matrix yields
\be\label{e:Psi_time} 
\frac{\partial \rho(x, x^\prime , t)}{\partial t} = \frac{\partial\psi(x, t)}{\partial t}\psi^\ast(x^\prime, t) + \psi(x, t)\frac{\partial\psi^\ast(x^\prime, t)}{\partial t};
\ee 
now, replacing the time derivatives of the wave function $\psi$ and its conjugate $\psi^\ast$ in Eq.~\eqref{e:Psi_time}, respectively, by the right hand side of Eq.~\eqref{e:GaussianSchrodinger} and its conjugate we obtain
\begin{widetext}
\be\label{e:Gaussian_vN}
\begin{split}
\frac{\partial \rho(x, x^\prime , t)}{\partial t} =& \frac{i}{\epsilon}\left\{\frac{1}{|\mathcal{A}|}\int \exp{\left[-\frac{\mathcal{H}(x, x^{\prime\prime})\epsilon}{\hbar} \right]}\rho(x^{\prime\prime},x^\prime, t)\mathrm{d} x^{\prime\prime}  - \rho(x, x^\prime , t) \right\} \\
& - \frac{i}{\epsilon}\left\{\frac{1}{|\mathcal{A}|}\int \exp{\left[-\frac{\mathcal{H}(x^\prime, x^{\prime\prime})\epsilon}{\hbar} \right]}\rho(x, x^{\prime\prime}, t)\mathrm{d} x^{\prime\prime}  - \rho(x, x^\prime , t) \right\}.
\end{split}
\ee
\end{widetext}

Clearly, the terms $\rho(x, x^\prime , t)$ in the right hand side cancel out, and we can write Eq.~\eqref{e:Gaussian_vN} in compact form as
\be\label{e:GaussianKernel_vN}
\frac{\partial \rho}{\partial t} = \frac{i}{\epsilon}\left[(\mathcal{K}\ast\rho)-(\rho\ast\mathcal{K})\right],
\ee
where we have introduced the kernel 
\be\label{e:K_T+V}
\mathcal{K}(x, x^{\prime}) = \frac{1}{|\mathcal{A}|}\exp{\left[-\frac{\mathcal{H}(x, x^{\prime})\epsilon}{\hbar} \right]},
\ee
and the convolutions 
\begin{widetext}
\BE
\left[\mathcal{K}\ast\rho\right](x,x^\prime, t) &= \frac{1}{|\mathcal{A}|}\int \exp{\left[-\frac{\mathcal{H}(x, x^{\prime\prime})\epsilon}{\hbar} \right]}\rho(x^{\prime\prime},x^\prime, t)\mathrm{d} x^{\prime\prime}, \\
\left[\rho\ast\mathcal{K}\right](x^\prime , x, t)  &=\frac{1}{|\mathcal{A}|}\int \exp{\left[-\frac{\mathcal{H}(x^\prime, x^{\prime\prime})\epsilon}{\hbar} \right]}\rho(x, x^{\prime\prime}, t)\mathrm{d} x^{\prime\prime}.
\EE
\end{widetext}
to represent, respectively, the first and second integrals in the right hand side of Eq.~\eqref{e:Gaussian_vN}. Notice that the integration variables in $(\mathcal{K}\ast\rho)$ and $(\rho\ast\mathcal{K})$ are, respectively, the first and second arguments of $\rho$, which yields the analogous of left and right matrix multiplication. 

\

\subsection{Particle in an electromagnetic field via asymmetric real kernels}\label{s:App_EM}

\subsubsection{Prelude: Hermitian kernels via complex Hamiltonian functions}

The Schr\"odinger equation of a particle of charge $e$ interacting with an electromagnetic field can be written as 
\be\label{e:SchrodingerEM}
\begin{split}
i\hbar\frac{\partial \psi(\bx,t)}{\partial t}=& -\frac{\hbar^2}{2m}\left(\nabla - i \frac{e}{\hbar c}\mathbf{A}\right)^2\psi(\bx,t)\\
 &+ e V(\bx,t)\psi(\bx,t),\\
\end{split}
\ee
%

\noindent where $\bx$ denotes the position vector in three dimensional space, while $V$ and $\mathbf{A}$ denote the scalar and vector fields respectively.  Notice that the Hamiltonian associated to Eq.~\eqref{e:SchrodingerEM} now contains an imaginary part given by the terms linear in $\mathbf{A}$ arising from the expansion of ${(\nabla - i e\mathbf{A} /\hbar c)^2\psi(\bx,t)}$.

Equation~\eqref{e:SchrodingerEM} can be derived via the path integral formulation from the extension of Eq.~\eqref{e:path} to three-dimensional space (i.e. by doing the substitution $x\to\bx$)
\begin{equation}\label{e:path}
\psi(\bx_{\ell+1}, t+\epsilon) = \frac{1}{\mathcal{A}_{EM}}\int\exp{\left[\frac{i}{\hbar}S_{EM}(\bx_{\ell+1},\bx_\ell)\right]}\psi(\bx_\ell,t)\mathrm{d}^3 \bx_\ell,
\end{equation}
with action
\begin{widetext}
\be\label{e:actionEM}
S_{EM}(\bx_{\ell+1}, \bx_\ell) = \frac{m\epsilon}{2}\left(\frac{\bx_{\ell + 1}- \bx_\ell}{\epsilon} \right)^2 + \frac{e \epsilon}{c}\left(\frac{\mathbf{x}_{\ell+1} - \mathbf{x}_{\ell}}{\epsilon}\right)\cdot\mathbf{A}\left(\frac{\bx_{\ell+ 1 }+\bx_\ell}{2},t\right) - \epsilon V\left(\frac{\bx_{\ell+ 1 }+\bx_\ell}{2},t\right),  
\ee
\end{widetext}
and 
\be\label{e:A_EM}
\mathcal{A} \to\mathcal{A}_{EM} \equiv (i 2\pi\hbar\epsilon / m)^{3/2};
\ee
here we are using a midpoint discretization for the action.

As in the previous section, it is possible to derive Eq.~\eqref{e:SchrodingerEM} by doing a replacement similar to that in Eq.~\eqref{e:replacement}, i.e.
\begin{widetext}
\be\label{e:replacementEM}
\frac{1}{\mathcal{A}_{EM}}\exp{\left[\frac{i}{\hbar}S_{EM}(\bx,\bx^\prime)\right]}\to\frac{1}{|\mathcal{A}_{EM}|} \exp{\left[-\frac{\epsilon}{\hbar}\widetilde{\mathcal{H}}_{EM}(\bx, \bx^\prime)\right] },
\ee
\end{widetext}
but with a complex Hamiltonian function
\begin{widetext}
\be\label{e:classicalH_EM}
\widetilde{\mathcal{H}}_{EM}(\bx, \bx^\prime) = \frac{m}{2}\left(\frac{\bx - \bx^\prime}{\epsilon}\right)^2 + V\left(\frac{\bx+\bx^\prime}{2}, t\right) - i\frac{e}{c}\left(\frac{\bx-\bx^\prime}{\epsilon}\right)\cdot\mathbf{A}\left(\frac{\bx+\bx^\prime}{2}, t\right).
\ee
\end{widetext}
As with Eq.~\eqref{e:replacement}, this corresponds to a Wick rotation $\epsilon\to -i\epsilon$. We can recognize that the real part is the three-dimensional version of the Hamiltonian function defined in Eq.~\eqref{e:classicalH}. We could think of $\mathcal{H}_{EM}$ as a Hamiltonian function with a complex interaction energy whose real and imaginary parts correspond to the electric and magnetic fields, respectively. This seems similar to writing the electromagnetic field as $\mathbf{E} + i\mathbf{B}$ which allows to write Maxwell's equations in compact form (see e.g. Eq. (7.10) in Ref.~\cite{doran2003geometric}). 

As in the previous section, we can derive Eq.~\eqref{e:SchrodingerEM} from an equation analogous to Eq.~\eqref{e:GaussianSchrodinger}, i.e. 
\begin{widetext}
\be\label{e:GaussianSchrodingerEM}
\epsilon\frac{\partial\psi(\bx_{\ell+1}, t)}{\partial t} = i \left\{\frac{1}{|\mathcal{A}_{EM}|}\int \exp{\left[-\frac{\epsilon}{\hbar}\widetilde{\mathcal{H}}_{EM}(\bx_{\ell + 1}, \bx_\ell) \right]}\psi(\bx_\ell, t)\mathrm{d}^3 \bx_\ell  - \psi(\bx_{\ell + 1}, t) \right\}.
\ee
\end{widetext}
Indeed, as in the previous section, the Gaussian factor in the complex kernel (see Eqs.\eqref{e:A_EM} and~\eqref{e:classicalH_EM})
\be\label{e:C}
\mathcal{C}(\bx , \bx^\prime) = \exp{\left[-\frac{\epsilon}{\hbar}\widetilde{\mathcal{H}}_{EM}(\bx, \bx^\prime) \right]} / {|\mathcal{A}_{EM}|},
\ee
associated to the kinetic term in Eq.~\eqref{e:classicalH_EM} allows us to expand the other factors in the integral in Eq.~\eqref{e:GaussianSchrodingerEM} around $\bx_{\ell +1}$ up to second order in $|\bx_{\ell +1} - \bx_\ell|$ or to first order in $\epsilon$. More precisely, by introducing the variable ${\bu = \bx_{\ell+1}-\bx_\ell}$, so ${(\bx_{\ell+1}+\bx_\ell)/2 = \bx_{\ell + 1} - \bu/2}$ as well as ${\bx_\ell = \bx_{\ell+1} - \bu}$, and doing ${\bx = \bx_{\ell+1}}$, $\bx^\prime = \bx_\ell$ to avoid cluttering the equations with indexes, we can write

\begin{widetext}
\be\label{e:C*psi}
[\mathcal{C}\ast\psi](\bx , t) =\frac{1}{|\mathcal{A}_{EM}|}\int \exp\left(-\frac{m \bu^2}{2 \hbar\epsilon } \right) f(\bx, \bu, t)\left[\psi(\bx ,t) - \bu\cdot\nabla\psi(\bx ,t) + \frac{1}{2}\bu\cdot \mathbf{H}\psi(\bx, t)\cdot\bu \right] + O(\epsilon^2),
\ee
\end{widetext}
where the convolution $\mathcal{C} \ast\psi$ denotes the integral in the right hand side of Eq.~\eqref{e:GaussianSchrodingerEM}, $\mathbf{H}\psi$ stands for the Hessian or matrix of second derivatives of $\psi$. Furthermore, the function

\begin{widetext}
\be
f(\bx, \bu, t) = \left\{1-\frac{\epsilon}{\hbar}V(\bx, t) + i\frac{e}{\hbar c}\bu\cdot\mathbf{A}(\bx,t) - i\frac{e}{2\hbar c}\bu\cdot\nabla\mathbf{A}(\bx,t)\cdot\bu -\frac{1}{2}\left[\frac{e}{\hbar c}\bu\cdot\mathbf{A}(\bx, t)\right]^2 \right\},
\ee
\end{widetext}
is the expansion up to first order in $\epsilon$ or second order in $\bu$ of the exponential factors in the complex kernel $\mathcal{C}$, which correspond to the interaction terms in Eq.~\eqref{e:classicalH_EM}, i.e. those containing $V$ and $\mathbf{A}$.
\newpage

Taking into account that the first two moments of $\bu$ are $\bra u_j\ket_\bu = 0$ and $\bra u_j u_k\ket_\bu = \delta_{j k}\hbar\epsilon/m$, where $\langle\cdot\rangle_\bu$ refers to the average taken with the Gaussian ${\exp(-m\bu^2/2\hbar\epsilon)/|\mathcal{A}_{EM}|}$, and that terms containing $\epsilon \bu^2$ and $\bu^3$ or higher can be neglected, the integral in Eq.~\eqref{e:C*psi} yields 
\begin{widetext}
\be
[\mathcal{C}\ast\psi](\bx, t) = \left(1-\frac{\epsilon}{\hbar} V\right)\psi +\frac{\hbar\epsilon}{2 m}\nabla^2\psi - i\frac{e\epsilon}{mc}\mathbf{A}\cdot\nabla\psi - i\frac{e \epsilon}{2 m c}\nabla\cdot\mathbf{A} \psi - \frac{e^2\epsilon}{2 \hbar m c^2 }\mathbf{A}^2\psi
\ee
\end{widetext}

Furthermore, taking into account that
\be 
\begin{split}
\left(\nabla - i\frac{e}{\hbar c}\mathbf{A}\right)^2\psi =& \nabla^2\psi - \left(\frac{e}{\hbar c}\right)^2\mathbf{A}^2\psi - \\
& i \frac{e}{\hbar c}\left[2 \mathbf{A}\cdot\nabla\psi + (\nabla\cdot\mathbf{A})\psi\right],
\end{split}
\ee
we obtain
\begin{widetext}
\be\label{e:C*psi_final}
[\mathcal{C}\ast\psi](\bx , t) =\psi(\bx ,t) + \frac{\epsilon}{\hbar}\left[\frac{\hbar^2}{2 m}\left(\nabla - i\frac{e}{\hbar c}\mathbf{A}\right)^2 \psi(\bx ,t) - V(\bx, t)\psi(\bx ,t)\right] + O(\epsilon^2).
\ee
\end{widetext}

So, we can indeed write the Schr\"odinger equation Eq.~\eqref{e:SchrodingerEM} as 
\be\label{e:partial}
\epsilon\frac{\partial\psi}{\partial t} = i [\mathcal{C}\ast\psi - \psi],
\ee
which is the analogous of Eq.~\eqref{e:GaussianSchrodinger}. To see this, we can replace $\mathcal{C}\ast\psi$ in Eq.~\eqref{e:partial} by the right hand side of Eq.~\eqref{e:C*psi_final}, cancel out the $\psi$ terms, and multiply the remaining equation by $i\hbar/\epsilon$, which yields Eq.~\eqref{e:SchrodingerEM}.

By doing the same analysis that led from Eq.~\eqref{e:GaussianSchrodinger} to Eqs.~\eqref{e:Gaussian_vN}~and~\eqref{e:GaussianKernel_vN} we get
\be\label{e:C*rho}
\frac{\partial\rho}{\partial t} = \frac{i}{\epsilon} [\mathcal{C}\ast\rho - \rho\ast\mathcal{C}]\equiv \frac{i}{\epsilon}[\mathcal{C}, \rho] .
\ee
Notice, however, that contrary to the kernel $\mathcal{K}$ in Eq.~\eqref{e:GaussianKernel_vN}, which is real and symmetric, the kernel $\mathcal{C}$ in Eq.~\eqref{e:C*rho} is still complex and asymmetric. Indeed, from Eqs.\eqref{e:classicalH_EM}~and~\eqref{e:C} we can see that by exchanging the two arguments of $\mathcal{C}$ we get
\be\label{e:C_asymmetric}
\mathcal{C}(\bx^\prime ,\bx ) = [\mathcal{C}(\bx ,\bx^\prime )]^\ast,
\ee
which, considering $\mathcal{C}$ as a matrix, also shows that $\mathcal{C}$ is Hermitean. We could also see that $\mathcal{C}$ is Hermitian by writing Eq.~\eqref{e:C} as $\mathcal{C} = \mathcal{K}_s + i\mathcal{K}_a$ with
\begin{widetext}
\BE
\mathcal{K}_s(\bx , \bx^\prime) &=& \frac{1}{|\mathcal{A}_{EM}|}\exp{\left[-\frac{m({\bx - \bx^\prime})^2}{2\hbar\epsilon} -\frac{\epsilon}{\hbar} V\left(\frac{\bx+\bx^\prime}{2}, t\right) \right]}\cos\left[\frac{e}{\hbar c}\left({\bx-\bx^\prime}\right)\cdot\mathbf{A}\left(\frac{\bx+\bx^\prime}{2}, t\right)\right],\label{e:Ks} \\
\mathcal{K}_a(\bx , \bx^\prime) &=& \frac{1}{|\mathcal{A}_{EM}|}\exp{\left[-\frac{m({\bx - \bx^\prime})^2}{2\hbar\epsilon} -\frac{\epsilon}{\hbar} V\left(\frac{\bx+\bx^\prime}{2}, t\right) \right]}\sin\left[\frac{e}{\hbar c}\left({\bx-\bx^\prime}\right)\cdot\mathbf{A}\left(\frac{\bx+\bx^\prime}{2}, t\right)\right], \label{e:Ka}
\EE
\end{widetext}
which are clearly symmetric and antisymmetric, respectively, due to the symmetry properties of the cosinusoidal and sinusoidal functions.

\subsubsection{Postlude: real asymmetric kernels via real Hamiltonian functions}

Since the kernel $\mathcal{C}$ defined in Eq.~\eqref{e:C} is Hermitian (see Eq.~\eqref{e:C_asymmetric}), Eq.~\eqref{e:C*rho} has the same structure of Eq.~\eqref{e:vN}. So, as described in Sec.~\ref{s:vNreal} (see also remark 1 therein), we can write Eq.~\eqref{e:C*rho} as a pair of real equations in terms of a real kernel $\mathcal{K} = \mathcal{K}_s + \mathcal{K}_a$, which in this case is given by the sum of the right hand sides of Eqs.~\eqref{e:Ks}~and~\eqref{e:Ka},
\begin{widetext}
\be
\mathcal{K}(\bx , \bx^\prime) = \frac{1}{|\mathcal{A}_{EM}|}\exp{\left[-\frac{m({\bx - \bx^\prime})^2}{2\hbar\epsilon} -\frac{\epsilon}{\hbar} V\left(\frac{\bx+\bx^\prime}{2}, t\right) \right]}[\cos z + \sin z],\label{e:Kpre}
\ee
\end{widetext}
where 
\be\label{e:z}
z = \frac{\epsilon }{\hbar }\frac{e}{ c}\left(\frac{\bx-\bx^\prime}{\epsilon}\right)\cdot\mathbf{A}\left(\frac{\bx+\bx^\prime}{2}, t\right).
\ee

Due to the very sharp Gaussian factor (since ${\epsilon\to 0}$), we can expand the sinusoidal and cosinusoidal functions up to second order in their argument since the rest gives contributions of order higher than $\epsilon$. Now, up to second order we have 
\be\label{e:cossinexp}
\cos z + \sin z = \exp(z-z^2) + O(z^3).
\ee

Eqs.~\eqref{e:Ks}-\eqref{e:z} shows that, even in the case of a charged particle in an electromagnetic field, whose Hamiltonian operator is complex (and so non-stoquastic), can be thought of as arising from a real kernel ${\mathcal{K}=\mathcal{K}_s+\mathcal{K}_a}$. More precisely, following Sec.~\ref{s:vNreal} the corresponding von Neumann equation can be understood as a pair of real matrix equations
\BE
\frac{\partial P_A}{\partial t} &=& -\frac{1}{\epsilon}[\mathcal{K}_a, P_A] + \frac{1}{\epsilon}[\mathcal{K}_s, P_B],\\
\frac{\partial P_B}{\partial t} &=& -\frac{1}{\epsilon}[\mathcal{K}_a, P_B] - \frac{1}{\epsilon}[\mathcal{K}_s, P_A],
\EE

\

\noindent where the two probability matrices satisfy $P_A = P$, $P_B = P^T$, and ${\rho = (P + P^T)/2 + i (P-P^T)/2}$. As we argue later these equations can be interpreted in probabilistic terms. 

However, the term $z^2$ in Eq.~\eqref{e:cossinexp} leads via Eq.~\eqref{e:z} to a term with a quadratic factor $(\epsilon / \hbar)^2$, which prevents us from writing $\mathcal{K}\propto\exp{(-\epsilon\mathcal{H}_{\rm EM}/\hbar)}$ with a Hamiltonian function $\mathcal{H}_{\rm EM} $ independent of $\epsilon$ and $\hbar$ (cf. Eq.~\eqref{e:C}). It is possible to go around this issue, though, by noticing that (remember that $\bu = \bx-\bx^\prime$, so $(\bx + \bx^\prime)/2 = \bx - \bu/2$)
\begin{widetext}
\be\label{e:averages}
\bra z^2\ket_\bu = \left(\frac{e}{\hbar c}\right)^2\bra\left[\bu\cdot\mathbf{A}\left(\bx-\frac{\bu}{2}, t\right)\right]^2\ket_\bu = \frac{\epsilon}{\hbar}\frac{e^2}{ m c^2}[\mathbf{A}(\bx, t)]^2 + O(\epsilon^2),
\ee
\end{widetext}
where $\langle\cdot\rangle_\bu$ refers to the average taken with the Gaussian ${\exp(-m\bu^2/2\hbar\epsilon)/|\mathcal{A}_{EM}|}$, and we have used the results $\bra u_j\ket_\bu = 0$ and $\bra u_j u_k\ket_\bu = (\hbar\epsilon / m) \delta_{j k} $. 

Since the first term in the right hand side of Eq.~\eqref{e:averages} is already of order $\epsilon$, we can safely replace
\be
z^2 \to \frac{\epsilon}{\hbar}\frac{e^2}{m c^2}\left[\mathbf{A}\left(\frac{\bx+\bx^\prime}{2}, t\right)\right]^2 + O(\epsilon^2),
\ee
so
\be
e^{-z^2}\to\exp\left(-\frac{\epsilon}{\hbar}\frac{e^2}{ m c^2} \mathbf{A}^2\right),
\ee
in Eq.~\eqref{e:cossinexp}, since when we expand in $\bu$ only the term independent of $\bu$ remains as the other terms in the expansion lead to terms of higher order in $\epsilon$.

So, we can indeed write the asymmetric real kernel ${\mathcal{K}=\mathcal{K}_s+\mathcal{K}_a}$ in Eq.~\eqref{e:Kpre} associated to the complex kernel $\mathcal{C}$ in Eq.~\eqref{e:C} as
\be\label{e:KrealEM}
\mathcal{K}(\bx , \bx^\prime) = \frac{1}{|\mathcal{A}_{EM}|}\exp{\left[-\frac{\epsilon}{\hbar}\mathcal{H}_{\rm EM}(\bx , \bx^\prime) \right]}
\ee
where the {\it real} electromagnetic Hamiltonian function (with no tilde) is given by
\begin{widetext}
\be\label{e:HrealEM}
\begin{split}
\mathcal{H}_{\rm EM}(\bx , \bx^\prime) &= \frac{m}{2}\left(\frac{\bx - \bx^\prime}{\epsilon}\right)^2 + V\left(\frac{\bx+\bx^\prime}{2}, t\right) - \frac{e}{c}\left(\frac{\bx-\bx^\prime}{\epsilon}\right)\cdot\mathbf{A}\left(\frac{\bx+\bx^\prime}{2}, t\right) +\frac{e^2}{ m c^2} \left[\mathbf{A}\left(\frac{\bx+\bx^\prime}{2}, t\right)\right]^2 
\end{split}
\ee
\end{widetext}

It is no completely clear at this point, though, how to interpret $\mathcal{H}_{\rm EM}$ defined in Eq.~\eqref{e:HrealEM} nor the real kernel $\mathcal{K}$ defined in Eq.~\eqref{e:KrealEM}. It seems to suggests a probabilistic interpretation of electromagnetic phenomena. We leave this for future work. 

\

\section{Effective kernels with negative entries}\label{s:neg_prob}

The derivation of genuine quantum-like dynamics in the main text (see, e.g., Secs.~\ref{s:imaginary_vN} and \ref{s:quantum}) was built on factors $F_\ell$ with non-negative entries. In principle, quantum mechanics does not have this restriction. Here we will argue that our approach is not restricted to factors with non-negative entries either.  

We will discuss the well-known example of a non-relativistic atom in a radiation field. This is an infinite-dimensional quantum system that, following the approach discussed in Sec.~\ref{s:QMrecasted} and Appendix~\ref{s:details}, can originally be described by factors with non-negative entries. After a standard truncation of the original model to its first two energy levels, however, it turns into an effective two-dimensional system described by factors with negative entries. The latter is known as a two-level atom interacting with a coherent radiation field~\cite{haken2005physics} (see Sec.~15.3 therein). Nevertheless, it is the former system which is actually implemented in the lab, where the experimenter has to make sure certain near resonance conditions are met for the effective two-dimensional model to be a good approximation of the actual system. 

This discussion is restricted to non-relativistic quantum mechanics. In relativistic quantum mechanics there may be some intrinsic sources of negative numbers in probabilistic expressions. We leave the study of relativistic quantum mechanics for future work.

Consider therefore the Hamiltonian of an atom modelled as an electron, described by the momentum operator $i\hbar\nabla_{\mathbf{r}}$, moving in a potential field $V(r)$ produced by the nucleus,
\be\label{e:H0_atom}
H_0 = -\frac{\hbar^2}{2m}\nabla_{\mathbf{r}}^2 + V(r) = \sum_{n=0}^\infty E_n \left.| n\ket\bra n\right|.
\ee
In the second equality we have expanded the Hamiltonian in terms of its eigenvalues $E_n$ and its eigenvectors $\left| n\ket$, where $n\in\mathbb{Z}$, $n\geq 0$. 

Now consider a perturbation 
\be\label{e:U(t)}
U(\mathbf{r},t) = e \mathbf{r}\cdot\mathbf{E}(t) = e\mathbf{r}\cdot\mathbf{E}_0\cos(\omega t),
\ee
where $\mathbf{E}(t) = \mathbf{E}_0\cos(\omega t)$ is an oscillating electric field with frequency $\omega$ and amplitude $\mathbf{E}_0$.  The perturbed Hamiltonian, ${H = H_0 + U}$, can be derived via a path integral with Lagrangian 
\be\label{e:L2}
L = \frac{m}{2}\dot{\mathbf{r}} - V(r) - U(\mathbf{r},t).
\ee

Following the approach discussed in Sec.~\ref{s:QMrecasted} and Appendix~\ref{s:details}, we can also derive $H$ via a real non-negative kernel 
\be
\mathcal{K}_\epsilon(\mathbf{r}^\prime, \mathbf{r}) = e^{-\epsilon\mathcal{H}(\mathbf{r}^\prime, \mathbf{r})/\hbar}\geq 0,\label{e:K2}
\ee
where the Hamiltonian function
\be
\mathcal{H}(\mathbf{r}^\prime, \mathbf{r}) = \frac{m}{2}\left(\frac{\mathbf{r}^\prime-\mathbf{r}}{\epsilon}\right)^2 + V\left(\frac{|\mathbf{r}^\prime + \mathbf{r}|}{2}\right) + U(t),\label{e:HK2}
\ee
is essentially the Wick rotation $\epsilon\to -i\epsilon$ of the Lagrangian $L$ in Eq.~\eqref{e:L2} (appart from an irrelevant global minus sign).

We will now see that, after a standard truncation of the full Hamiltonian $H=H_0+U$ (see Eqs.~\eqref{e:H0_atom} and \eqref{e:U(t)}) into an effective two-level system, we lose the equivalence with the positive kernel given by Eqs.~\eqref{e:K2} and \eqref{e:HK2}. Indeed, in the derivation of the Hamiltonian of a two-level atom it is usually assumed that the perturbation defined in Eq.~\eqref{e:U(t)} is near resonance with two relevant energy levels of the Hamiltonian $H_0$, say $E_0$ and $E_1$, i.e. {$|\omega - \omega_0|\lll\omega_0$}, where ${\hbar\omega_0 = E_1 - E_0}$. In this case, it is usually assumed that only the dynamics of these two energy levels matter. So, we can write
\be\label{e:H_R}
\begin{split}
H =& E_0\left| 0 \ket\bra 0\right| + E_1\left| 1 \ket\bra 1\right| + U_{01}\left(\left| 0 \ket\bra 1\right| + \left| 1 \ket\bra 0\right|\right)\\
&+ H_\mathcal{R},
\end{split}
\ee
where the first three terms in the right hand side of Eq.~\eqref{e:H_R} correspond to the transitions taking place within the subspace spanned by $\{\left|0\ket ,\left|1\ket\}$, and 
\be\label{e:Residual}
\begin{split}
H_\mathcal{R} =& \sum_{n=2}^\infty E_n \left.| n\ket\bra n\right|\\
&+ \sum_{m=0}^\infty\sum_{n> m, n\neq 0, 1}^\infty \left(U_{mn}\left| m \ket\bra n\right| + U_{nm}\left| n \ket\bra m\right|\right),
\end{split}
\ee
collects all the remaining transitions. Here we have written 
\be 
U_{mn} = U_{nm}^\ast =  \bra m\right| U(t) \left| n \ket,
\ee
for $m, n\in\mathbb{Z}$, $m,n\geq 0$. For simplicity, we are restricting here to the case where $U_{01}=U^\ast_{01}$ and $U_{00} = U_{11} = 0$~\cite{haken2005physics} (see Sec.~15.3 therein); this explains the form of Eq.~\eqref{e:H_R}.

At this point it is argued that we can neglect $H_\mathcal{R}$ since the system is near resonance.  This yields the effective two-level Hamiltonian 
\be\label{e:H2eff}
H_{\rm eff} = \overline{E}\id^{(01)} -\frac{\hbar\omega_0}{2} \sigma_Z^{(01)} + D\cos(\omega t)\sigma_X^{01}, 
\ee
where $\overline{E}=(E_0+E_1)/2$, $D = \bra 0\right| \mathbf{r}\cdot\mathbf{E}_0\left| 1 \ket$, and
\BE
\id^{(01)} &=& \left| 0 \ket\bra 0\right| + \left| 1 \ket\bra 1\right|,\\
\sigma_X^{(01)} &=& \left| 0 \ket\bra 1\right| + \left| 1 \ket\bra 0\right|,\\
\sigma_Z^{(01)} &=& \left| 0 \ket\bra 0\right| - \left| 1 \ket\bra 1\right|.
\EE

If we now try to write this as a real kernel in the way we did in Sec.~\ref{s:QMrecasted} and Appendix~\ref{s:details}
\be\label{e:F_eff_atom}
F_{\rm eff} = \id - \epsilon J_{\rm eff},
\ee
with $J_{\rm eff} = H_{\rm eff}/\hbar$, we end up with off-diagonal negative entries due to the factor $\cos(\omega t)$ accompanying $\sigma_X$ in Eq.~\eqref{e:H2eff}. So, the full Hamiltonian $H=H_0 + U$ in Eq.~\eqref{e:H_R} can be represented in terms of the real positive kernel given by  Eqs.~\eqref{e:K2} and \eqref{e:HK2}, but the truncated effective Hamiltonian $H_{\rm eff}$ in Eq.~\eqref{e:H2eff} cannot. What happened? The effective factor 
\be\label{e:F_R_atom}
F_{\rm eff} =  F - \mathcal{R}
\ee
in Eq.~\eqref{e:F_eff_atom} can be interpreted as the full factor $F = \id-\epsilon H/\hbar$ associated to kernel $\mathcal{K}_\epsilon $ (see Eqs.~\eqref{e:K2} and \eqref{e:H_R}), which has only positive entries, relative to the reference factor $\mathcal{R}= -\epsilon H_\mathcal{R}/\hbar$.

Key in this example is the neglect of part of the original system, which is then completely forgot. If we keep in mind this fact, we can interpret the negative entries as indicators that we are describing probabilistic expressions---e.g., probability distributions, transition probabilities, or factors---relative to a similar reference probabilistic expression~\cite{realpe2019negative}. The presence of negative numbers in general probabilistic expressions do not need to be an issue. It is in principle possible to provide an operational interpretation to linear algebra manipulations of probabilistic expressions without the usual restrictions that all numbers involved must be in the interval $[0,1]$ nor that all probabilistic expressions need be properly normalized~\cite{burgin2010interpretations,abramsky2014operational,realpe2019negative}. Some of these ideas have recently led~\cite{realpe2019quantum} to classical quantum-inspired algorithms for stochastic simulation that require much less memory than the best classical algorithms known to date.

\section{Phase-less quantum-like formulation of Markov processes}\label{s:Q-MP}
Here we show in more detail that standard Markov processes can be written as imaginary-time quantum mechanics, but in terms of phaseless wave functions. 

\subsection{Time-symmetric evolution equations}
To begin let us notice that using Bayes rule in Eq.~\eqref{e:MarkovBayes} we can change $\mathcal{P}^-_\ell$ in Eq.~\eqref{e:K_P+*P-} for $\mathcal{P}^+_\ell$, i.e.
\be\label{e:KEQM+}
\begin{split}
K_\ell(x_\ell, x_{\ell+1}) =& \mathcal{P}_\ell^+(x_{\ell+1}|x_\ell)\sqrt{\frac{p_\ell(x_\ell)}{p_{\ell+1}(x_{\ell+1})}}\\
=& \mathcal{P}_\ell^+(x_{\ell+1}|x_\ell){\frac{\theta_\ell(x_\ell)}{\theta_{\ell+1}(x_{\ell+1})}},
\end{split}
\ee
which is the analogous of Eq. (2.12) in Ref.~\cite{Zambrini-1987}, where $K_\ell$ here plays the role of $h$ there. So, since $p_{\ell+1}(x_{\ell+1})= \int \mathcal{P}_\ell^+(x_{\ell+1}|x_\ell)p_\ell(x_\ell) \mathrm{d} x_\ell$, we have
\be\label{e:quantum-like}
\theta_{\ell+1}(x_{\ell+1})=\int K_\ell(x_\ell, x_{\ell+1})\theta_\ell(x_\ell)\mathrm{d} x_\ell.
\ee
which has the same {\em structure} of Eq.~\eqref{e:path} in Sec.~\ref{s:path} (see also Eq.~\eqref{e:GaussianSchrodinger}); indeed, after carrying out a Wick rotation ($t\to i t$) and restricting to (real) Wick-rotated wave functions with no phase, i.e. ${\psi_\ell(x_\ell) = \sqrt{p_\ell(x_\ell)}}$, Eq.~\eqref{e:path} can be considered as an instance of Eq.~\eqref{e:quantum-like}.

Notice that, like the real kernels we obtained in Sec.~\ref{s:QMrecasted} (see e.g. Eqs.~\eqref{e:Main_K_T+V} and~\eqref{e:Main_KrealEM}), the kernel $K_i$ above, which is defined in Eq.~\eqref{e:K_P+*P-}, is in general {\em not} normalized. Indeed, except for special cases like ${\mathcal{P}^+_{i}(x_{\ell+1}|x_\ell) = \mathcal{P}^-_{\ell}(x_{\ell}|x_{\ell+1})} $, when ${K_\ell(x_\ell, x_{\ell+1}) = \mathcal{P}^+_{\ell}(x_{\ell+1}|x_\ell)}$ is actually a probability distribution, the integral of the square root of the product of two {\em different} probability distributions is {\em not} one in general. We will discuss this further in Appendix~\ref{s:KnotNormalized}, where we will also show there is a close analogy with quantum mechanics, more specifically with the final remark in Sec.~\ref{s:path}.

Let us first notice that by changing $\mathcal{P}^-_\ell$ in Eq.~\eqref{e:K_P+*P-} for $\mathcal{P}^+_\ell$ we obtain 
\be\label{e:KEQM-}
\begin{split}
K_\ell(x_\ell, x_{\ell+1}) =& \mathcal{P}_\ell^-(x_{\ell+1}|x_\ell)\sqrt{\frac{p_{\ell+1}(x_{\ell+1})}{p_\ell(x_\ell)}}\\
=& \mathcal{P}_\ell^-(x_\ell|x_{\ell+1}){\frac{\theta^\ast_{\ell+1}(x_{\ell+1})}{\theta^\ast_\ell(x_\ell)}},
\end{split}
\ee
where we have introduced the asterisc notation $\theta^\ast_\ell$ just to emphasize that this equation is related to the backward transition probability $\mathcal{P}^-_\ell$. Indeed, this leads to the the reversed version of Eq.~\eqref{e:quantum-like}, i.e.
\be\label{e:quantum-like*}
\theta^\ast_{\ell}(x_{\ell})=\int K_\ell(x_\ell, x_{\ell+1})\theta^\ast_{\ell+1}(x_{ \ell+1})\mathrm{d} x_\ell,
\ee
which is the analogous of Eq. (2.11) in Ref.~\cite{Zambrini-1987}. 

In this particular case we have $\theta_\ell = \theta^\ast_\ell$, so we can equally write $p_\ell(x) = \theta^2_\ell(x) = \theta^\ast_\ell(x)\theta_\ell(x)$. We will emphasize the latter, however, which can be interpreted as the product of the solutions of an initial value problem $\theta_t(x)$ and a final value problem $\theta^\ast_\ell(x)$, much as the situation in Eqs.~\eqref{e:mu->*<-mu}, \eqref{e:EQBP->} and~\eqref{e:EQBP<-}.

\subsection{Time-symmetric kernels are not normalized}\label{s:KnotNormalized}

To see this analogy more closely, let us work with the continuous time version of Eq.~\eqref{e:KEQM+}, i.e. let us change $\ell\to t$ and $\ell+1\to t+\epsilon$ with $\epsilon\to 0$. Furthermore, since $\epsilon$ is very small $\mathcal{P}^+_{t+\epsilon| t}(x^\prime|x)$ is very close to a Dirac delta $\delta(x^\prime-x)$, so we can expand the fraction $\theta_t(x)/\theta_{t+\epsilon}(x^\prime)$ around $t$ and $x = x^\prime - \xi$ to yield
\begin{widetext}
\be\label{e:intK}
\int K_t(x, x^\prime)\mathrm{d} x = 1 - \epsilon\left\{\frac{\dot{\theta}_t(x)}{\theta_t(x)} + \frac{D_t^+(x)}{2}\frac{\theta_t^{\prime\prime}(x)}{\theta_t(x)} + b_t^+(x)\frac{\theta_t^{\prime}(x)}{\theta_t(x)} - D_t^+(x)\left[\frac{\theta_t^{\prime}(x)}{\theta_t(x)}\right]^2\right\} + O(\epsilon^2),
\ee
\end{widetext}
where 
\BE
b_t^+(x) &=& \lim_{\epsilon\to 0}\frac{1}{\epsilon}\int \xi \mathcal{P}^+_{t+\epsilon | t}(x+\xi | x) \mathrm{d}\xi,\label{e:drift+}\\
D_t^+(x) &=& \lim_{\epsilon\to 0}\frac{1}{\epsilon}\int \xi^2 \mathcal{P}^+_{t+\epsilon|t}(x+\xi | x) \mathrm{d}\xi \label{e:diff+},
\EE
are the forward drift and difussion terms respectively; here we are representing derivatives with respect to $x$ and $t$ with prime and dot operators respectively. 

Let us now consider the special case where the diffusion coefficeint ${D_t^+(x) = D}$ is constant and the forward drift satisfies ${b_t^+(x) = D\theta^{\prime}_t(x)/\theta_t(x)}$ (cf. Eq.~(2.12') in Ref.~\cite{Zambrini-1987}; notice the prime), which will be motivated and further discussed below in Appendix~\ref{s:Q-MP_symmetry}. In this special case the last two terms in Eq.~\eqref{e:intK} cancel out and we get
\be\label{e:intKepsilon}
\int K_t(x, x^\prime)\mathrm{d} x = 1 - \epsilon V_t(x)/ D,
\ee
where we have defined the function
\be
V_t(x) = D\frac{\dot{\theta}_t(x)}{\theta_t(x)} + \frac{D^2}{2}\frac{\theta_t^{\prime\prime}(x)}{\theta_t(x)},
\ee
which can be rewritten as
\be\label{e:imaginarySchrodinger}
D{\dot{\theta}_t(x)} = - \frac{D^2}{2}\theta_t^{\prime\prime}(x) + V_t(x) {\theta_t(x)}. 
\ee
Eq.~\eqref{e:imaginarySchrodinger} is the exact analogous of the (adjoint) imaginary-time Schr\"odinger equation (see e.g. Eqs.~(2.1)~and~(2.17) in Ref.~\cite{Zambrini-1987}), and similar also to Eq.~\eqref{e:intK_T+V} in the final remark of Sec.~\ref{s:path}.

\subsection{Future-past symmetry and quantum-like Markov processes}\label{s:Q-MP_symmetry}
Consider a Markov process with distribution $\mathcal{P}$, such that the joint distribution of the past and future states, $x^\prime$ and $x^{\prime\prime}$ respectively, conditioned on the present state $x$ is symmetric, i.e.
\be
\mathcal{P}^+_{t+\epsilon|t}(x^{\prime\prime} | x)\mathcal{P}^-_{t-\epsilon|t}(x^\prime | x) = \mathcal{P}^+_{t+\epsilon|t}(x^{\prime} | x)\mathcal{P}^-_{t-\epsilon|t}(x^{\prime\prime} | x) .
\ee
Assume such a distribution can be written as the symmetric part of a generic Markov process $Q$, i.e.
\begin{widetext}
\be\label{e:future-past}
\mathcal{P}^+_{t+\epsilon|t}(x^{\prime\prime} | x)\mathcal{P}^-_{t-\epsilon|t}(x^\prime | x) = \frac{1}{2}\left[Q^+_{t+\epsilon|t}(x^{\prime\prime}|x)Q^-_{t-\epsilon|t}(x^{\prime}|x) + Q^+_{t+\epsilon|t}(x^{\prime}|x)Q^-_{t-\epsilon|t}(x^{\prime\prime}|x)\right],
\ee
\end{widetext}
where $Q^+$ and $Q^-$ refer to the corresponding forward and backward transition probabilities of the generic Markov process. Notice the exchange of $x^\prime$ and $x^{\prime\prime}$ in the two terms in the right hand side. 

Marginalizing Eq.~\eqref{e:future-past} over $x^\prime$ we get
%
\be \label{e:P+}
\mathcal{P}^+_{t+\epsilon|t}(x^{\prime\prime} | x) = \frac{1}{2}\left[Q^+_{t+\epsilon|t}(x^{\prime\prime}|x) + Q^-_{t-\epsilon|t}(x^{\prime\prime}|x)\right],
\ee
and marginalizing over Eq.~\eqref{e:future-past} over $x^\prime$ instead we get
\be\label{e:P+=P-}
\mathcal{P}^-_{t-\epsilon|t}(x^{\prime} | x) = \mathcal{P}^+_{t+\epsilon|t}(x^{\prime} | x).
\ee

Using Eqs.~\eqref{e:drift+} and \eqref{e:P+} we see that 
\be\label{e:b+=c+-c-}
b^+_t(x) = \frac{1}{2}\left[c^+_t(x) - c^-_t(x) \right],
\ee
where
\BE 
c^+_t(x) &=& \lim_{\epsilon\to 0}\frac{1}{\epsilon}\int (x^\prime-x) Q^+_{t+\epsilon|t}(x^\prime | x)\mathrm{d}x^\prime,\label{e:c+_DEF}\\
c^-_t(x) &=& \lim_{\epsilon\to 0}\frac{1}{\epsilon}\int (x-x^\prime) Q^-_{t-\epsilon|t}(x^\prime | x)\mathrm{d}x^\prime,\label{e:c-_DEF}
\EE
are the forward and backward drifts corresponding to Markov process $Q$. It is known that the difference between forward and backward drifts satisfy (see e.g. Eq.~(30) in Ref~\cite{caticha2009entropic,caticha2011entropic} and references therein).  
\be\label{e:c+-c-}
c^+_t(x) - c^-_t(x) = D\frac{q_t^\prime(x)}{q_t(x)},
\ee
where $q_t(x)$ is the single variable marginal associated to the generic Markov process described by $Q^+$ and $Q^-$, and $q^\prime_t(x)$ its derivative with respect to $x$.

In Eq.~\eqref{e:c+-c-} we can replace $q_t(x)$ by the single variable marginal $p_t(x)$ corresponding to the process described by distributions $\mathcal{P}^+$ and $\mathcal{P}^-$, up to a correction that vanishes when $\epsilon\to 0$, i.e. ${p_t(x) = q_t(x) + O(\epsilon)}$. Indeed, multiplying Eq.~\eqref{e:P+} by $q_t(x)$ and using Bayes rule to invert $Q^+$ and $Q^-$ we get
\begin{widetext}
\be
\mathcal{P}^+_{t+\epsilon|t}(x^{\prime} | x) q_t(x) = \frac{1}{2}\left[Q^-_{t|t+\epsilon}(x|x^\prime) q_{t+\epsilon}(x^\prime)  + Q^+_{t|t-\epsilon}(x|x^\prime) q_{t-\epsilon}(x^\prime) \right].
\ee
\end{widetext}
By expanding in $t$ the functions $q_{t-\epsilon}(x^\prime)$ around $t+\epsilon$ and $Q^+_{t|t-\epsilon}(x|x^\prime)$ around $t+2\epsilon$ we obtain 
\begin{widetext}
\be
\mathcal{P}^+_{t+\epsilon|t}(x^{\prime} | x) q_t(x) = \frac{1}{2}\left[Q^-_{t|t+\epsilon}(x|x^\prime)   + Q^+_{t+\epsilon|t}(x|x^\prime) \right] q_{t+\epsilon}(x^\prime) + O(\epsilon) = \mathcal{P}^-_{t|t+\epsilon}(x|x^{\prime}) q_{t+\epsilon}(x^\prime) + O(\epsilon),
\ee
\end{widetext}
which is Bayes rule for $\mathcal{P}^+$ and $\mathcal{P}^-$ up to an error term of order $\epsilon$.

So, replacing $q_t$ by $p_t$ in Eq.~\eqref{e:c+-c-}, Eq.~\eqref{e:b+=c+-c-} can be written as
\be
b^+(x) = \frac{D}{2}\frac{p^\prime_t(x)}{p_t(x)} = D\frac{\theta^\prime_t(x)}{\theta_t(x)},
\ee
where $\theta_t(x) = \sqrt{p_t(x)}$, which is the analogous of Eq.~(2.12') in Ref.~\cite{Zambrini-1987}. Notice also that Eq.~\eqref{e:P+=P-} implies that the backward drift, obtained from Eq.~\eqref{e:drift+} by replacing the integrand ${(x^\prime - x) \mathcal{P}^+_{t+\epsilon}(x^\prime|x)}$ with ${(x-x^\prime ) \mathcal{P}^-_{t-\epsilon}(x^\prime|x)}$ (see Eqs.~\eqref{e:c+_DEF} and \eqref{e:c-_DEF}), is given by 
\be
b^-_{t}(x) = -b^+_t(x) = -D{\theta^\ast}^\prime_t(x)/\theta^\ast_t(x).
\ee
(Cf. Eqs. (2.11') and (2.12') in Ref.~\cite{Zambrini-1987}.) Here we have used $\theta^\ast_t$ instead of $\theta_t=\theta^\ast_t$ to emphasize that we are dealing with the backward process. Such a distinction becomes relevant when we deal with cycles rather than chains, since then a phase arises and ${\theta_t\neq\theta_t^\ast}$ in general.
%



\newpage 


\end{document}